\documentclass[longauth,traditabstract]{aa} 
\usepackage{amsmath}
\usepackage{amssymb}
\usepackage{longtable}
\usepackage{multirow}
\usepackage{overpic}
\usepackage{array}
\usepackage{changepage}

\def\setsymbol#1#2{\expandafter\def\csname #1\endcsname{#2}}
\def\getsymbol#1{\csname #1\endcsname}

\def\Planck{\textit{Planck}}

\newbox\tablebox    \newdimen\tablewidth
\def\leaderfil{\leaders\hbox to 5pt{\hss.\hss}\hfil}
\def\endPlancktable{\tablewidth=\columnwidth 
    $$\hss\copy\tablebox\hss$$
    \vskip-\lastskip\vskip -2pt}
\def\endPlancktablewide{\tablewidth=\textwidth 
    $$\hss\copy\tablebox\hss$$
    \vskip-\lastskip\vskip -2pt}
\def\tablenote#1 #2\par{\begingroup \parindent=0.8em
    \abovedisplayshortskip=0pt\belowdisplayshortskip=0pt
    \noindent
    $$\hss\vbox{\hsize\tablewidth \hangindent=\parindent \hangafter=1 \noindent
    \hbox to \parindent{$^#1$\hss}\strut#2\strut\par}\hss$$
    \endgroup}
\def\doubleline{\vskip 3pt\hrule \vskip 1.5pt \hrule \vskip 5pt}

\def\L2{\ifmmode L_2\else $L_2$\fi}

\def\DeltaT{\ifmmode \Delta T\else $\Delta T$\fi}
\def\deltat{\ifmmode \Delta t\else $\Delta t$\fi}
\def\fknee{\ifmmode f_{\rm knee}\else $f_{\rm knee}$\fi}
\def\Fmax{\ifmmode F_{\rm max}\else $F_{\rm max}$\fi}
\def\solar{\ifmmode{\rm M}_{\mathord\odot}\else${\rm M}_{\mathord\odot}$\fi}
\def\Msolar{\ifmmode{\rm M}_{\mathord\odot}\else${\rm M}_{\mathord\odot}$\fi}
\def\Lsolar{\ifmmode{\rm L}_{\mathord\odot}\else${\rm L}_{\mathord\odot}$\fi}

\def\inv{\ifmmode^{-1}\else$^{-1}$\fi}
\def\mo{\ifmmode^{-1}\else$^{-1}$\fi}
\def\sup#1{\ifmmode ^{\rm #1}\else $^{\rm #1}$\fi}
\def\expo#1{\ifmmode \times 10^{#1}\else $\times 10^{#1}$\fi}
\def\,{\thinspace}
\def\lsim{\mathrel{\raise .4ex\hbox{\rlap{$<$}\lower 1.2ex\hbox{$\sim$}}}}
\def\gsim{\mathrel{\raise .4ex\hbox{\rlap{$>$}\lower 1.2ex\hbox{$\sim$}}}}

\def\simprop{\mathrel{\raise .4ex\hbox{\rlap{$\propto$}\lower 1.2ex\hbox{$\sim$}}}}
\def\deg{\ifmmode^\circ\else$^\circ$\fi}
\def\pdeg{\ifmmode $\setbox0=\hbox{$^{\circ}$}\rlap{\hskip.11\wd0 .}$^{\circ}
          \else \setbox0=\hbox{$^{\circ}$}\rlap{\hskip.11\wd0 .}$^{\circ}$\fi}
\def\arcs{\ifmmode {^{\scriptstyle\prime\prime}}
          \else $^{\scriptstyle\prime\prime}$\fi}
\def\arcm{\ifmmode {^{\scriptstyle\prime}}
          \else $^{\scriptstyle\prime}$\fi}
\newdimen\sa  \newdimen\sb
\def\parcs{\sa=.07em \sb=.03em
     \ifmmode \hbox{\rlap{.}}^{\scriptstyle\prime\kern -\sb\prime}\hbox{\kern -\sa}
     \else \rlap{.}$^{\scriptstyle\prime\kern -\sb\prime}$\kern -\sa\fi}
\def\parcm{\sa=.08em \sb=.03em
     \ifmmode \hbox{\rlap{.}\kern\sa}^{\scriptstyle\prime}\hbox{\kern-\sb}
     \else \rlap{.}\kern\sa$^{\scriptstyle\prime}$\kern-\sb\fi}
\def\ra[#1 #2 #3.#4]{#1\sup{h}#2\sup{m}#3\sup{s}\llap.#4}
\def\dec[#1 #2 #3.#4]{#1\deg#2\arcm#3\arcs\llap.#4}
\def\deco[#1 #2 #3]{#1\deg#2\arcm#3\arcs}
\def\rra[#1 #2]{#1\sup{h}#2\sup{m}}

\def\dots{\relax\ifmmode \ldots\else $\ldots$\fi}
\def\WHzsr{\ifmmode $W\,Hz\mo\,sr\mo$\else W\,Hz\mo\,sr\mo\fi}
\def\mHz{\ifmmode $\,mHz$\else \,mHz\fi}
\def\GHz{\ifmmode $\,GHz$\else \,GHz\fi}
\def\mKs{\ifmmode $\,mK\,s$^{1/2}\else \,mK\,s$^{1/2}$\fi}
\def\muKs{\ifmmode \,\mu$K\,s$^{1/2}\else \,$\mu$K\,s$^{1/2}$\fi}
\def\muKRJs{\ifmmode \,\mu$K$_{\rm RJ}$\,s$^{1/2}\else \,$\mu$K$_{\rm RJ}$\,s$^{1/2}$\fi}
\def\muKHz{\ifmmode \,\mu$K\,Hz$^{-1/2}\else \,$\mu$K\,Hz$^{-1/2}$\fi}
\def\MJysr{\ifmmode \,$MJy\,sr\mo$\else \,MJy\,sr\mo\fi}
\def\MJysrmK{\ifmmode \,$MJy\,sr\mo$\,mK$_{\rm CMB}\mo\else \,MJy\,sr\mo\,mK$_{\rm CMB}\mo$\fi}
\def\microns{\ifmmode \,\mu$m$\else \,$\mu$m\fi}

\def\muK{\ifmmode \,\mu$K$\else \,$\mu$\hbox{K}\fi}
\def\microK{\ifmmode \,\mu$K$\else \,$\mu$\hbox{K}\fi}
\def\muW{\ifmmode \,\mu$W$\else \,$\mu$\hbox{W}\fi}
\def\kms{\ifmmode $\,km\,s$^{-1}\else \,km\,s$^{-1}$\fi}
\def\kmsMpc{\ifmmode $\,\kms\,Mpc\mo$\else \,\kms\,Mpc\mo\fi}
\usepackage{natbib}
\bibpunct{(}{)}{;}{a}{}{,} %
\usepackage{graphicx}
\usepackage{txfonts}

\usepackage[table,usenames,dvipsnames]{xcolor}
\usepackage{fixltx2e}
\definecolor{RoyalPurple}{cmyk}{0.75,0.9,0,0.1}
\definecolor{MyBlue}{rgb}{0.025,0.025,0.8} 
\definecolor{MyBlue2}{rgb}{0.05,0.025,0.8} 
\usepackage[breaklinks, colorlinks, citecolor=MyBlue, linkcolor=MyBlue2, urlcolor=RoyalPurple, 
            colorlinks=true, debug, baseurl=' ']{hyperref}

\newcommand{\cm}[0]{cm$^{-1}$}

\makeatletter
\newif\ifref
\@ifclasswith{aa}{referee}{\reftrue}{\reffalse}
\makeatother

\newcommand{\reftabstrt}{\ifref \begin{adjustwidth}{3.5cm}{}\else \begin{center}\fi}
\newcommand{\reftabend}{\ifref \end{adjustwidth}\else \end{center}\fi}

\begin{document}
%
\author{\small
Planck Collaboration:
P.~A.~R.~Ade\inst{82}
\and
N.~Aghanim\inst{57}
\and
C.~Armitage-Caplan\inst{86}
\and
M.~Arnaud\inst{69}
\and
M.~Ashdown\inst{66, 6}
\and
F.~Atrio-Barandela\inst{18}
\and
J.~Aumont\inst{57}
\and
C.~Baccigalupi\inst{81}
\and
A.~J.~Banday\inst{89, 10}
\and
R.~B.~Barreiro\inst{63}
% ...
}
\institute{\small
APC, AstroParticule et Cosmologie, Universit\'{e} Paris Diderot, CNRS/IN2P3, CEA/lrfu, Observatoire de Paris, Sorbonne Paris Cit\'{e}, 10, rue Alice Domon et L\'{e}onie Duquet, 75205 Paris Cedex 13, France\\
\and
Aalto University Mets\"{a}hovi Radio Observatory, Mets\"{a}hovintie 114, FIN-02540 Kylm\"{a}l\"{a}, Finland\\
\and
African Institute for Mathematical Sciences, 6-8 Melrose Road, Muizenberg, Cape Town, South Africa\\
\and
Agenzia Spaziale Italiana Science Data Center, Via del Politecnico snc, 00133, Roma, Italy\\
\and
Agenzia Spaziale Italiana, Viale Liegi 26, Roma, Italy\\
\and
Astrophysics Group, Cavendish Laboratory, University of Cambridge, J J Thomson Avenue, Cambridge CB3 0HE, U.K.\\
\and
Astrophysics \& Cosmology Research Unit, School of Mathematics, Statistics \& Computer Science, University of KwaZulu-Natal, Westville Campus, Private Bag X54001, Durban 4000, South Africa\\
\and
Atacama Large Millimeter/submillimeter Array, ALMA Santiago Central Offices, Alonso de Cordova 3107, Vitacura, Casilla 763 0355, Santiago, Chile\\
\and
CITA, University of Toronto, 60 St. George St., Toronto, ON M5S 3H8, Canada\\
\and
CNRS, IRAP, 9 Av. colonel Roche, BP 44346, F-31028 Toulouse cedex 4, France\\
\and
California Institute of Technology, Pasadena, California, U.S.A.\\
\and
Centre for Theoretical Cosmology, DAMTP, University of Cambridge, Wilberforce Road, Cambridge CB3 0WA, U.K.\\
\and
Centro de Estudios de F\'{i}sica del Cosmos de Arag\'{o}n (CEFCA), Plaza San Juan, 1, planta 2, E-44001, Teruel, Spain\\
\and
Computational Cosmology Center, Lawrence Berkeley National Laboratory, Berkeley, California, U.S.A.\\
\and
DSM/Irfu/SPP, CEA-Saclay, F-91191 Gif-sur-Yvette Cedex, France\\
\and
DTU Space, National Space Institute, Technical University of Denmark, Elektrovej 327, DK-2800 Kgs. Lyngby, Denmark\\
\and
D\'{e}partement de Physique Th\'{e}orique, Universit\'{e} de Gen\`{e}ve, 24, Quai E. Ansermet,1211 Gen\`{e}ve 4, Switzerland\\
\and
Departamento de F\'{\i}sica Fundamental, Facultad de Ciencias, Universidad de Salamanca, 37008 Salamanca, Spain\\
\and
Department of Astronomy and Astrophysics, University of Toronto, 50 Saint George Street, Toronto, Ontario, Canada\\
\and
Department of Astrophysics/IMAPP, Radboud University Nijmegen, P.O. Box 9010, 6500 GL Nijmegen, The Netherlands\\
\and
Department of Electrical Engineering and Computer Sciences, University of California, Berkeley, California, U.S.A.\\
\and
Department of Physics \& Astronomy, University of British Columbia, 6224 Agricultural Road, Vancouver, British Columbia, Canada\\
\and
Department of Physics and Astronomy, Dana and David Dornsife College of Letter, Arts and Sciences, University of Southern California, Los Angeles, CA 90089, U.S.A.\\
\and
Department of Physics and Astronomy, University College London, London WC1E 6BT, U.K.\\
\and
Department of Physics, Florida State University, Keen Physics Building, 77 Chieftan Way, Tallahassee, Florida, U.S.A.\\
\and
Department of Physics, Gustaf H\"{a}llstr\"{o}min katu 2a, University of Helsinki, Helsinki, Finland\\
\and
Department of Physics, Princeton University, Princeton, New Jersey, U.S.A.\\
\and
Department of Physics, University of California, One Shields Avenue, Davis, California, U.S.A.\\
\and
Department of Physics, University of California, Santa Barbara, California, U.S.A.\\
\and
Department of Physics, University of Illinois at Urbana-Champaign, 1110 West Green Street, Urbana, Illinois, U.S.A.\\
\and
Dipartimento di Fisica e Astronomia G. Galilei, Universit\`{a} degli Studi di Padova, via Marzolo 8, 35131 Padova, Italy\\
\and
Dipartimento di Fisica e Scienze della Terra, Universit\`{a} di Ferrara, Via Saragat 1, 44122 Ferrara, Italy\\
\and
Dipartimento di Fisica, Universit\`{a} La Sapienza, P. le A. Moro 2, Roma, Italy\\
\and
Dipartimento di Fisica, Universit\`{a} degli Studi di Milano, Via Celoria, 16, Milano, Italy\\
\and
Dipartimento di Fisica, Universit\`{a} degli Studi di Trieste, via A. Valerio 2, Trieste, Italy\\
\and
Dipartimento di Fisica, Universit\`{a} di Roma Tor Vergata, Via della Ricerca Scientifica, 1, Roma, Italy\\
\and
Discovery Center, Niels Bohr Institute, Blegdamsvej 17, Copenhagen, Denmark\\
\and
European Southern Observatory, ESO Vitacura, Alonso de Cordova 3107, Vitacura, Casilla 19001, Santiago, Chile\\
\and
European Space Agency, ESAC, Planck Science Office, Camino bajo del Castillo, s/n, Urbanizaci\'{o}n Villafranca del Castillo, Villanueva de la Ca\~{n}ada, Madrid, Spain\\
\and
European Space Agency, ESTEC, Keplerlaan 1, 2201 AZ Noordwijk, The Netherlands\\
\and
Helsinki Institute of Physics, Gustaf H\"{a}llstr\"{o}min katu 2, University of Helsinki, Helsinki, Finland\\
\and
INAF - Osservatorio Astrofisico di Catania, Via S. Sofia 78, Catania, Italy\\
\and
INAF - Osservatorio Astronomico di Padova, Vicolo dell'Osservatorio 5, Padova, Italy\\
\and
INAF - Osservatorio Astronomico di Roma, via di Frascati 33, Monte Porzio Catone, Italy\\
\and
INAF - Osservatorio Astronomico di Trieste, Via G.B. Tiepolo 11, Trieste, Italy\\
\and
INAF Istituto di Radioastronomia, Via P. Gobetti 101, 40129 Bologna, Italy\\
\and
INAF/IASF Bologna, Via Gobetti 101, Bologna, Italy\\
\and
INAF/IASF Milano, Via E. Bassini 15, Milano, Italy\\
\and
INFN, Sezione di Bologna, Via Irnerio 46, I-40126, Bologna, Italy\\
\and
INFN, Sezione di Roma 1, Universit\`{a} di Roma Sapienza, Piazzale Aldo Moro 2, 00185, Roma, Italy\\
\and
IPAG: Institut de Plan\'{e}tologie et d'Astrophysique de Grenoble, Universit\'{e} Joseph Fourier, Grenoble 1 / CNRS-INSU, UMR 5274, Grenoble, F-38041, France\\
\and
IUCAA, Post Bag 4, Ganeshkhind, Pune University Campus, Pune 411 007, India\\
\and
Imperial College London, Astrophysics group, Blackett Laboratory, Prince Consort Road, London, SW7 2AZ, U.K.\\
\and
Infrared Processing and Analysis Center, California Institute of Technology, Pasadena, CA 91125, U.S.A.\\
\and
Institut N\'{e}el, CNRS, Universit\'{e} Joseph Fourier Grenoble I, 25 rue des Martyrs, Grenoble, France\\
\and
Institut Universitaire de France, 103, bd Saint-Michel, 75005, Paris, France\\
\and
Institut d'Astrophysique Spatiale, CNRS (UMR8617) Universit\'{e} Paris-Sud 11, B\^{a}timent 121, Orsay, France\\
\and
Institut d'Astrophysique de Paris, CNRS (UMR7095), 98 bis Boulevard Arago, F-75014, Paris, France\\
\and
Institute for Space Sciences, Bucharest-Magurale, Romania\\
\and
Institute of Astronomy and Astrophysics, Academia Sinica, Taipei, Taiwan\\
\and
Institute of Astronomy, University of Cambridge, Madingley Road, Cambridge CB3 0HA, U.K.\\
\and
Institute of Theoretical Astrophysics, University of Oslo, Blindern, Oslo, Norway\\
\and
Instituto de F\'{\i}sica de Cantabria (CSIC-Universidad de Cantabria), Avda. de los Castros s/n, Santander, Spain\\
\and
Jet Propulsion Laboratory, California Institute of Technology, 4800 Oak Grove Drive, Pasadena, California, U.S.A.\\
\and
Jodrell Bank Centre for Astrophysics, Alan Turing Building, School of Physics and Astronomy, The University of Manchester, Oxford Road, Manchester, M13 9PL, U.K.\\
\and
Kavli Institute for Cosmology Cambridge, Madingley Road, Cambridge, CB3 0HA, U.K.\\
\and
LAL, Universit\'{e} Paris-Sud, CNRS/IN2P3, Orsay, France\\
\and
LERMA, CNRS, Observatoire de Paris, 61 Avenue de l'Observatoire, Paris, France\\
\and
Laboratoire AIM, IRFU/Service d'Astrophysique - CEA/DSM - CNRS - Universit\'{e} Paris Diderot, B\^{a}t. 709, CEA-Saclay, F-91191 Gif-sur-Yvette Cedex, France\\
\and
Laboratoire Traitement et Communication de l'Information, CNRS (UMR 5141) and T\'{e}l\'{e}com ParisTech, 46 rue Barrault F-75634 Paris Cedex 13, France\\
\and
Laboratoire de Physique Subatomique et de Cosmologie, Universit\'{e} Joseph Fourier Grenoble I, CNRS/IN2P3, Institut National Polytechnique de Grenoble, 53 rue des Martyrs, 38026 Grenoble cedex, France\\
\and
Laboratoire de Physique Th\'{e}orique, Universit\'{e} Paris-Sud 11 \& CNRS, B\^{a}timent 210, 91405 Orsay, France\\
\and
Lawrence Berkeley National Laboratory, Berkeley, California, U.S.A.\\
\and
Max-Planck-Institut f\"{u}r Astrophysik, Karl-Schwarzschild-Str. 1, 85741 Garching, Germany\\
\and
McGill Physics, Ernest Rutherford Physics Building, McGill University, 3600 rue University, Montr\'{e}al, QC, H3A 2T8, Canada\\
\and
National University of Ireland, Department of Experimental Physics, Maynooth, Co. Kildare, Ireland\\
\and
Niels Bohr Institute, Blegdamsvej 17, Copenhagen, Denmark\\
\and
Observational Cosmology, Mail Stop 367-17, California Institute of Technology, Pasadena, CA, 91125, U.S.A.\\
\and
Optical Science Laboratory, University College London, Gower Street, London, U.K.\\
\and
SB-ITP-LPPC, EPFL, CH-1015, Lausanne, Switzerland\\
\and
SISSA, Astrophysics Sector, via Bonomea 265, 34136, Trieste, Italy\\
\and
School of Physics and Astronomy, Cardiff University, Queens Buildings, The Parade, Cardiff, CF24 3AA, U.K.\\
\and
Space Sciences Laboratory, University of California, Berkeley, California, U.S.A.\\
\and
Special Astrophysical Observatory, Russian Academy of Sciences, Nizhnij Arkhyz, Zelenchukskiy region, Karachai-Cherkessian Republic, 369167, Russia\\
\and
Stanford University, Dept of Physics, Varian Physics Bldg, 382 Via Pueblo Mall, Stanford, California, U.S.A.\\
\and
Sub-Department of Astrophysics, University of Oxford, Keble Road, Oxford OX1 3RH, U.K.\\
\and
Theory Division, PH-TH, CERN, CH-1211, Geneva 23, Switzerland\\
\and
UPMC Univ Paris 06, UMR7095, 98 bis Boulevard Arago, F-75014, Paris, France\\
\and
Universit\'{e} de Toulouse, UPS-OMP, IRAP, F-31028 Toulouse cedex 4, France\\
% ...
}

   \title{\textit{Planck} 2013 results. IX. HFI spectral response}

%
 \date{Received 22 March, 2013; revised 26 September, 2013; accepted 25 November, 2013}



 \abstract{
The \Planck\ High Frequency Instrument (HFI) spectral response was determined through a series of ground based tests conducted with the HFI focal plane in a cryogenic environment prior to launch.  The main goal of the spectral transmission tests was to measure the relative spectral response (including the level of out-of-band signal rejection) of all HFI detectors to a known source of electromagnetic radiation individually. This was determined by measuring the interferometric output of a continuously scanned Fourier transform spectrometer with all HFI detectors.  As there is no on-board spectrometer within HFI, the ground-based spectral response experiments provide the definitive data set for the relative spectral calibration of the HFI\@.  Knowledge of the relative variations in the spectral response between HFI detectors allows for a more thorough analysis of the HFI data.  The spectral response of the HFI is used in \Planck\ data analysis and component separation, this includes extraction of CO emission observed within \Planck\ bands, dust emission, Sunyaev-Zeldovich sources, and intensity to polarization leakage.  The HFI spectral response data have also been used to provide unit conversion and colour correction analysis tools. 

While previous papers describe the pre-flight experiments conducted on the \Planck\ HFI, this paper focuses on the analysis of the pre-flight spectral response measurements and the derivation of data products, e.g. band-average spectra, unit conversion coefficients, and
colour correction coefficients, all with related uncertainties.  Verifications of the HFI spectral response data are provided through comparisons with photometric HFI flight data.  This validation includes use of HFI zodiacal emission observations to demonstrate out-of-band spectral signal rejection better than $10^8$. The accuracy of the HFI relative spectral response data is verified through comparison with complementary flight-data based unit conversion coefficients and colour correction coefficients.  These coefficients include those based upon HFI observations of CO, dust, and Sunyaev-Zeldovich emission.  General agreement is observed between the ground-based spectral characterization of HFI and corresponding in-flight observations, within the quoted uncertainty of each; explanations are provided for any discrepancies.}



\authorrunning{Planck Collaboration}
\titlerunning{\textit{Planck} HFI spectral response}
   \maketitle
%


\section{Introduction}
\label{sec:Intro}

This paper, one of a set associated with the 2013 release of data from the \Planck
\footnote{\Planck\ (\url{http://www.esa.int/Planck}) is a project of the European Space Agency (ESA) with instruments provided by two scientific consortia funded by ESA member states (in particular the lead countries France and Italy), with contributions from NASA(USA) and telescope reflectors provided by a collaboration between ESA and a scientific consortium led and funded by Denmark.}
mission \citep{planck2013-p01}, describes the determination and verification of the \Planck\ High Frequency Instrument (HFI) spectral response.  As the HFI employs a series of broad-band photometric receivers, an accurate understanding of the relative spectral response of each detector within a frequency channel, and that of each frequency channel within the instrument, is important in data processing and analysis \citep{planck2013-p03}; this is particularly important for \Planck\ component separation \citep{planck2013-p06} where the magnitude of foreground components is often much greater than the associated cosmic microwave background (CMB) signals and uncertainties.  

\begin{figure*}
\begin{centering}\includegraphics[width=180mm]{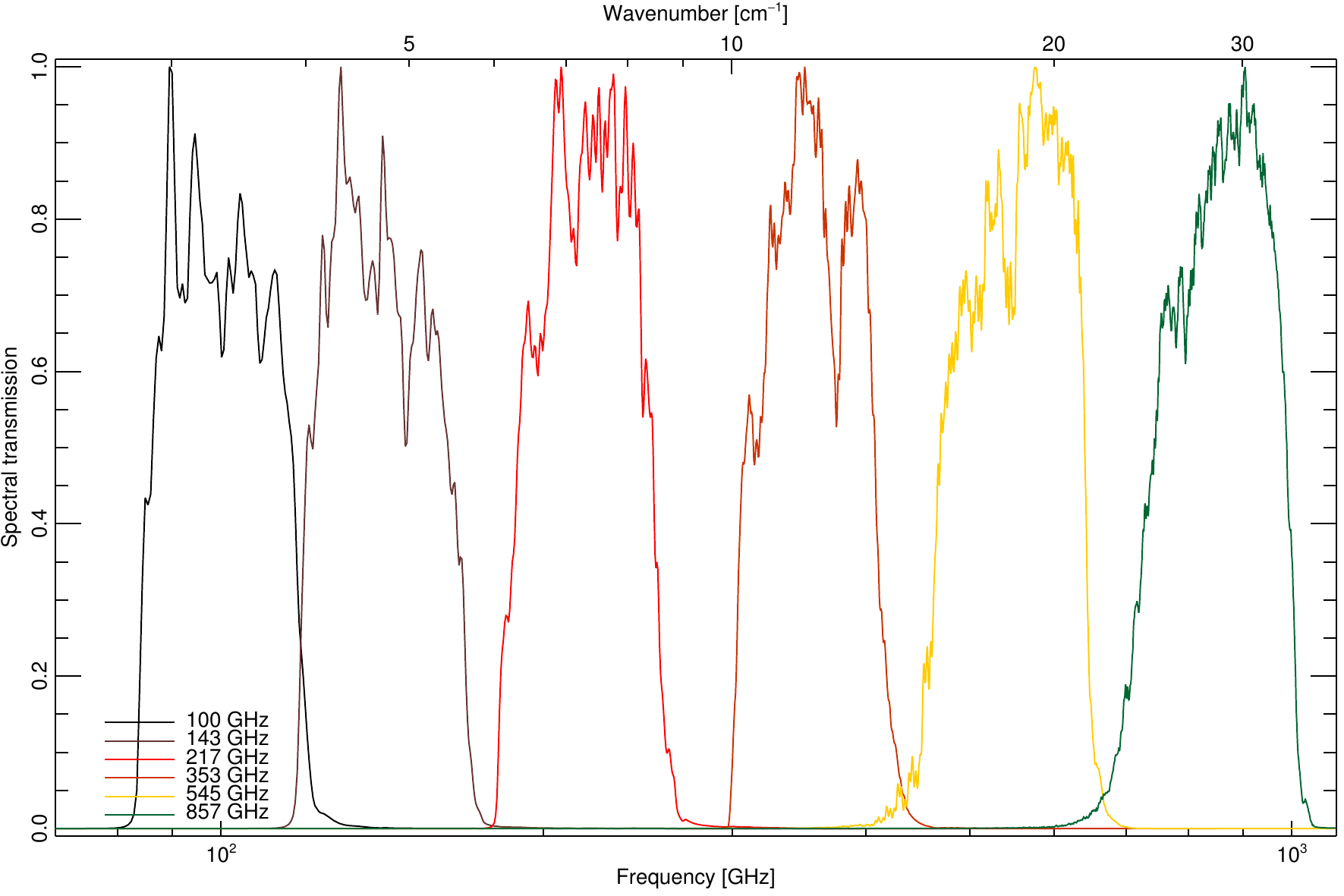}\par\end{centering}
\caption[HFI Band-average spectra]{\label{fig:AvgSpec} Band-average spectral transmission for each of the HFI frequency channels.}
\end{figure*}

The \Planck\ High Frequency Instrument (HFI) spectral response was determined through a series of ground based tests conducted with the HFI focal plane in a cryogenic environment prior to launch.  One of the main goals of pre-flight calibration testing was to measure the relative spectral response (including the level of out-of-band signal rejection) of all HFI detectors to a known source of electromagnetic (EM) radiation individually. This was determined by measuring the interferometric output of a continuously scanned Fourier transform spectrometer (FTS) with all HFI detection channels. As all in-flight HFI observations are photometric, i.e., there is no on-board spectrometer within HFI, the ground-based spectral response experiments provide the definitive HFI spectral calibration.  These ground-based spectral calibration results are then compared with in-flight photometric calibration observations to confirm the HFI spectral response. Pre-flight component level spectral characterization testing is described in detail in \cite{ade2010}.  The pre-flight system level spectral characterization testing is described in detail in \cite{pajot2010}.  This paper will discuss the testing itself as needed to provide context, but will primarily concentrate on the spectral characterization measurements, their analysis, and their utility within the HFI consortium and the \Planck\ legacy data archive (PLA)\footnote{see \url{http://www.sciops.esa.int/index.php?project=planck&page=Planck_Legacy_Archive}}. While many of the HFI detectors are polarization sensitive, the discussion in this paper is primarily limited to the intensity response of the HFI detectors.  Details concerning the polarization sensitivity of the HFI detectors are presented in \citep{rosset2010}.  Additional details of the HFI pre-flight calibration are found in \cite{lamarre2010} and \cite{maffei2010}.  The average spectrum for each of the HFI bands is illustrated in Figure \ref{fig:AvgSpec} (see Sect.~\ref{sec:avg} for details).  

An accurate understanding of the spectral response of HFI is critical in HFI data processing and analysis.  Information derived from the spectral calibration is important in many aspects of component separation (\citealt{planck2013-p06} and \citealt{planck2013-pip56}).  Unit conversion factors and colour corrections, which are important when dealing with signals of varying spectral profiles within a photometric channel, are derived for HFI using the transmission spectra.  Spectral mismatch between detectors within a given band must be understood in order to accurately interpret multi-detector averages and differences that otherwise yield systematic errors and increased uncertainty in data products.  This is especially important in the evaluation of weak components including polarized signal.

This paper presents the propagation of the raw spectral characterization data through its processing and analysis to yield detector level spectral response data in Sect.~\ref{sec:Obs}.  Sect.~\ref{sec:Res} then describes the analysis of the detector spectral transmission data to provide advanced spectral data products such as band-average spectra, and unit conversion and colour correction coefficients.  
Sect.~\ref{sec:Disc} presents the evaluation of the HFI spectral response and data products through comparisons with flight data.  This includes comparisons with HFI observations of zodiacal light \citep{planck2013-pip88}, CO emission \citep{planck2013-p03a}, Sunyaev-Zeldovich (SZ) sources (\citealt{planck2013-p05}, \citealt{planck2013-p05a}, and \citealt{planck2013-p05b}), and dust emission (\citealt{planck2013-p06b}, \citealt{planck2011-6.4b}, and \citealt{planck2011-7.0}). 

Details of the spectral response of the \Planck\ LFI instrument may be found in \cite{planck2013-p02} and in \cite{zonca2009}. Further details on technical aspects of the HFI spectral response data are also available in the \Planck\ Explanatory Supplement \citep{planck2013-p28}.  

\section{Measurements}
\label{sec:Obs}

This section outlines the pre-flight data required for the HFI spectral response determination.  The processing of the recorded data is outlined, and the resultant spectra are presented.
\subsection{Spectral Response Data Collection}
\label{sec:meas}

The \Planck\ HFI detector spectral response data were measured using a broadband mercury arc lamp radiation source, a polarizing FTS, an integrating sphere, and a rotating filter wheel, all coupled with the HFI focal plane in its evacuated cryostat.  A reference bolometer (at approximately 300\,mK) was mounted within the integrating sphere as an external measure of the radiation incident on HFI (see Fig.\ 3, \citealt{pajot2010}).  With this configuration, the entire HFI focal plane and the reference bolometer were exposed to the FTS modulated spectral signal synchronously.  The ratio of the HFI detector spectra with the reference bolometer spectra provides a relative spectral response with the systematics due to the test configuration removed.  The HFI spectral response measurements were collected during June and July of 2006 at the Institut d'Astrophysique Spatiale (IAS) laboratories in Orsay, France.  For each HFI detector, roughly 100 interferograms were recorded with consistent scan parameters to allow a spectral resolution of approximately 0.5\,GHz and a Nyquist frequency (\citealt{Nyquist1928} and \citealt{Shannon1948}) of approximately 12\,THz in the resultant spectra.  The rotating filter wheel was placed before the entrance to the integrating sphere with two settings used in these observations \citep{pajot2010}.  A 10\,\cm (approximately 300\,GHz)\footnote{The term wavenumber will be used to refer to units of \cm.} low-pass edge (LPE) filter was used for observations of the 100--217\,GHz detectors while a 36\,\cm (approximately 11\,THz) LPE filter was used for the 353--857\,GHz observations.  A more generic discussion of this filter technology is found in \cite{Ade2006Filters}. This LPE filtering being external to HFI and the integrating sphere exposed the reference bolometer and HFI detectors to the same input source while obtaining better reference bolometer performance over the low frequency channel spectral range.  Further details of the experimental setup, including diagrams of the FTS, integrating sphere, and HFI focal plane locations, are found in \cite{pajot2010}.  Additional relevant information is also found in \cite{ade2010}.

There were two significant additional tests in the derivation of the HFI detector spectral response beyond the scope of the IAS HFI FTS measurements.  Optical efficiency experiments, also conducted at IAS, provide optical efficiency estimates for each HFI detector.  When coupled with the FTS spectra this allows an estimate of the absolute spectral transmission.  More details on the optical efficiency tests are in \cite{ACATphd} and \cite{EFF}.  The other additional test is comprised of the filter measurements recorded at the Cardiff astronomical instrumentation group (AIG) facilities during filter stack production and verification \citep{ade2010}.  These measurements extend the IAS FTS spectral measurements far beyond the HFI spectral passband up to approximately 20\,THz\@.  



\subsection{Spectral Response Data Processing}
\label{sec:proc}

\begin{figure}
\begin{centering}\includegraphics[width=88mm]{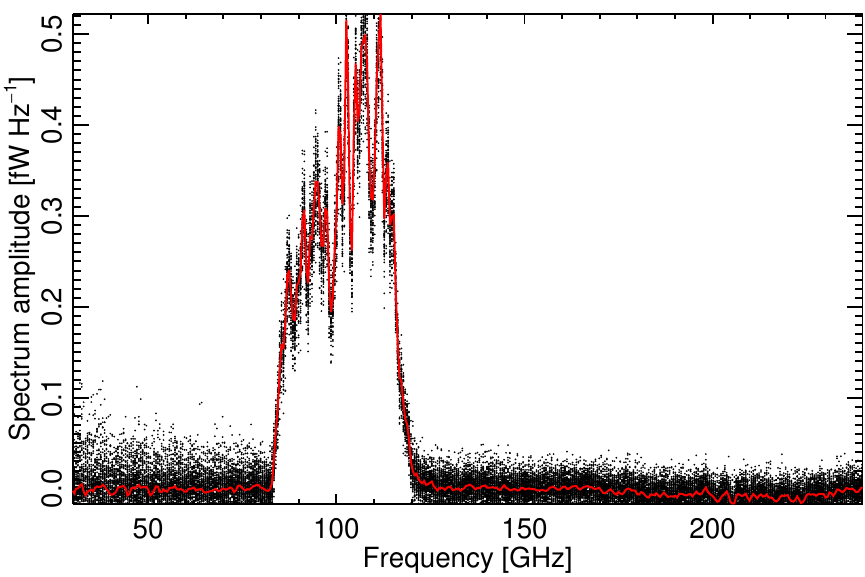}\par\end{centering}
\caption[Sample bolometer spectra]{\label{fig:SPECex} Sample bolometer spectra from HFI 100\,GHz detector 1a (the HFI detector naming scheme is discussed in Sect.~\ref{sec:Data}). The black dots represent data from individual spectra while the coloured solid line represents the average of approximately 100 individual spectra.}
\end{figure}

\begin{figure}
\centering
\includegraphics[width=88mm]{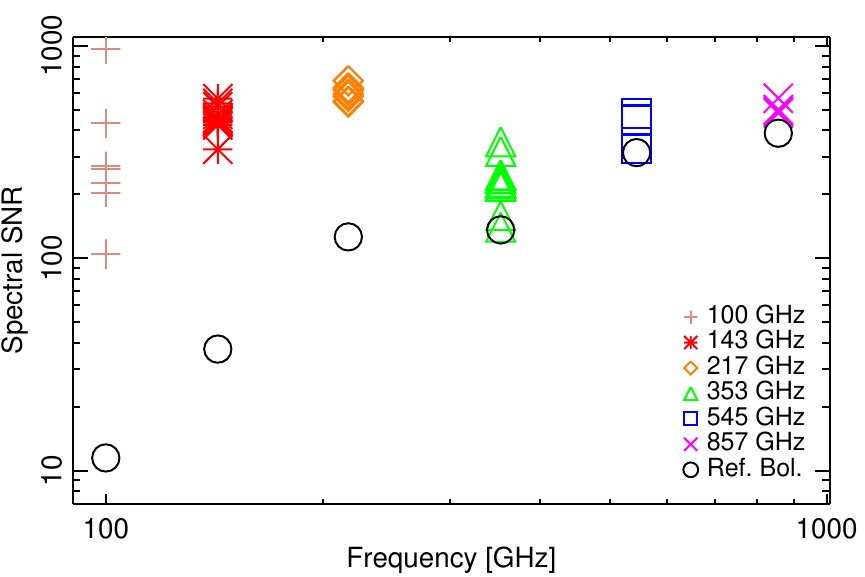}
\caption[HFI detector and reference bolometer SNR comparison]{\label{fig:SNall}HFI detector spectral SNR for the scan-averaged HFI detector spectra prior to taking the ratio with the scan-averaged reference spectra.  Also shown is the SNR for the scan-averaged reference bolometer spectra over the same spectral region.}
\end{figure}

The raw detector signals were combined with a bolometer model (\citealt{holmes2008}, \citealt{lamarre2010}, and \citealt{planck2011-1.7}) to both convert the signal into physical units and perform a detector nonlinearity correction \citep{SJVancouver}.  The recorded interferograms were processed and Fourier transformed individually (\citealt{Bell} and \citealt{davisFTS}), including phase correction (\citealt{For66} and \citealt{Brault87}) and apodization \citep{NaylorApod07}.  The resultant spectra were then averaged together to provide a mean and standard deviation for every independent spectral data point.  Similar analysis was conducted for the reference bolometer measurements. The ratio of each detector average spectrum against the corresponding reference bolometer average spectrum was taken to obtain the relative spectral transmission.  These ratio spectra represent the transmission of the entire HFI optical path tested, in its pseudo-flight configuration, including any standing waves within the feed horns, cold plate, focal plane unit, etc.; some of the structure observed within the in-band portion of the spectral response profiles demonstrates the presence of standing waves.  Fig.~\ref{fig:SPECex} demonstrates an example set of raw and average spectra for the HFI 100\,GHz 1a detector.  The signal-to-noise ratio (SNR) for the average spectra for each HFI detector and the relevant spectral region of the reference bolometer spectra are shown in Fig.\ \ref{fig:SNall}.  In terms of the noise of the HFI detector relative transmission spectra, the reference bolometer is the limiting case, especially for the 100\,GHz detectors.  It is important to note, however, that each HFI detector spectrum within a given frequency band is divided by the same reference bolometer spectrum.  Therefore, the \emph{relative} uncertainty between HFI detectors is indicated by the HFI-only marks of Fig.~\ref{fig:SNall}, even though the \emph{absolute} uncertainty is dictated by the limiting SNR of the reference bolometer. 
The detector relative transmission spectra are normalized to have a maximum value of unity, with the optical efficiency test results \citep{ACATphd} providing a multiplicative term to obtain the absolute spectral response, i.e., the product of the normalized spectral response and the optical efficiency factor provides an estimate of the absolute spectral response of a given detector.  As the reference bolometer accepts 2$\pi$\,sr.\ of incident radiation within the integrating sphere, the relative transmission spectrum for each HFI detector is also throughput
\footnote{Throughput, i.e., $A~\Omega$, is defined as the area ($A$) - solid angle ($\Omega$) product of a diffraction limited system \citep{BornWolf}.}
 normalized by virtue of the reference bolometer ratio.

\begin{figure}
\centering
\includegraphics[width=88mm]{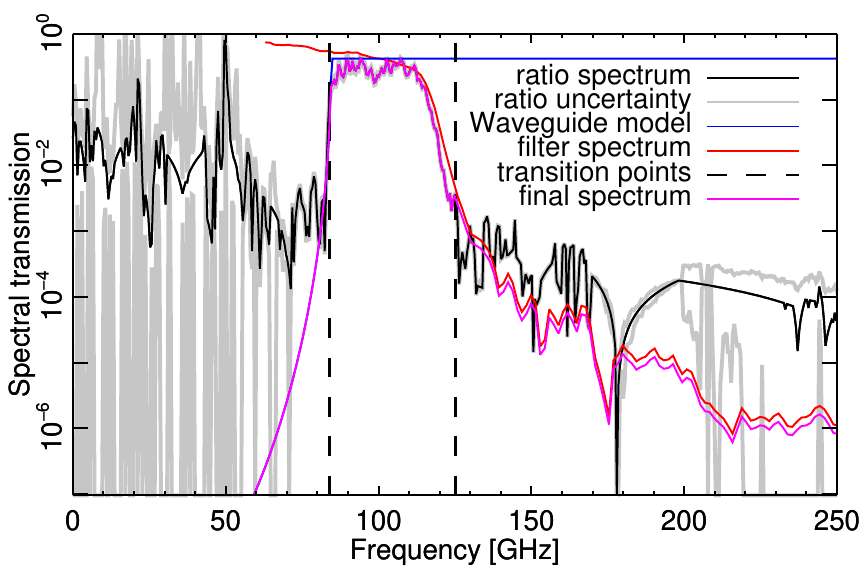}
\caption[Sample ratioed spectrum, waveguide model, and filter spectrum]{\label{fig:OOBex}Example ratioed spectrum (black), filter spectrum (red), waveguide model (blue), and final output spectral response (violet) for HFI 100 GHz detector 1a.  The transition frequencies are shown by the vertical dashed lines.  While the in-band spectral response is determined by the ratioed spectrum, the out-of-band transmission is determined by the waveguide model and filter data.}
\end{figure}

Once the detector and reference spectra ratios have been obtained, the out-of-band spectral regions are modified to improve the overall data quality.  A waveguide model is used for frequencies below the high-pass filter edge and component level filter stack spectra are used for frequencies above the low-pass edge of the pass-band. For the 545 and 857\,GHz channels, the filter spectra are used in conjunction with the waveguide model for frequencies below the high-pass edge. These ancilliary spectra are used where it is deemed to be of better quality than the IAS FTS measurements (i.e., for frequencies outside of the HFI detector pass-band).  The filter stacks for the low-pass edges of each frequency band are comprised of five filters.  There is an additional high-pass filter included in the 545 and 857\,GHz bands as the waveguide high-pass edge is too low for the desired multi-mode performance \citep{MurphyHorn}.  Further details on the HFI detector spectral filters are provided in \cite{ade2010}. Fig.~\ref{fig:OOBex} illustrates an example composite spectrum showing the relative FTS spectrum, the waveguide model, the component level filter spectra, and the corresponding spectral transition regions where the external spectra are spliced onto the in-band spectrum.  

As each detector signal is processed independently, the frequency sampling for a given detector may be slightly different than those within the same frequency channel.  To allow easier intra-channel comparisons, all spectra are interpolated onto a common frequency grid (i.e., one common frequency sampling per HFI channel). As all measurements within a frequency channel are conducted synchronously, and all detectors are referenced against the same reference bolometer, uncertainties introduced as a result of the common frequency interpolation are expected to be negligible.  The largest deviations in frequency sampling occur as a result of differences in band-edge location within a channel.  Thus, the transition frequency where the component level filter spectra (at a lower spectral resolution) are used, in place of the IAS spectra, varies within an HFI channel.  The common frequency sampling adopted selects the limiting case within a frequency band to avoid interpolation to an increased spectral resolution.  This may result in a minor degradation of the spectral resolution for some detectors, but avoids presenting data at a spectral resolution that is higher than that of the original measurement.\footnote{The CO interpolation, discussed below, is an exception to this.}

To assist with the use of the HFI spectral response data with HFI CO studies (see \citealt{planck2013-p03a}), spectral regions within the HFI bands with sensitivity to CO emission have been identified \citep{JPLspec}. These interpolation regions have also been extended to include the frequency range of other CO isotopes (i.e., CO, $^{13}$CO, C$^{17}$O, and C$^{18}$O) over a range of radial velocities (i.e., $\pm 300 $km\,s$^{-1}$). For each HFI detector, the spectral response data in the CO-sensitive regions (see Table \ref{tab:CO}) is interpolated by a factor of ten using the FTS instrument spectral line shape (ILS, \citealt{Bell}).  While this interpolation provides a more accurate estimate of the spectral transmission than, e.g., linear interpolation, it is important to note that this step does not increase the fundamental spectral resolution of the data. Thus, an interpolation flag has been created to indicate whether a data point within the correcponding spectral profile is original or interpolated.  

The material presented in this section is in supplement to that of \cite{ade2010} and \cite{pajot2010}.  Following the steps outlined above, spectral response profiles are created for each of the HFI detectors.
%
%
%
%

\begin{table}[tmb]                 
\begin{center}
\begingroup
\newdimen\tblskip \tblskip=5pt
\caption{CO rotational transmission lines within the HFI bands. The final column indicates the ranges of the spectral bandpass data that have been over-sampled for the HFI CO analysis.}                          
\label{tab:CO}                            
\nointerlineskip
\vskip -3mm
\footnotesize
\setbox\tablebox=\vbox{
   \newdimen\digitwidth 
   \setbox0=\hbox{\rm 0} 
   \digitwidth=\wd0 
   \catcode`*=\active 
   \def*{\kern\digitwidth}
   \newdimen\signwidth 
   \setbox0=\hbox{+} 
   \signwidth=\wd0 
   \catcode`!=\active 
   \def!{\kern\signwidth}
   \def\leaderfi1{\leaders\hbox to 5pt{\hss.\hss}\hfil}
\halign{\hbox to 0.5in{#\leaderfil}\tabskip=1em& 
  \tabskip=1em\hfil#\hfil\tabskip=1em&
  \tabskip=1em\hfil#\hfil\tabskip=1em plus 2em minus 1em&
  \tabskip=1em plus 2em minus 1em\hfil#\hfil\tabskip=0em plus 2em\cr                           
\noalign{\doubleline}
\omit Band& CO transition& $\nu_{\mbox{\scalebox{0.75}{CO}}}$\tablefootmark{a}& Oversampled\cr
\omit [GHz]& ($J_{\mbox{\tiny{upper}}} \rightarrow J_{\mbox{\tiny{lower}}}$)& [GHz]& region [GHz]\cr
\noalign{\vskip 3pt\hrule\vskip 5pt}
100& 1 $\rightarrow$ 0& *115.27& 109.67 -- *115.39\cr
\noalign{\vskip 4pt}
217& 2 $\rightarrow$ 1& *230.54& 219.34 -- *230.77\cr
\noalign{\vskip 4pt}
353& 3 $\rightarrow$ 2& *345.80& 329.00 -- *346.15\cr
\noalign{\vskip 4pt}
545& 4 $\rightarrow$ 3& *461.04& 438.64 -- *461.51\cr
545& 5 $\rightarrow$ 4& *576.27& 548.28 -- *576.85\cr
\noalign{\vskip 4pt}
857& 6 $\rightarrow$ 5& *691.47& 657.89 -- *692.17\cr
857& 7 $\rightarrow$ 6& *806.65& 767.48 -- *807.46\cr
857& 8 $\rightarrow$ 7& *921.80& 877.04 -- *922.73\cr
857& 9 $\rightarrow$ 8& 1036.91& 986.57 -- 1037.95\cr 
\noalign{\vskip 5pt\hrule\vskip 3pt}}}
\endPlancktable                 
\tablefoottext{a}{These frequencies are known to much finer precision (see \url{spec.jpl.nasa.gov}) but are shown truncated here for clarity.} 
\endgroup
\end{center}
\end{table}                        



\subsection{Spectral Response Data Products}
\label{sec:Data}

\begin{figure}
\centering
\begin{overpic}[width=88mm]{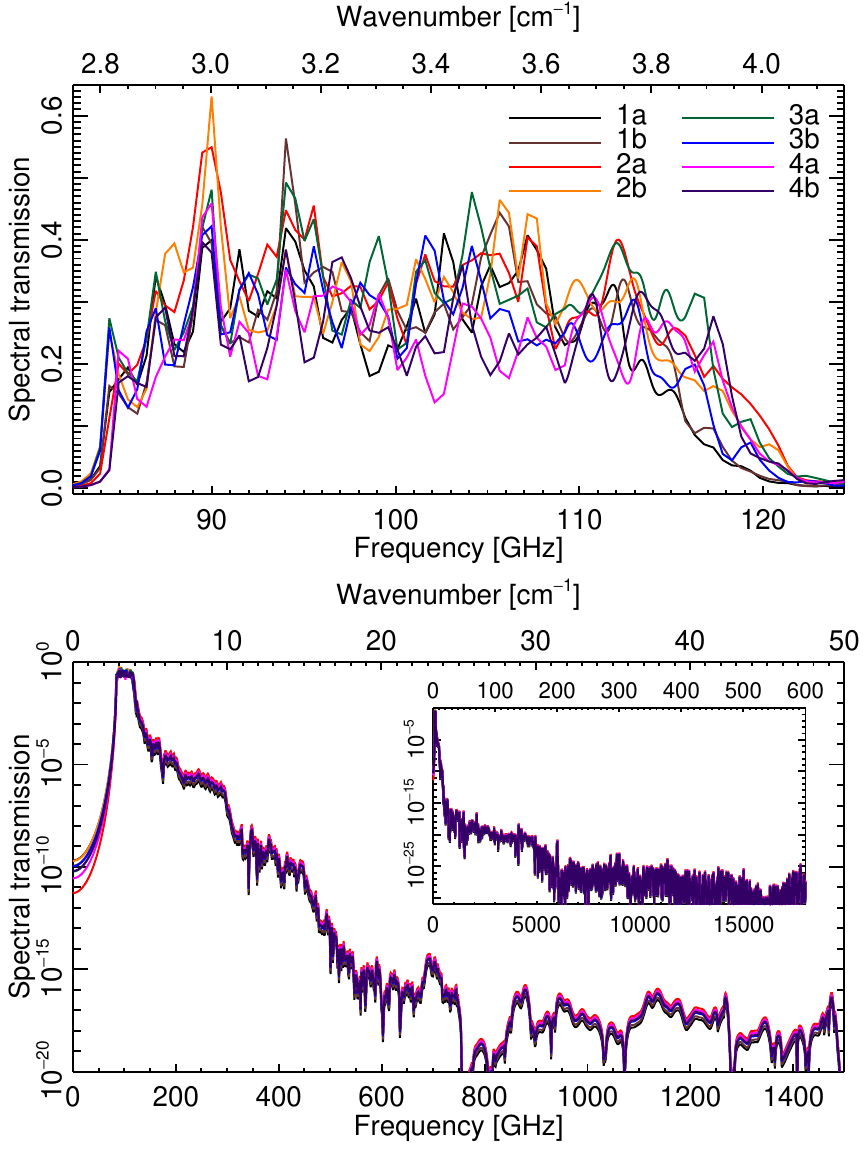}
\put(8,90){a}
\put(8,40){b}
\end{overpic}
\caption{\label{fig:HFI100}The detector spectral transmission profiles for the HFI 100 GHz detectors. Plot (a) shows the in-band region on a linear vertical scale with plot (b) showing a wider spectral region on a logarithmic vertical scale.  The inset within plot (b) shows the same spectra over the full spectral range available with the same units as the main plot for all axes.  The same conventions hold for Fig.~\ref{fig:HFI143}--Fig.~\ref{fig:HFI857}.}
\end{figure}

\begin{figure}
\centering
\begin{overpic}[width=88mm]{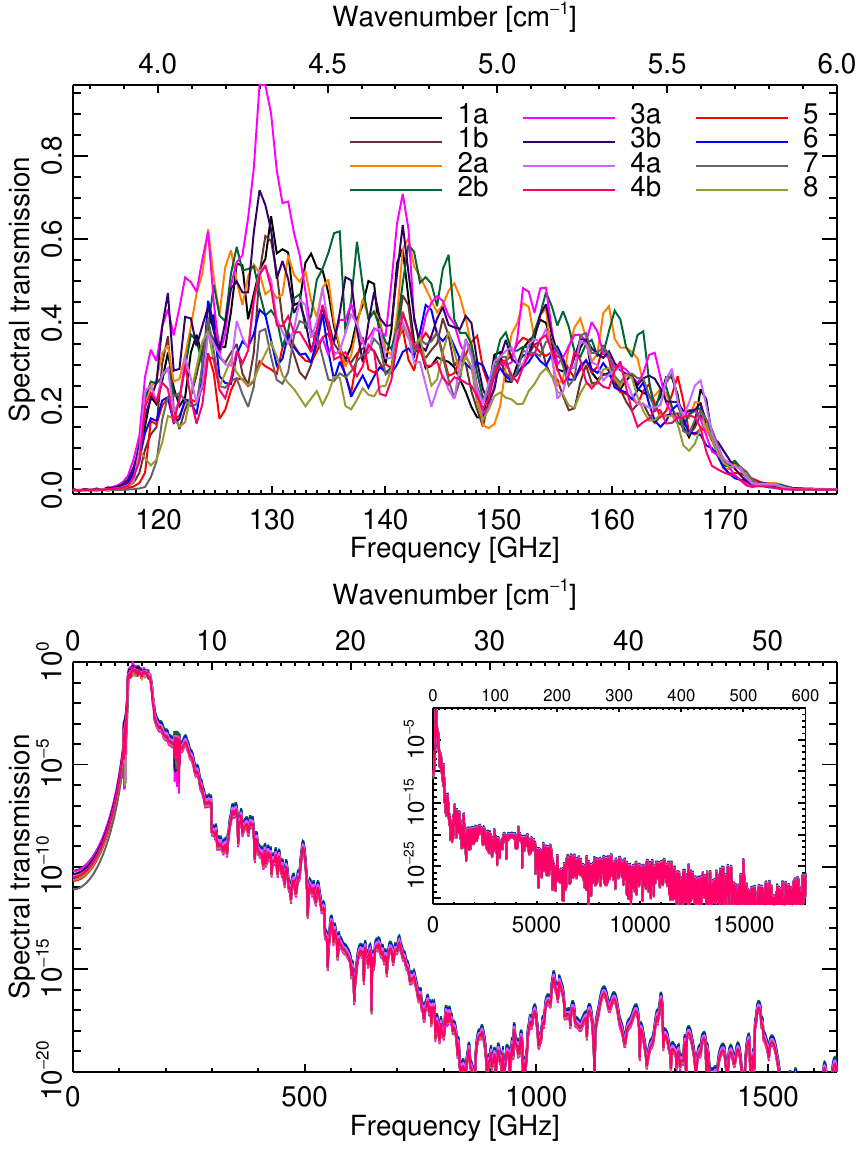}
\put(8,90){a}
\put(8,40){b}
\end{overpic}
\caption{\label{fig:HFI143}The detector spectral transmission profiles for the HFI 143 GHz detectors.}
\end{figure}

\begin{figure}
\centering
\begin{overpic}[width=88mm]{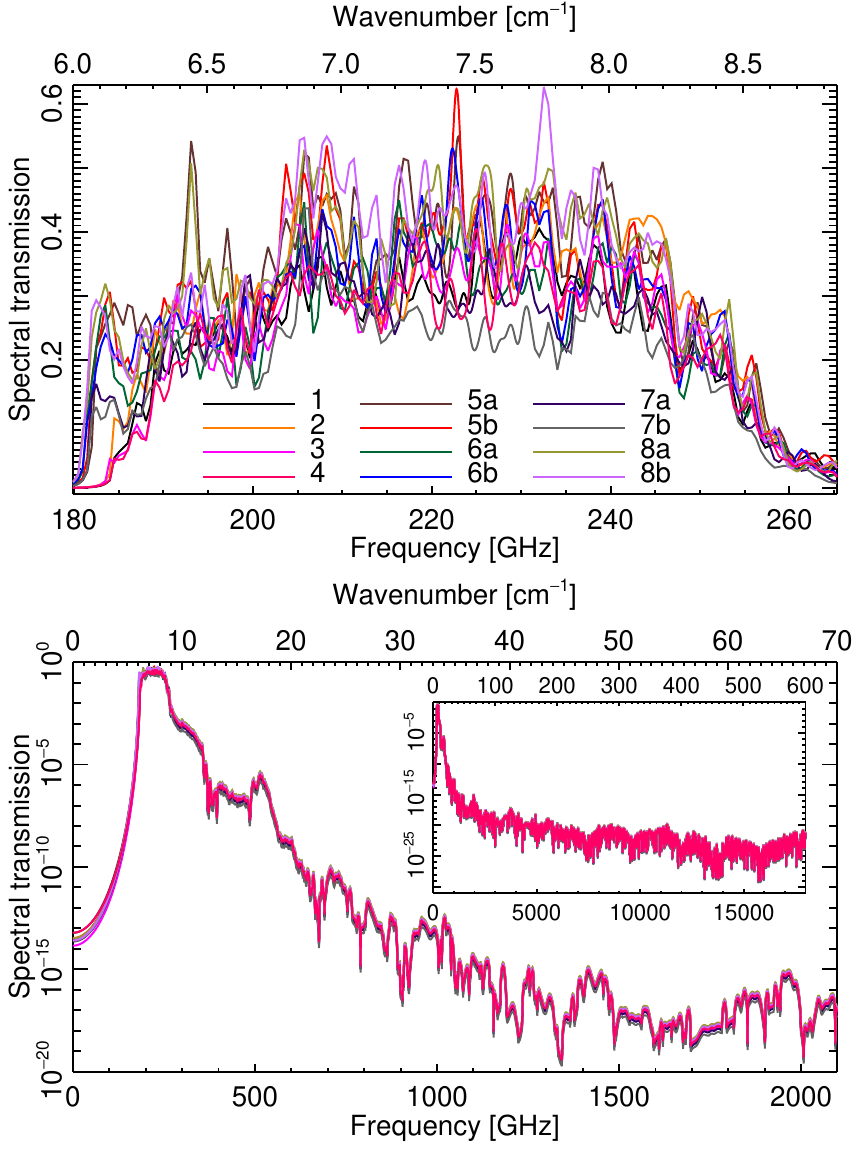}
\put(8,90){a}
\put(8,40){b}
\end{overpic}
\caption{\label{fig:HFI217}The detector spectral transmission profiles for the HFI 217 GHz detectors.}
\end{figure}

\begin{figure}
\centering
\begin{overpic}[width=88mm]{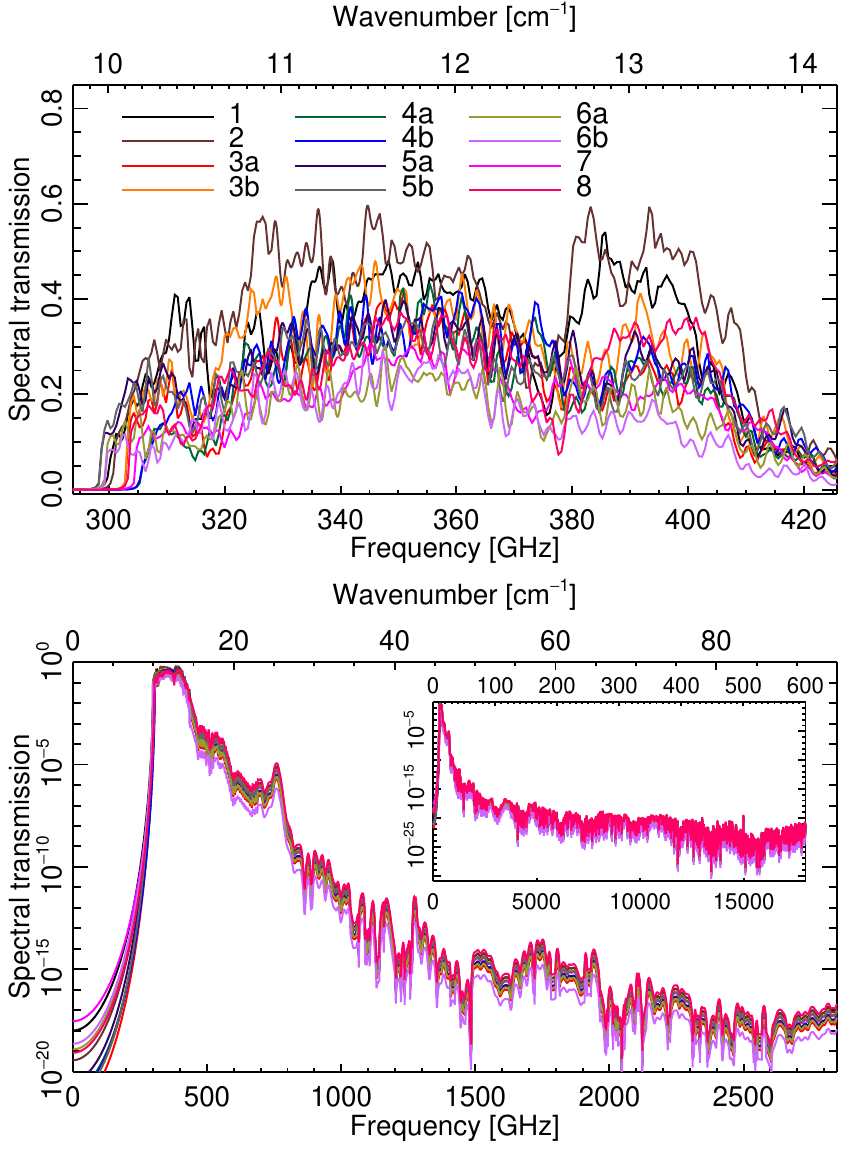}
\put(8,90){a}
\put(8,40){b}
\end{overpic}
\caption{\label{fig:HFI353}The detector spectral transmission profiles for the HFI 353 GHz detectors.}
\end{figure}

\begin{figure}
\centering
\begin{overpic}[width=88mm]{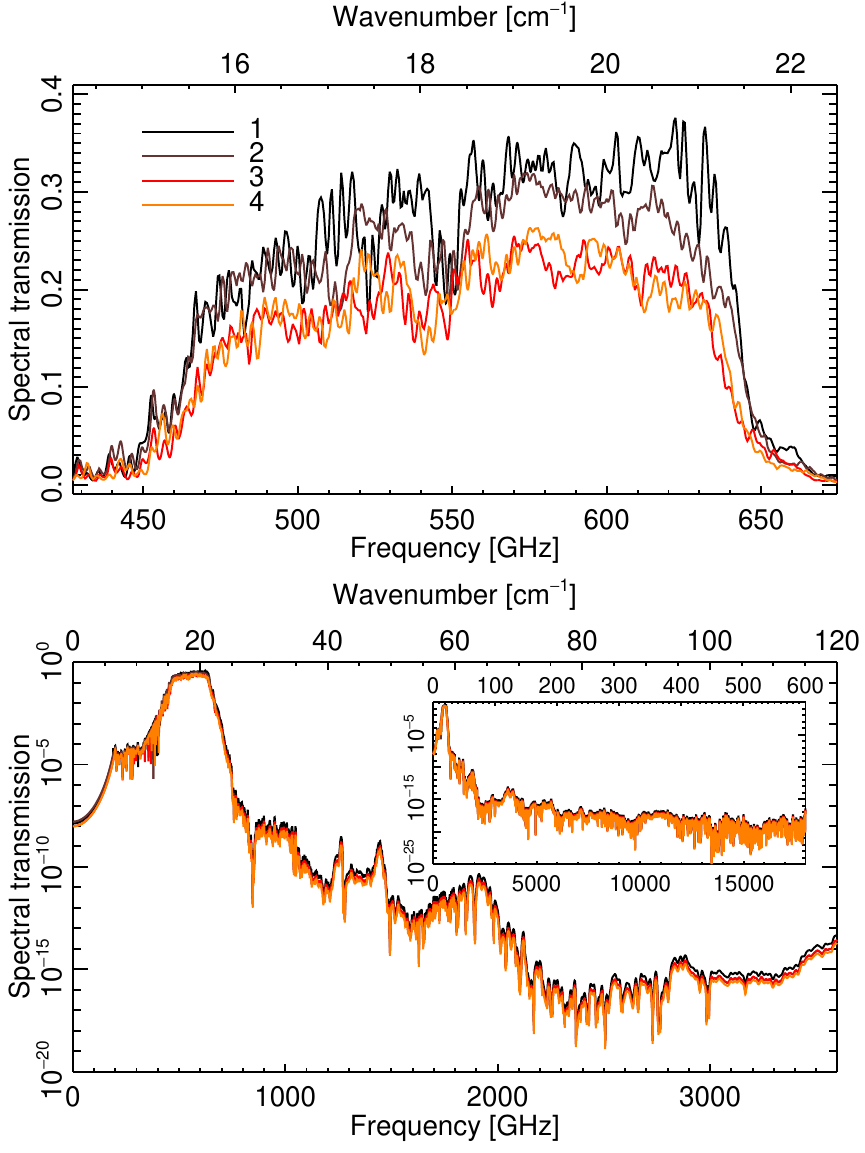}
\put(8,90){a}
\put(8,40){b}
\end{overpic}
\caption{\label{fig:HFI545}The detector spectral transmission profiles for the HFI 545 GHz detectors.}
\end{figure}

\begin{figure}
\centering
\begin{overpic}[width=88mm]{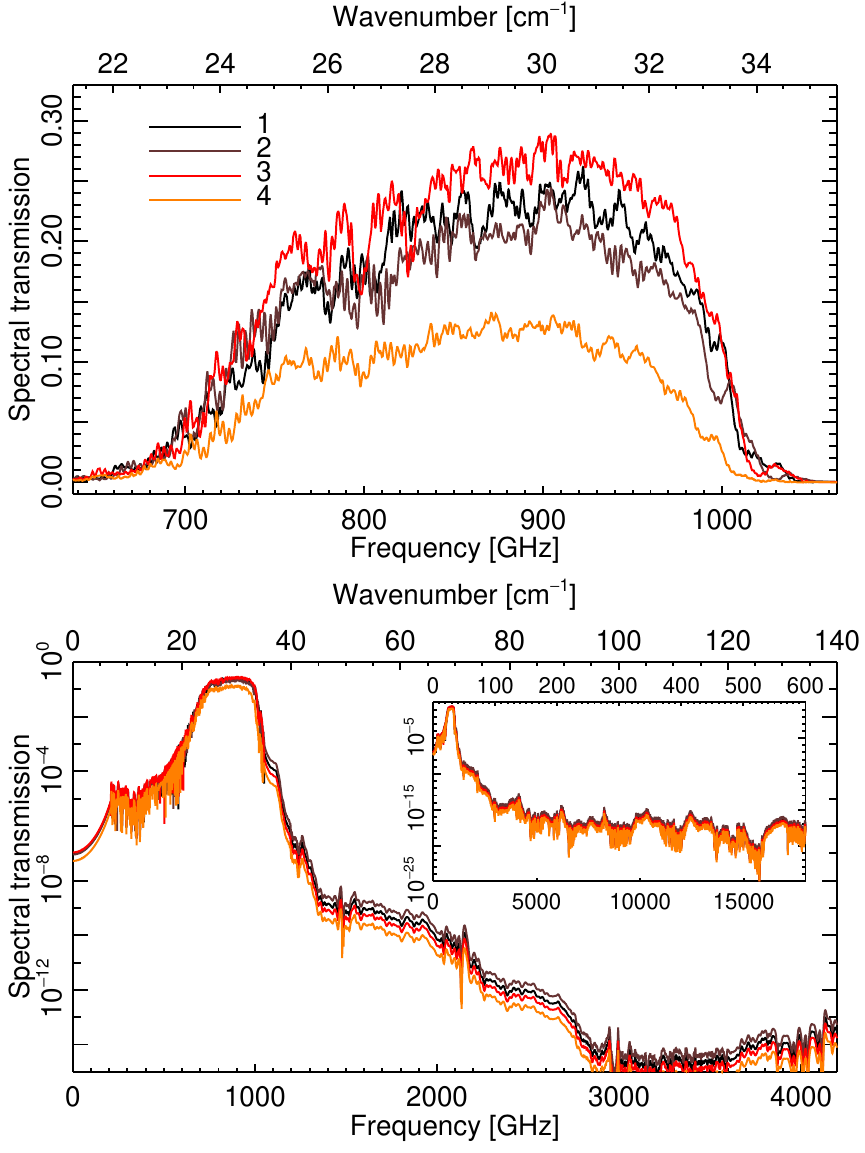}
\put(8,90){a}
\put(8,40){b}
\end{overpic}
\caption{\label{fig:HFI857}The detector spectral transmission profiles for the HFI 857 GHz detectors.}
\end{figure}

The spectral response profiles for each of the HFI detectors are shown in Figs.~\ref{fig:HFI100}--\ref{fig:HFI857}, grouped by common frequency bands.  While all detectors are labelled by the nominal band frequency (in GHz) and with a feed horn number, some are also labelled with a letter suffix to indicate that they are polarization sensitive. The polarization sensitive bolometer (PSB) pairs are labelled with an \emph{a} or \emph{b} feed horn number suffix, indicating orthogonal  linear polarization sensitivity.  The HFI spider-web bolometers (SWBs) are ideally insensitive to the incident radiation's polarization, and thus the SWB labels do not have a suffix.  Further details on the HFI focal plane layout are provided in \cite{planck2013-p03} and \cite{planck2013-p03c}.  Furthermore, band-average and detector sub-set spectra are also computed with the detector spectral response data; these are abbreviated as avg and DetSet, respectively. Table \ref{tab:DetSet} lists the various sub-band detector groupings, including the SWB-only detector subsets; the derivation of the averaging methods used to determine these spectra is presented in Sect.~\ref{sec:avg}.  

\begin{table}[tmb]                 
\begin{center}
\begingroup
\newdimen\tblskip \tblskip=5pt
\caption{HFI detector groupings used to determine the sub-band-average data products (e.g., sub-band-average maps, spectra, etc.).}                          
\label{tab:DetSet}                            
\nointerlineskip
\vskip -3mm
\footnotesize
\setbox\tablebox=\vbox{
   \newdimen\digitwidth 
   \setbox0=\hbox{\rm 0} 
   \digitwidth=\wd0 
   \catcode`*=\active 
   \def*{\kern\digitwidth}
   \newdimen\signwidth 
   \setbox0=\hbox{+} 
   \signwidth=\wd0 
   \catcode`!=\active 
   \def!{\kern\signwidth}
   \def\leaderfi1{\leaders\hbox to 5pt{\hss.\hss}\hfil}
%
\halign{\hbox to 1.0in{#\leaderfil}\tabskip=1em&
\tabskip=2em\hfil#\hfil\tabskip=2em&
\tabskip=2em\hfil#\hfil\tabskip=2em&
\tabskip=2em\hfil#\hfil\tabskip=0em\cr                           
\noalign{\doubleline}
\omit Band$\quad$\,[GHz] & DetSet1 & DetSet2 & SWB \cr                                    
\noalign{\vskip 3pt\hrule\vskip 5pt}
100& 1a/b, 4a/b& 2a/b, 3a/b& \dots\cr
143& 1a/b, 3a/b& 2a/b, 4a/b& 5,6,7 \cr
217& 5a/b, 7a/b& 6a/b, 8a/b& 1,2,3,4 \cr
353& 3a/b, 5a/b& 4a/b, 6a/b& 1,2,6,7 \cr
545& 1, 2&       4&          \dots\cr
857& 1, 2&       3, 4&       \dots\cr 
\noalign{\vskip 5pt\hrule\vskip 3pt}}}
\endPlancktable                 
\endgroup
\end{center}
\end{table}                        

The HFI detector spectral response data products are available within the database instrument model (see \citealt{planck2013-p28}) and within the PLA\@. The data is comprised of the spectral frequency in units of both GHz and \cm, the normalized spectral response (with its associated uncertainty), a CO interpolation flag, and meta-data including waveguide and filter transition regions, optical efficiency, and housekeeping information such as date and version.  The spectral normalization is such that the maximum value of any given spectrum is unity.  An estimate of the absolute spectral transmission is obtained through the product of the optical efficiency parameter and the normalized spectrum for a given detector or frequency band (see \citealt{planck2013-p28} for further details).  Several diagnostic parameters are determined for each HFI detector spectrum.  These parameters are defined below and include the cut-on and cut-off frequency, the effective bandwidth, the central frequency, various effective frequencies, and the integrated optical efficiency.  
\begin{itemize}
\item{\emph{Cut-on, $\nu_{\mbox{\tiny{on}}}$:}} The Cut-on frequency defines where the spectral band or the high-pass filter frequency dependence goes from the minimum to the maximum value. The most general definition is therefore where the smooth varying function reaches half of the maximum amplitude. This definition encounters a problem when spectra oscillate above and below the half-maximum amplitude.  The cut-on frequency in this work is defined as the lowest frequency occurrence of half-maximum amplitude. 

\item{\emph{Cut-off, $\nu_{\mbox{\tiny{off}}}$:}} Similar arguments apply to the cut-off as for the cut-on.  The cut-off frequency in this work is defined as the highest frequency occurrence of half-maximum amplitude. 

\item{\emph{Bandwidth, $\Delta\nu$:}} The full-width at half-maximum (FWHM) of the optical band, given by
\begin{equation}
\label{eq:BW}
\Delta\nu = \nu_{\mbox{\tiny{off}}} - \nu_{\mbox{\tiny{on}}} \ .
\end{equation}
\item{\emph{Central Frequency, $\nu_{\mbox{\tiny{cen}}}$:}} The central frequency is defined as the average of $\nu_{\mbox{\tiny{on}}}$ and $\nu_{\mbox{\tiny{off}}}$.  A more useful parameter, however, is the \emph{effective frequency}, $\nu_{\mbox{\tiny{eff}}}$.
\item{\emph{Effective frequency}, $\nu_{\mbox{\tiny{eff}}}$:} Alternatively to $\nu_{\mbox{\tiny{cen}}}$, the effective frequency is defined by weighing the spectra by the frequency itself.  This is analogous to the most-probable frequency for a given spectral response and is determined by 
\begin{equation}
\label{eq:nueff}
\nu_{\mbox{\tiny{eff}}} = \displaystyle \frac{\int{\nu\tau'(\nu)d\nu}}{\int{\tau'(\nu)d\nu}} \ ,
\end{equation}
where $\tau'(\nu)$ is the spectral transmission including the optical efficiency term $\varepsilon$.
\item{\emph{Integrated optical efficiency}, $\varepsilon_{\mbox{\tiny{Int}}}$:} The integrated optical efficiency is obtained by integrating the spectral transmission across the entire measured frequency range, and dividing by the detector bandwidth as follows
\begin{equation}
\label{eq:IntEff}
\varepsilon_{\mbox{\tiny{Int}}} = \displaystyle\frac{\epsilon\displaystyle\int{\tau(\nu)d\nu}}{\Delta\nu} \ ,
\end{equation}
where $\epsilon$ is the relative optical efficiency discussed in Sect.~\ref{sec:proc} above, and $\tau(\nu)$ is the normalized spectral transmission.
%
\item{\emph{Spectral index effective frequencies}, $\nu_{\alpha,\mbox{\tiny{eff}}}$:} As in the effective frequency case, this is equivalent to weighing the transmission spectrum by the frequency; in this case for sources following a power-law spectral profile with intensity proporitonal to $\nu^\alpha$, with spectral index defined as $\alpha$.  This is given by
\begin{equation}
\label{eq:nueffSpecInd}
\nu_{\alpha,\mbox{\tiny{eff}}} = \displaystyle\frac{\displaystyle\int{\nu\left(\displaystyle\frac{\nu}{\nu_{\mbox{\tiny{c}}}}\right)^\alpha\tau'(\nu)d\nu}}{\displaystyle\int{\left(\displaystyle\frac{\nu}{\nu_{\mbox{\tiny{c}}}}\right)^\alpha\tau'(\nu)d\nu}} \ ,
\end{equation}
where $\nu_{\mbox{\tiny{c}}}$ is the nominal band reference frequency.
\footnote{This frequency is somewhat arbitrarily defined as it does not have to be equal to $\nu_{\mbox{\tiny{eff}}}$ above, but is a matter of definition.  The choice of $\nu_{\mbox{\tiny{c}}}$ for \Planck\ detectors and frequency channels is discussed further in Sect.~\ref{sec:UcCCPhil}.}
%

\end{itemize}
There is an important distinction between $\epsilon$ and $\varepsilon_{\mbox{\tiny{Int}}}$ in that the former is a scaling term, which accompanies the normalized transmission spectra (and is meaningless on its own), and the latter is intended to represent an effective optical efficiency over the specified bandwidth, i.e., an equivalent spectral rectangle or tophat function.  Tables \ref{tab:SpecProd1} and \ref{tab:SpecProd2} report these parameters derived from the HFI detector spectral transmission profile data products, for the HFI band-average spectra; similar results for all HFI detectors are available in \cite{planck2013-p28}.  The sub-band average spectra corresponding to the additional data in these tables are introduced Sect.~\ref{sec:avg} and Table \ref{tab:DetSet}.  The spectral indices chosen for Table~\ref{tab:SpecProd2} correspond to the Infrared Astronomical Satellite (IRAS) spectral energy distribution (SED) convention ($\alpha=-1$, see Sect.~\ref{sec:UcCCPhil} and Eq.\ \ref{eq:IRAS}), a planetary SED ($\alpha=2$), and a dust SED ($\alpha=4$).  It is important to note the variation of the parameter uncertainty within Tables \ref{tab:SpecProd1} and \ref{tab:SpecProd2}.  The sub-band-average spectra are comprised of fewer individual detector spectra; and typically have greater uncertainty than the band-average spectra as a result.

In addition to the individual detector spectra, several other data products have been prepared for distribution.  These include band-average spectra and sub-band-average spectra (see Sect.~\ref{sec:avg}, and also Table \ref{tab:DetSet}), unit conversion and colour correction coefficients (see Sect.~\ref{sec:UcCC}), and a unit conversion and colour correction, {\tt UcCC}, software package to accompany HFI data (see \citealt{planck2013-p28}).  



\begin{table*}[tmb]                 
\reftabstrt 
\begingroup
\newdimen\tblskip \tblskip=5pt
\caption{HFI spectral response diagnostic parameters for the band-average, and sub-band-average (see Sect.~\ref{sec:avg}, Table~\ref{tab:DetSet}), spectra (see also Table~\ref{tab:SpecProd2}).  The parameters shown here are introduced in Sect.~\ref{sec:Data}.}                          
\label{tab:SpecProd1}                            
\nointerlineskip
\vskip -3mm
\footnotesize
\setbox\tablebox=\vbox{
   \newdimen\digitwidth 
   \setbox0=\hbox{\rm 0} 
   \digitwidth=\wd0 
   \catcode`*=\active 
   \def*{\kern\digitwidth}
   \newdimen\signwidth 
   \setbox0=\hbox{.} 
   \signwidth=\wd0 
   \catcode`!=\active 
   \def!{\kern\signwidth}
   \def\leaderfi1{\leaders\hbox to 5pt{\hss.\hss}\hfil}
%
\halign{\hbox to 1.00in{#\leaderfil}\tabskip=1em&
\tabskip=1em\hfil#\hfil\tabskip=1em&
\tabskip=1em\hfil#\hfil\tabskip=1em&
\tabskip=1em\hfil#\hfil\tabskip=1em&
\tabskip=1em\hfil#\hfil\tabskip=1em&
\tabskip=1em\hfil#\hfil\tabskip=1em&
\tabskip=1em\hfil#\hfil\tabskip=0em\cr                           
\noalign{\doubleline}
\omit Spectrum & 
$\nu_{\mbox{\tiny{on}}}$ [GHz]& 
$\nu_{\mbox{\tiny{off}}}$ [GHz]& 
$\Delta\nu$ [GHz]& 
$\nu_{\mbox{\tiny{cen}}}$ [GHz]& 
$\nu_{\mbox{\tiny{eff}}}$ [GHz]& 
$\varepsilon_{\mbox{\tiny{Int}}}$\cr 
\noalign{\vskip 3pt\hrule\vskip 5pt}
100-avg&      *84.4** $\pm$ 0.3**& 117.36* $\pm$ 0.05*& *32.9** $\pm$ 0.3**& 100.89* $\pm$ 0.14*& 101.31* $\pm$ 0.05*& 0.304** $\pm$ 0.003**\cr
100-DetSet1&  *84.77* $\pm$ 0.09*& 117.81* $\pm$ 0.05*& *33.03* $\pm$ 0.11*& 101.29* $\pm$ 0.05*& 101.43* $\pm$ 0.07*& 0.265** $\pm$ 0.002**\cr
100-DetSet2&  *84.29* $\pm$ 0.18*& 117.14* $\pm$ 0.05*& *32.85* $\pm$ 0.19*& 100.72* $\pm$ 0.09*& 101.25* $\pm$ 0.06*& 0.321** $\pm$ 0.003**\cr
\noalign{\vskip 4pt}
143-avg&      119.994 $\pm$ 0.018& 165.76* $\pm$ 0.04*& *45.76* $\pm$ 0.05*& 142.875 $\pm$ 0.020& 142.709 $\pm$ 0.015& 0.3669* $\pm$ 0.0006*\cr
143-DetSet1&  120.05* $\pm$ 0.03*& 160.18* $\pm$ 0.09*& *40.13* $\pm$ 0.10*& 140.12* $\pm$ 0.05*& 141.45* $\pm$ 0.03*& 0.4614* $\pm$ 0.0017*\cr
143-DetSet2&  118.95* $\pm$ 0.08*& 164.9** $\pm$ 0.8**& *45.9** $\pm$ 0.8**& 141.9** $\pm$ 0.4**& 142.27* $\pm$ 0.02*& 0.379** $\pm$ 0.007**\cr
143-SWBs&     120.17* $\pm$ 0.03*& 166.308 $\pm$ 0.018& *46.14* $\pm$ 0.04*& 143.238 $\pm$ 0.018& 143.96* $\pm$ 0.03*& 0.3123* $\pm$ 0.0007*\cr
\noalign{\vskip 4pt}
217-avg&      188.892 $\pm$ 0.011& 253.419 $\pm$ 0.007& *64.527 $\pm$ 0.013& 221.156 $\pm$ 0.006& 221.914 $\pm$ 0.005& 0.33850 $\pm$ 0.00012\cr
217-DetSet1&  183.3** $\pm$ 0.3**& 253.606 $\pm$ 0.020& *70.3** $\pm$ 0.3**& 218.46* $\pm$ 0.13*& 220.548 $\pm$ 0.009& 0.3053* $\pm$ 0.0011*\cr
217-DetSet2&  182.159 $\pm$ 0.013& 253.592 $\pm$ 0.007& *71.433 $\pm$ 0.016& 217.875 $\pm$ 0.007& 220.614 $\pm$ 0.009& 0.34838 $\pm$ 0.00018\cr
217-SWBs&     189.02* $\pm$ 0.03*& 253.247 $\pm$ 0.013& *64.22* $\pm$ 0.04*& 221.136 $\pm$ 0.017& 222.957 $\pm$ 0.008& 0.3226* $\pm$ 0.0002*\cr
\noalign{\vskip 4pt}
353-avg&      306.8** $\pm$ 0.6**& 408.22* $\pm$ 0.02*& 101.4** $\pm$ 0.6**& 357.5** $\pm$ 0.3**& 361.289 $\pm$ 0.008& 0.335** $\pm$ 0.002**\cr
353-DetSet1&  303.582 $\pm$ 0.015& 406.333 $\pm$ 0.017& 102.75* $\pm$ 0.02*& 354.957 $\pm$ 0.011& 359.156 $\pm$ 0.011& 0.29902 $\pm$ 0.00014\cr
353-DetSet2&  318.885 $\pm$ 0.014& 407.86* $\pm$ 0.02*& *88.97* $\pm$ 0.03*& 363.372 $\pm$ 0.013& 360.870 $\pm$ 0.012& 0.28730 $\pm$ 0.00015\cr
353-SWBs&     306.3** $\pm$ 0.4**& 408.81* $\pm$ 0.03*& 102.5** $\pm$ 0.4**& 357.56* $\pm$ 0.18*& 361.921 $\pm$ 0.011& 0.3575* $\pm$ 0.0013*\cr
\noalign{\vskip 4pt}
545-avg&      469.5** $\pm$ 0.5**& 640.81* $\pm$ 0.03*& 171.3** $\pm$ 0.5**& 555.2** $\pm$ 0.3**& 557.54* $\pm$ 0.03*& 0.2612* $\pm$ 0.0008*\cr
545-DetSet1&  466.44* $\pm$ 0.02*& 642.36* $\pm$ 0.04*& 175.91* $\pm$ 0.04*& 554.40* $\pm$ 0.02*& 557.86* $\pm$ 0.03*& 0.28031 $\pm$ 0.00013\cr
545-DetSet2&  470.9** $\pm$ 0.3**& 638.52* $\pm$ 0.11*& 167.6** $\pm$ 0.4**& 554.73* $\pm$ 0.17*& 556.85* $\pm$ 0.05*& 0.2143* $\pm$ 0.0005*\cr
\noalign{\vskip 4pt}
857-avg&      743.9** $\pm$ 0.4**& 989.78* $\pm$ 0.08*& 245.9** $\pm$ 0.4**& 866.8** $\pm$ 0.2**& 862.68* $\pm$ 0.05*& 0.2165* $\pm$ 0.0004*\cr
857-DetSet1&  736.9** $\pm$ 0.7**& 990.38* $\pm$ 0.06*& 253.4** $\pm$ 0.7**& 863.7** $\pm$ 0.4**& 863.42* $\pm$ 0.06*& 0.2121* $\pm$ 0.0006*\cr
857-DetSet2&  741.79* $\pm$ 0.13*& 987.01* $\pm$ 0.10*& 245.22* $\pm$ 0.17*& 864.40* $\pm$ 0.08*& 861.74* $\pm$ 0.07*& 0.21419 $\pm$ 0.00017\cr
\noalign{\vskip 5pt\hrule\vskip 3pt}}}
\endPlancktablewide                 
\endgroup
\reftabend 
\end{table*}                        

\begin{table}[tmb]                 
\begin{center}
\begingroup
\newdimen\tblskip \tblskip=5pt
\caption{HFI spectral response effective frequencies for the band-average, and sub-band-average, spectra (see also Table~\ref{tab:SpecProd1}).  The effective frequencies shown here are calculated using Eq.~\ref{eq:nueffSpecInd}.}                          
\label{tab:SpecProd2}                            
\nointerlineskip
\vskip -3mm
\footnotesize
\setbox\tablebox=\vbox{
   \newdimen\digitwidth 
   \setbox0=\hbox{\rm 0} 
   \digitwidth=\wd0 
   \catcode`*=\active 
   \def*{\kern\digitwidth}
   \newdimen\signwidth 
   \setbox0=\hbox{.} 
   \signwidth=\wd0 
   \catcode`!=\active 
   \def!{\kern\signwidth}
   \def\leaderfi1{\leaders\hbox to 5pt{\hss.\hss}\hfil}
%
\halign{\hbox to 0.85in{#\leaderfil}\tabskip=0em&
\tabskip=0em\hfil#\hfil\tabskip=1em&
\tabskip=1em\hfil#\hfil\tabskip=1em&
\tabskip=1em\hfil#\hfil\tabskip=0em\cr                           
\noalign{\doubleline}
\omit Spectrum & 
$\nu_{\alpha\mbox{\tiny{=$-$1}}}$ [GHz]&
$\nu_{\alpha\mbox{\tiny{=2}}}$ [GHz]&  
$\nu_{\alpha\mbox{\tiny{=4}}}$ [GHz]\cr                                    
\noalign{\vskip 3pt\hrule\vskip 5pt}%
100-avg&      100.36* $\pm$ 0.05*& 103.24* $\pm$ 0.05*& 105.25* $\pm$ 0.04*\cr
100-DetSet1&  100.49* $\pm$ 0.07*& 103.35* $\pm$ 0.06*& 105.34* $\pm$ 0.06*\cr
100-DetSet2&  100.31* $\pm$ 0.07*& 103.19* $\pm$ 0.06*& 105.21* $\pm$ 0.05*\cr
\noalign{\vskip 4pt}
143-avg&      141.362 $\pm$ 0.015& 145.457 $\pm$ 0.014& 148.234 $\pm$ 0.013\cr
143-DetSet1&  140.11* $\pm$ 0.03*& 144.22* $\pm$ 0.02*& 147.05* $\pm$ 0.02*\cr
143-DetSet2&  140.91* $\pm$ 0.02*& 145.05* $\pm$ 0.02*& 147.90* $\pm$ 0.02*\cr
143-SWBs&     142.64* $\pm$ 0.03*& 146.63* $\pm$ 0.02*& 149.28* $\pm$ 0.02*\cr
\noalign{\vskip 4pt}
217-avg&      220.111 $\pm$ 0.005& 225.517 $\pm$ 0.006& 229.096 $\pm$ 0.007\cr
217-DetSet1&  218.666 $\pm$ 0.009& 224.312 $\pm$ 0.009& 228.038 $\pm$ 0.010\cr
217-DetSet2&  218.697 $\pm$ 0.009& 224.429 $\pm$ 0.009& 228.200 $\pm$ 0.010\cr
217-SWBs&     221.241 $\pm$ 0.008& 226.395 $\pm$ 0.008& 229.834 $\pm$ 0.010\cr
\noalign{\vskip 4pt}
353-avg&      358.563 $\pm$ 0.008& 366.763 $\pm$ 0.009& 372.192 $\pm$ 0.010\cr
353-DetSet1&  356.386 $\pm$ 0.011& 364.744 $\pm$ 0.012& 370.302 $\pm$ 0.013\cr
353-DetSet2&  358.409 $\pm$ 0.012& 365.850 $\pm$ 0.012& 370.837 $\pm$ 0.013\cr
353-SWBs&     359.158 $\pm$ 0.011& 367.455 $\pm$ 0.012& 372.930 $\pm$ 0.014\cr
\noalign{\vskip 4pt}
545-avg&      552.22* $\pm$ 0.05*& 567.596 $\pm$ 0.017& 576.778 $\pm$ 0.014\cr
545-DetSet1&  552.43* $\pm$ 0.06*& 568.12* $\pm$ 0.02*& 577.458 $\pm$ 0.017\cr
545-DetSet2&  551.76* $\pm$ 0.08*& 566.48* $\pm$ 0.03*& 575.32* $\pm$ 0.02*\cr
\noalign{\vskip 4pt}
857-avg&      854.69* $\pm$ 0.11*& 877.724 $\pm$ 0.018& 891.462 $\pm$ 0.016\cr
857-DetSet1&  855.33* $\pm$ 0.16*& 878.67* $\pm$ 0.02*& 892.59* $\pm$ 0.02*\cr
857-DetSet2&  853.89* $\pm$ 0.17*& 876.53* $\pm$ 0.03*& 890.03* $\pm$ 0.02*\cr
\noalign{\vskip 5pt\hrule\vskip 3pt}}}
\endPlancktable                 
\endgroup
\end{center}
\end{table}                        

\section{Results}
\label{sec:Res}

This section presents data products derived from the HFI detector spectra.  This includes band-average spectra, along with unit conversion and colour correction algorithms and coefficients.


\subsection{Frequency Channel-Average Spectra}
\label{sec:avg}

Frequency channel-average transmission spectra, i.e., band-average spectra, are derived to complement the various HFI frequency channel maps and component maps \citep{planck2013-p06}.  To produce band-average spectra, individual detector spectra of a given frequency channel are weighted by a detector scaling factor to mimic the proportional weighting applied in the HFI mapmaking algorithms (\citealt{planck2013-p03}, \citealt{planck2013-p03f}, and \citealt{planck2013-p01a}).  This scaling factor, i.e., $w_i$, is based on relative noise levels, spectral response, and instrument scan strategy, all of which are described below.  Although efforts were made to duplicate the mapmaking routines, the determination of the band-average transmission spectra is similar, but not identical, to its mapmaking counterpart.  The individual detector weights described in Sect.~\ref{sec:NET} are identical to those used in the mapmaking scripts.  The divergence lies in the hit-map normalization (Eq.\ \ref{eq:HitNoiseNorm}), and the CMB normalization (Sect.~\ref{sec:CMBnorm}).  The hit map normalization could fully reproduce the approach of the mapmaking routines if an average were produced for each map-pixel (see Sect.~\ref{sec:CoeffMaps}), rather than producing a scalar $w_i$ coefficient for each detector.  The CMB normalization described here is analogous to the dipole calibration done in the standard HFI mapmaking (but, again, not identical). 


\subsubsection{Noise Weighting}
\label{sec:NET}

The HFI detector noise equivalent temperature (NET) estimates (\citealt{planck2013-p03} and \citealt{planck2013-p03f}) are used to weight the individual detector signals in averaging during data processing.  An attempt at duplicating this behaviour is made to obtain multi-detector average spectra.  To determine the relative weights of individual detectors within an average, the inverse square of the detector NETs is normalized such that the sum total within the desired detector grouping is unity as follows
\begin{equation}
\label{eq:NoiseNorm}
w_{\mbox{\scalebox{0.75}{NET}}~i} = \displaystyle\frac{1/(\mbox{NET}_i)^2}{W} \quad \mbox{where } W = \sum_i{(1/(\mbox{NET}_i)^2)} \ .
\end{equation}
The detector NETs and the $w_{\mbox{\scalebox{0.75}{NET}}~i}$ factors can be found in \cite{planck2013-p28}. Two detectors have been omitted from contributing towards the band-average spectra due to random telegraphic signal (RTS), i.e., popcorn noise: 143\,GHz-8 and 545\,GHz-3.  The $w_i$ factor introduced above is a general concept, with the $w_{\mbox{\scalebox{0.75}{NET}}~i}$ factor in this section representing a special case of the concept.  Other special cases of the $w_i$ factor will be introduced later.  


\subsubsection{Detector Channel-Map Contribution Weighting}
\label{sec:wi}

\begin{figure}
\begin{center}
\begin{tabular}{cc}
Nominal survey& Survey 1\\
\begin{overpic}[trim= 1 21 1 2, clip, width=44mm]{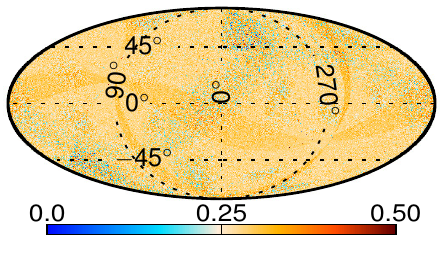} 
\put(3,40){a}
\end{overpic}&
\begin{overpic}[trim= 1 21 1 2, clip, width=44mm]{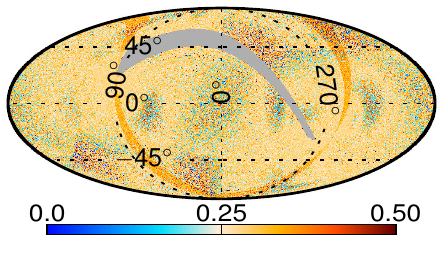} 
\put(3,40){b}
\put(-8,3){857-1}
\end{overpic}\\
%
%
\begin{overpic}[trim= 1 21 1 2, clip, width=44mm]{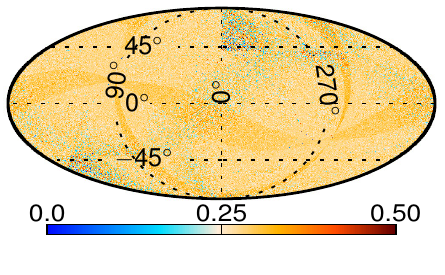} 
\put(3,40){c}
\end{overpic}&
\begin{overpic}[trim= 1 21 1 2, clip, width=44mm]{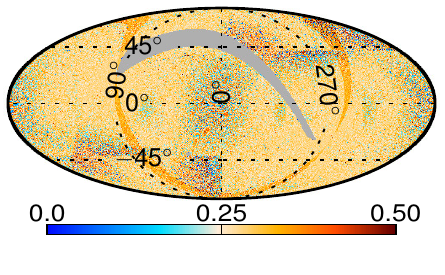} 
\put(3,40){d}
\put(-8,3){857-2}
\end{overpic}\\
%
%
\begin{overpic}[trim= 1 21 1 2, clip, width=44mm]{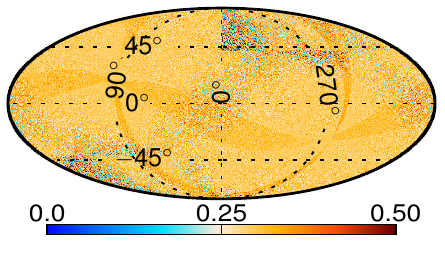} 
\put(3,40){e}
\end{overpic}&
\begin{overpic}[trim= 1 21 1 2, clip, width=44mm]{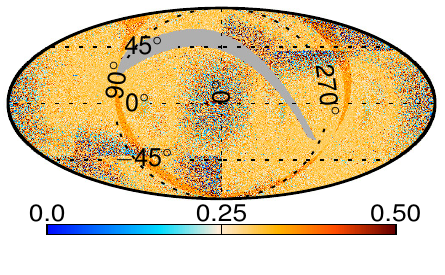} 
\put(3,40){f}
\put(-8,3){857-3}
\end{overpic}\\
%
%
\begin{overpic}[trim= 1 21 1 2, clip, width=44mm]{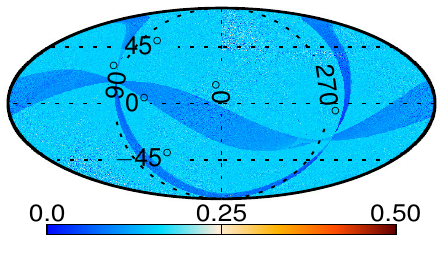} 
\put(3,40){g}
\end{overpic}&
\begin{overpic}[trim= 1 21 1 2, clip, width=44mm]{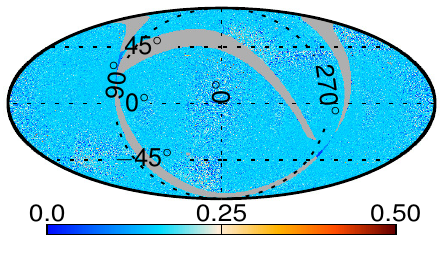} 
\put(3,40){h}
\put(-8,3){857-4}
\end{overpic}\\
%
%
\multicolumn{2}{c}{\includegraphics[trim= 0 12 0 119, clip,width=88mm]{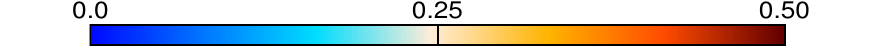}}\\
\end{tabular}
\caption{\label{fig:857wmaps} Detector relative contribution weight factor maps in Galactic coordinates for the 857 GHz detectors, i.e., $w_{m,i}$ as defined in Eq.\ \ref{eq:HitNoiseNorm}.  The left half represent the nominal survey (a, c, e, and g), while the right half represent the survey 1 subset of the data (b, d, f, and h).  Data corresponding to detectors 1 through 4 are grouped in rows, i.e., 857-1 is illustrated in (a,b), and 857-2,-3, and -4 are found in (c,d), (e,f), and (g,h), respectively.  The grey shaded regions indicate where there was no sky coverage.}
\end{center}
\end{figure}

The NET can be scan-normalized using the individual-detector pixel-hit maps available as standard data products (these will be made publicly available in the final release of \Planck\ data if not earlier, further details on the hit-maps can be found in \citealt{planck2013-p28}), i.e., 
\begin{equation}
\label{eq:HitNoiseNorm}
w_{m,i} = \displaystyle\frac{(\sum_{\theta,\phi} H_{m,i}(\theta,\phi))/(\mbox{NET}_i)^2}{W} \ ,
\end{equation}
where $W$ is a normalization term as described above (see Eq.~\ref{eq:NoiseNorm}), $H_{m,i}(\theta,\phi)$ represents the hit-map counts for a given detector, sky position, and a given map, $m$ (e.g., full-survey, nominal-survey, survey 1, etc.); the $\sum_{\theta,\phi}$ term represents summing over the entire map.  A similar approach could be taken where the $\sum_{\theta,\phi}$ summation is omitted; instead of a single $w_i$ factor for a given map, $m$, this would result in a map of weighting factors of the same spatial resolution as the map, i.e., $w_i(\theta,\phi)$.  An example of this, using the nominal survey and survey 1 results for the 857 GHz band, is shown in Fig.~\ref{fig:857wmaps}\footnote{This figure, and subsequent map figures within this paper, were produced using a modified version of the HEALPix software routines \citep{Healpix}.}, where it is clear from the colour-scale that the 857-4 detector contributes less to the band-average maps.

\renewcommand{\tabcolsep}{0em}

\begin{figure*}
\begin{center}
\begin{tabular}{c}
\begin{overpic}[width=180mm]{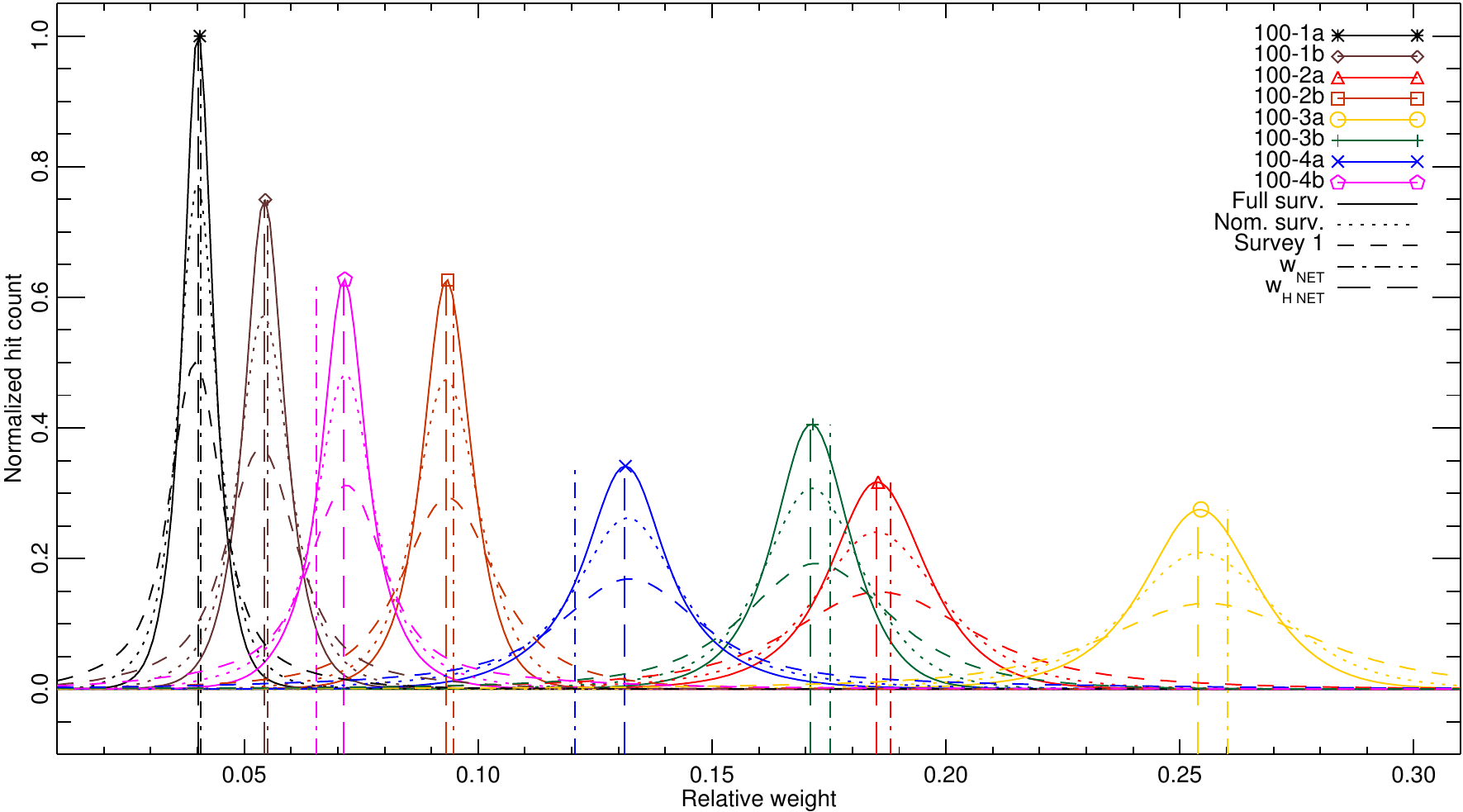} 
\put(7,52){a}
\end{overpic}\\
\begin{overpic}[width=180mm]{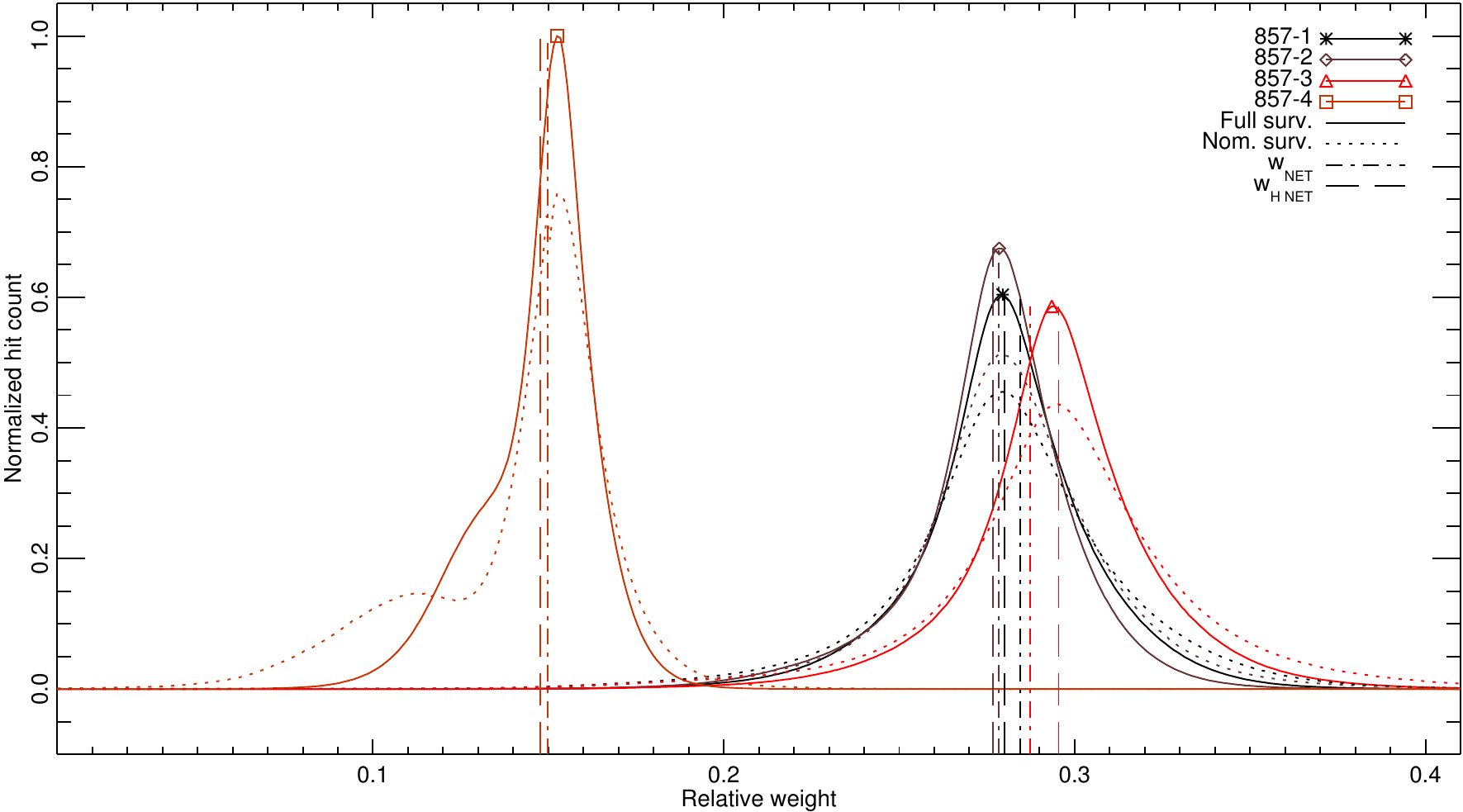} 
\put(7,52){b}
\end{overpic} 
\end{tabular}
\caption{\label{fig:Sfullwhist} Histograms of the $w_{m,i}$ band-average spectra scaling factors. Values are shown for 100\,GHz (a) and 857\,GHz (b) detectors, including the full, nominal, and individual surveys (some are omitted for clarity).  The vertical bars represent the resultant weight factor both with and without the \Planck\ sky coverage and hit-maps taken into consideration. }
\end{center}
\end{figure*}

A global factor, $w_{\mbox{\scalebox{0.6}{H~NET}}~i}$, can be obtained from each $w_i(\theta,\phi)$ map by choosing the statistical mean, or median, of the map, by taking the peak of a histogram of the map values, or by choosing some other diagnostic method.  
Fig.\ \ref{fig:Sfullwhist} illustrates histograms based on the detector $w_{m,i}(\theta,\phi)$ maps for the HFI full survey, nominal survey\footnote{The full survey is comprised of all five full sky surveys and the nominal survey is comprised of the first three individual full sky surveys.}, and, in the 100 GHz case, survey 1.  Similar histograms were also computed for individual sky surveys, masked surveys (masking varying percentages of the galactic plane and bright sources, see \citealt{planck2013-p01a}), and detector sub-set maps (see Table \ref{tab:DetSet}); these have been omitted from Fig.\ \ref{fig:Sfullwhist} for clarity.  For each HFI detector included in the plot, the dashed-dotted vertical lines (marked as NET) indicate the $w_{\mbox{\scalebox{0.75}{NET}}~i}$ factors resultant from Eq.~\ref{eq:NoiseNorm}, while the long-dashed vertical bars (marked as H~NET) indicate those resultant from Eq.~\ref{eq:HitNoiseNorm}; these coefficients are generated from the full survey $w_{m,i}(\theta,\phi)$ map histograms.  Similar results are available for various subsets, all converging towards the full survey values (more detailed figures for each of the HFI bands, and various data sub-sets, are found in \citealt{planck2013-p28}).  Although the 2013 \Planck\ data release only contains the nominal survey data, the full survey values are displayed here as they present stronger convergence than any given subset of the data.  The incorporation of the detector hit count into the band-average scaling factors ensures that the resultant frequency-channel spectra best represent the corresponding effective transmission spectra for a given data-subset.  This is demonstrated by the difference in position of the NET and H\,NET markings on Fig.\ \ref{fig:Sfullwhist}. In other words, the \Planck\ scanning strategy is an important consideration in determining the band-average spectra due to relative hit-counts and integration time changing for different detectors and sky positions.  

The 857\,GHz example shown is a special case, as the 857-4 bolometer exhibits higher noise properties than its counterparts.  The 857-4 detector contribution relative weight is thus much less than the other 857\,GHz detectors.  The 857-4 histogram (Fig.~\ref{fig:Sfullwhist}.b) is not symmetric, with an inflated tail towards zero weight.  As a result, the 857 $w_{m.i}$ scalar value is intentionally off-peak by a small factor.

The band-average spectra made available for distribution are thus based on the $w_{m,i}$ values from the full survey maps.  While the nominal band-average spectra are produced using the detector-weight histogram peak values, the effects of using off-peak histogram weights on the resultant band-average spectra were investigated.  Using compatible weight factors (i.e., $\sum w_i=1$) from the histogram tails produce band-average spectra that vary from the nominal band-average spectra at the percent level. 


\subsubsection{Photometric Bandpass Averaging}
\label{sec:CMBnorm}

A map from detector $i$, $m_i$, is given by
\begin{equation}
\label{eq:mi}
m_i = K_i \frac{1 + \eta_i}{2} \epsilon_i \displaystyle\int \!\!d\nu (A \Omega)_{\nu}\tau_i(\nu) dI_\nu \quad [\mbox{K}_{\mbox{\scalebox{0.6}{CMB}}}] \ ,
\end{equation}
where $K_i$ is the photometric CMB dipole calibration factor (e.g., K$_{\mbox{\scalebox{0.6}{CMB}}}$\,W$^{-1}$), the $(1 + \eta_i)/2$ fraction is used to distinguish between SWBs and PSBs, $(A\Omega)_{\nu}$ represents the telescope throughput at frequency $\nu$, $\tau_i(\nu)$ represents the normalized spectral transmission, $\epsilon_i$ represents the optical efficiency, and $dI_\nu$ represents the differential source intensity (see Eq.\ \ref{eq:Inu}).  As $\tau(\nu)$ is throughput normalized, by virtue of the ratio of the HFI detector spectra against a reference bolometer spectrum within an integrating sphere (see \citealt{pajot2010}), the $(A\Omega)_{\nu}\tau(\nu)$ term is further reduced to $(A\Omega)_{\nu_{\mbox{\tiny{c}}}}\tau(\nu)$, i.e., the product of the throughput at the nominal reference frequency and the throughput normalized transmission spectra. For SWBs, the $(1 + \eta_i)/2$ coefficient is 1, and for PSBs it is $1/2$.  

As described in \citet[see Eq.~3]{planck2013-p03f}, the local motion of the \Planck\ telescope in our Solar system with respect to the CMB can be aproximated by the temperature derivative of the Planck function evaluated at the CMB temperature, i.e., the CMB dipole; the principal calibration source for HFI\@.  This is used as the calibration source for the 100--353\,GHz channels, while FIRAS data \citep{Mather1994}, accompanied by a planet-based re-normalization, is the calibrator used for the 545 and 857\,GHz channels \citep{planck2013-p03}.  Based on the well known Planck function, $B_\nu(T,\nu)$, the CMB dipole signal is assumed to have the following form \citep{Fixsen1994}:
\begin{equation}
\label{eq:InuDeriv}
\begin{array}{rcl}
b_\nu' & = & \left.\displaystyle\frac{\partial B_\nu(T,\nu)}{\partial T} \right|_{T = 2.7255\mbox{{\sc ~k}}} \\
 & & \\
 & = & \left[ \displaystyle\frac{2 h \nu^3}{c^2(\exp{[h \nu / (K T)]} - 1 )} \right]
 \left( \displaystyle\frac{\exp{[h\nu/(K T )]}}{\exp{[h\nu/(K T)]} - 1 } \right) \\
 & & \\
 & & \times \left.\left( \displaystyle\frac{h\nu}{K T^2} \right)\right|_{T = 2.7255\mbox{{\sc ~k}}}  
 \quad \left[\frac{\mbox{W}}{\mbox{m$^2$\,sr\,Hz\,K}}\right] \\ 
\end{array} \ ,
\end{equation}
and the differential source intensity is thus given by
\begin{equation}
\label{eq:Inu}
dI_\nu = b_\nu' dT_{\mbox{\scalebox{0.6}{CMB}}} \quad  \left[\frac{\mbox{W}}{\mbox{m$^2$\,sr\,Hz}}\right] \ .
\end{equation}
All of the terms outside of the integral in Eq.\ \ref{eq:mi} can be replaced by a single constant, $A_i$, as follows
\begin{equation}
\label{eq:Ai}
A_i = K_i \frac{1 + \eta_i}{2}(A \Omega)_{\nu_{\mbox{\tiny{c}}}} \epsilon_i \quad \left[\frac{\mbox{K}_{\mbox{\scalebox{0.6}{CMB}}} \mbox{m}^2\mbox{sr}}{\mbox{W}}\right] \ .
\end{equation}
If the source under study, $dI_\nu$, is a dipole CMB spectrum, then the map itself should be given by $m_i = dT_{\mbox{\scalebox{0.6}{CMB}}}$, therefore
\begin{equation}
\label{eq:Ai2}
A_i = \left( \displaystyle\int \!\!d\nu \tau_i(\nu) b_\nu' \right)^{-1}\!\!\!\! = K_i \frac{1 + \eta_i}{2}(A \Omega)_{\nu_{\mbox{\tiny{c}}}} \epsilon_i \quad \left[\frac{\mbox{K}_{\mbox{\scalebox{0.6}{CMB}}} \mbox{m}^2\mbox{sr}}{\mbox{W}}\right] \ .
\end{equation}
The corresponding channel map, $M$, is given as a weighted combination of the detectors comprising that channel.  For individual weightings of $w_i$, the map, $M$, neglecting polarization effects for now, is given by
\begin{equation}
\label{eq:M}
M = \displaystyle\frac{\displaystyle\Sigma_i(w_i m_i)}{\Sigma_i w_i} \quad [\mbox{K}_{\mbox{\scalebox{0.6}{CMB}}}] \ .
\end{equation}
Expanding this expression to include Eqs.\ \ref{eq:mi}, \ref{eq:Ai}, and \ref{eq:Ai2}, we get
\begin{equation}
\label{eq:M2}
\begin{array}{rcl}
M & = & \left( \displaystyle\frac{1}{\Sigma_iw_i} \right) \displaystyle\sum_i\left( w_i A_i \displaystyle\int\!\!d\nu \tau_i(\nu) dI_\nu \right) \\
 & & \\
 & = & \displaystyle \int \!\!d\nu \left( \displaystyle\frac{\displaystyle\Sigma_i(w_i A_i \tau_i(\nu))}{\displaystyle\Sigma_i w_i} \right) dI_\nu \quad [\mbox{K}_{\mbox{\scalebox{0.6}{CMB}}}]
\end{array} \ .
\end{equation}
It is clear that the transmission and source components of the spectrum are separable, with the effective spectral response given by
\begin{equation}
\label{eq:AHnu}
A \tau(\nu) = \displaystyle\frac{\displaystyle\Sigma_i(w_iA_i\tau_i(\nu))}{\displaystyle\Sigma_iw_i} \quad \left[\frac{\mbox{K}_{\mbox{\scalebox{0.6}{CMB}}}\mbox{m}^2\mbox{sr}}{\mbox{W}}\right] \ ,
\end{equation}
where $A$ is an arbitrary scaling factor such that
\begin{equation}
\label{eq:M3}
M = A\displaystyle\int \!\!d\nu \tau(\nu) dI_\nu \quad [\mbox{K}_{\mbox{\scalebox{0.6}{CMB}}}] \ .
\end{equation}
Similar to the derivation of Eq.\ \ref{eq:Ai2}, for a CMB dipole source, $M = dT_{\mbox{\scalebox{0.6}{CMB}}}$ and
\begin{equation}
\label{eq:A}
A = \left( \displaystyle\int \!\!d\nu \tau(\nu) b_\nu' \right)^{-1} \quad \left[\frac{\mbox{K}_{\mbox{\scalebox{0.6}{CMB}}}\mbox{m}^2\mbox{sr}}{\mbox{W}}\right] \ .
\end{equation}
Equation \ref{eq:AHnu} can be rearranged to solve for $\tau(\nu)$ as follows
\begin{equation}
\label{eq:Hnu}
\begin{array}{rcccc}
\tau(\nu) & = & \left( \displaystyle\frac{1}{A} \right) & \left( \displaystyle\frac{1}{\displaystyle\Sigma_iw_i} \right) & \left[ \displaystyle\sum_i(w_i A_i \tau_i(\nu)) \right] \\
 & & & & \\
  & = & \left( \displaystyle\int \!\!d\nu \tau(\nu) b_\nu' \right) & \left( \displaystyle\frac{1}{\displaystyle\Sigma_iw_i} \right) & \left[ \displaystyle\sum_i\left( \displaystyle\frac{w_i \tau_i(\nu)}{\displaystyle\int \!\!d\nu \tau_i(\nu)b_\nu'} \right) \right] \\
\end{array} \ .
\end{equation}
The above expression, however, contains the desired $\tau(\nu)$ on both sides of the equation.  Since the right hand instance of $\tau(\nu)$ is within a frequency integral, and thus will only result in a single scaling factor being applied to the average transmission spectrum, the CMB-normalized channel average transmission spectrum is be defined with the remaining portions of Eq.\ \ref{eq:Hnu}.  This CMB-normalized channel average transmission spectrum, $\tau_{\mbox{\scalebox{0.6}{CMB}}}'(\nu)$, is defined as
\begin{equation}
\label{eq:HnuCMBnorm}
\tau_{\mbox{\scalebox{0.6}{CMB}}}'(\nu) = \mbox{Norm}\left[ \left( \displaystyle\frac{1}{\displaystyle\Sigma_iw_i} \right) \displaystyle\sum_i\left( \displaystyle\frac{w_i \tau_i(\nu)}{\displaystyle\int \!\!d\nu \tau_i(\nu)b_\nu'} \right) \right] \ ,
\end{equation}
where Norm[$f(x)$] is defined as $f(x)/\mbox{max}[f(x)]$.  A CMB-weight factor, $w_{\mbox{\scalebox{0.6}{CMB}}~i}$, is introduced to define the contribution of each detector to the bandpass average.  The above $w_i$ factors (Eqs.\ \ref{eq:NoiseNorm} and \ref{eq:HitNoiseNorm}) are coupled with the derivative of the CMB spectral function to determine the noise and CMB normalized scaling factors as follows
\begin{equation}
\label{eq:wCMB}
w_{\mbox{\scalebox{0.6}{CMB}}~i} = \displaystyle
  \frac{
    \left[
      \displaystyle\frac{w_i}{ (\Sigma_i w_i ) \left( \displaystyle\int \!\!d\nu\tau_i(\nu)b_{\nu}' \right) }
    \right]}
    {\mbox{max} 
      \left\{ \displaystyle\sum_i 
        \left[ 
          \displaystyle\frac{w_i \tau_i(\nu)}
          { \left( \displaystyle\Sigma_iw_i \right)\left( \displaystyle\int \!\!d\nu \tau_i(\nu)b_{\nu}'|_{T_{\mbox{\scalebox{0.6}{CMB}}}} \right) } 
        \right] 
      \right\} 
    } \ .
\end{equation}
In the above expression, $w_i$ is a scalar factor unique for each HFI detector.  This need not be the case, however, as this case may be more further generalized by allowing the detector weight factor $w_i$ to vary across the sky as in Eq.~\ref{eq:HitNoiseNorm}.  This generalization results in a $w_{\mbox{\scalebox{0.6}{CMB}}~i}(\theta,\phi)$ photometric weighting factor, i.e., the relative weights vary across the sky, and with relative integration time, etc.  Fig.~\ref{fig:Sfullwhist} illustrates histograms of the detector weight factors across the sky for the HFI full mission data, and various survey sub-sets of this data.  The difference between the static $w_i$ factor and the histogram peak value is demonstrated by the vertical bars in the figure.  
Thus, the frequency band average transmission spectra are comprised of the individual detector spectra proportionately scaled for both the relative response to the CMB spectrum, and the relative noise level within a given channel.  The resultant band-average transmission spectra are shown in Fig.~\ref{fig:AvgSpec} above.


As the 545 and 857\,GHz channels are calibrated using FIRAS data, and subsequently renormalized using planet observations \citep{planck2013-p03}, rather than using the CMB dipole directly, it is important to investigate the use of 
Eq.~\ref{eq:wCMB} in deriving the band-average spectra for these channels.  A comparison using both the $w_i$ and $w_{\mbox{\scalebox{0.6}{CMB}}~i}$ scaling factors for the 545 and 857\,GHz channels was thus conducted.  While these differences for the 100--353\,GHz channels are at the level of a few percent, they are at the 0.3--0.5\,\% level for the 545 and 857\,GHz channels.  Furthermore, the differences do not exceed the respective uncertainty of the corresponding detector spectra.  As this normalization removes the dependence of the individual $\tau_i(\nu)$ values on an absolute calibration, i.e., the optical efficiency, this CMB-normalization in the band-average spectra is maintained for all of the HFI bands.



\subsection{Unit Conversion and Colour Correction}
\label{sec:UcCC}

This section presents the formulae used to obtain unit conversion and colour correction coefficients for use with the HFI data, and the method used to derive the uncertainties on these coefficients.

\subsubsection{Unit Conversion and Colour Correction Philosophy}
\label{sec:UcCCPhil}

Broad-band detection instruments, including photometric instruments using band-defining filters such as HFI, measure power collected via an instrument collecting area for unresolved point-like sources, and power collected within a given throughput
for extended sources.  Although such an instrument directly measures power absorbed by the detectors, it is convenient to relate this power measurement to either flux density (for unresolved sources), or to specific intensity or brightness (for extended sources), such that the combined spectral and throughput integrated signal is equal to the measured power.  Thus, observation data expressed in units of brightness, specific intensity, or flux density, are intrinsically associated with an assumed reference frequency and SED profile.  

Spectral calibration of broad-band photometric instruments is performed by observation of a source of known SED\@.  Provided that an observed source has a similar SED (within the spectral band) to that of the calibration source, a measurement is calibrated by the ratio of the two observations.  The general case, however, is that observed sources have a different SED to that of the calibration source(s).  Any instrument observation is then related to a calibration observation by expression of both in terms of an equivalent intensity at a specified reference frequency.  The equivalent intensity is defined by knowledge of the source SED for a given observation.  A colour correction \citep{GriffinCC} is used to relate measurements of one SED to those of another.  There are two equivalent approaches to astronomical colour correction.  One approach is to determine the effective frequency that corresponds to the assumed SED and measured intensity, and determine a different reference frequency, based upon a different SED, for any other SEDs of interest.  In converting between SED types, the intensity remains the same, but at a different reference frequency.  The other approach is to determine the relative intensity for a given reference frequency, so the reference frequency remains the same for various SEDs, but the intensity will vary.  

The \Planck\ HFI uses two calibration schemes \citep{planck2013-p03f}, one based on the differential CMB dipole spectrum, and another based upon more local astrophysical sources.  It is thus important to express observation data in multiple formats for various aspects of data analysis.  This involves both unit conversion and colour correction.  In unit conversion, data are presented in a different unit, but remain consistent with a given SED (e.g., MJy\,sr$^{-1}$ can be expressed as an equivalent brightness in K).  With colour correction, data are expressed with respect to a different assumed SED at the same reference frequency (e.g., changing from K$_{\mbox{\scalebox{0.6}{CMB}}}$ to MJy\,sr$^{-1}$ with a different spectral index involves both a unit conversion and a colour correction). 

The HFI 100--353\,GHz channels are calibrated on the CMB dipole, which follows a $b_\nu'$ SED profile (see Eq.\ \ref{eq:InuDeriv}), where data are provided in units of differential CMB temperature, i.e., K$_{\mbox{\scalebox{0.6}{CMB}}}$.  Many astrophysical sources may be characterized photometrically by assuming that their emission follows a spectral power law and an associated spectral index (see Sec.\ \ref{sec:Data}).  The 545 and 857\,GHz channels are calibrated on a combination of galactic emission (typically dust following an approximate $\alpha=4$ power-law SED profile) and planetary emission (of an approximate $\alpha=2$ SED profile), where data are provided in units of brightness or intensity, i.e. MJy\,sr$^{-1}$.  Furthermore, the 545 and 857\,GHz data are scaled to equate with an SED following the IRAS convention (see Eq.\ \ref{eq:IRAS}), which has a SED profile with $\alpha$~=~$-$1. 

In the millimetre - sub millimetre region of the electro-magnetic spectrum, the diffuse emission is made of several components whose summation represents the observed signal. Component separation algorithms \citep{planck2013-p06}, when separating physical components with different SED profiles (e.g., emission from the CMB, thermal dust, spinning dust, free-free sources, synchrotron sources, CO lines, etc.), use models or templates for these components.  The component separation then must resolve an inverse problem going from several maps of broad-band measurements at different frequencies to component maps. The model is adjusted to minimize residuals through fitting the sum of the components to the measured intensity. It is therefore impractical to have the various components all expressed at different effective frequencies. Thus, \Planck\ adopts a fixed reference frequency for unit conversion and colour correction where the intensity is corrected for the assumed (or measured) SED of the source or component. 

Using K$_{\mbox{\scalebox{0.6}{CMB}}}$ calibration for the submillimeter channels, especially 857 GHz, should be avoided.  This is because the submillimetre  IRAS to CMB unit conversion depends heavily on the low-frequency region of the bandpass spectrum (see Fig.~\ref{fig:CMBprod}), which is known with less confidence than the main band.  Therefore, the conversion of the 100--353\,GHz data from a CMB to IRAS SED is less error-prone.  

Details on the derivation of the \Planck\ unit conversion and colour correction coefficients are provided in the following section.


\subsubsection{Coefficient Formula Derivation}
\label{sec:UcFccMath}

The following conversion factors are derived for the individual HFI detectors and the frequency-channel average spectra:
\begin{enumerate}
\item \label{UC:iKCMB_MJysr} Convert [MJy\,sr$^{-1}$] (IRAS) $\rightleftharpoons$ [K$_{\mbox{\scalebox{0.6}{CMB}}}$].
\item \label{UC:iKRJ_MJysr} Convert [MJy\,sr$^{-1}$] (IRAS) $\rightleftharpoons$ [K$_{\mbox{\tiny{b}}}$]. 
\item \label{UC:iKCMB_ySZ} Convert [y$_{\mbox{\scalebox{0.6}{SZ}}}$] $\rightleftharpoons$ [K$_{\mbox{\scalebox{0.6}{CMB}}}$].
\item \label{UC:iCC} Colour correction (power-law spectra and modified blackbody spectra).
\item \label{UC:iCO} CO correction.
\end{enumerate}

In general, the unit conversion terms are arrived at by equating changes in intensity, expressed in various forms.  Starting with the general expression $dI_\nu = dI_\nu$; e.g.,
\begin{equation}
\label{eq:nuInuBasic}
dI_\nu = (dI_\nu/dX_i)(dX_i) = (dI_\nu/dX_j)(dX_j) \ .
\end{equation}
Each side may also be multiplied by the spectral transmission, $\tau(\nu)$, and integrated across the frequency band as follows
\begin{equation}
\label{eq:nuInuBasic2}
\int \!\! d\nu ~\tau(\nu) \left(\frac{dI_\nu}{dX_i}\right)dX_i = \int \!\! d\nu ~\tau(\nu) \left(\frac{dI_\nu}{dX_j}\right)dX_j \ .
\end{equation}
This can be simplified to the form of a unit conversion coefficient
\begin{equation}
\label{eq:nuInuBasic3}
\displaystyle\frac{dX_i}{dX_j} = \displaystyle\frac{\displaystyle\int \!\! d\nu ~\tau(\nu) \left(\frac{dI_\nu}{dX_j}\right)}{\displaystyle\int \!\! d\nu ~\tau(\nu) \left(\frac{dI_\nu}{dX_i}\right)} \ .
\end{equation}
In converting to/from K$_{\mbox{\scalebox{0.6}{CMB}}}$, the derivative of the Planck function at the CMB monopole temperature ($T_{\mbox{\scalebox{0.6}{CMB}}} = 2.7255$\,K, see \citealt{fixsen2009}) will be used (see Eq.~\ref{eq:InuDeriv} and \ref{eq:Inu}). 


Conversion to MJy\,sr$^{-1}$ (IRAS) is accomplished using the IRAS convention \citep{IRASExSup}\footnote{\url{http://lambda.gsfc.nasa.gov/product/iras/docs/exp.sup/ch6/C3.html}}, $\nu ~dI_\nu = \mbox{constant}$, such that
\begin{equation}
\label{eq:IRAS}
 dI_{\nu~\mbox{\scalebox{0.6}{IRAS}}} \equiv \left(\frac{\nu_c}{\nu}\right)~dI_c \quad \left[\frac{\mbox{W}}{\mbox{m$^2$\,sr\,Hz}}\right] \ ,
\end{equation}
where $dI_c$ is the effective intensity at the specified frequency for a source of spectral index $\alpha=-$1.

As Rayleigh-Jeans brightness temperature units may be considered more convenient than that of W\,m$^{-2}$\,sr$^{-1}$\,Hz$^{-1}$ or even MJy\,sr$^{-1}$, a flux density to brightness temperature unit conversion is provided using the following relation \citep{1986RybickiLightman}:
\begin{equation}
\label{eq:IRAS2Tb}
 dT_b = \displaystyle\frac{c^2}{2\nu^2 k}dI_{\nu} \quad \left[{\rm K}_{\mbox{\tiny{b}}}\right] \ ,
\end{equation}
where K$_{\mbox{\tiny{b}}}$ is the temperature expression of flux density.  It is important to note that by this definition the use of brightness temperature units does not imply a Rayleigh-Jeans spectral profile. The brightness temperature unit of K$_{\mbox{\tiny{b}}}$ is selected over K$_{\mbox{\scalebox{0.6}{RJ}}}$ to avoid any confusion between this definition, and that of a source exhibiting a Rayleigh-Jeans SED profile. 
%

For SZ conversion the following intensity expression, based on the Kompaneets non-relativistic SZ formula (\citealt{Kompaneets1957}, \citealt{SZ1980}, and \citealt{GraingerSZ}), is used
\begin{equation}
\label{eq:ISZ}
\begin{array}{rcl}
dI_{\nu~\mbox{\scalebox{0.6}{SZ}}} & = & \left.\left(b_\nu'\right)(T)\left[\left(\displaystyle\frac{h\nu}{k T}\right)\displaystyle\frac{\exp{[h\nu/(k T)]} + 1}{\exp{[h\nu/(k T)]} - 1} - 4\right] (y_{\mbox{\scalebox{0.6}{SZ}}}) \right|_{T_{\mbox{\scalebox{0.6}{CMB}}}} \\
 & & \\
 & & \quad \left[\frac{\mbox{W}}{\mbox{m$^2$\,sr\,Hz}}\right] 
\end{array} \ ,
\end{equation}
where, again, $b_\nu'$ is the temperature derivative of the Planck function.

In power-law colour correction, it is assumed that the source intensity follows a power law over the spectral region of interest, i.e., $dI_\nu \propto \nu\hspace{1pt}^\alpha$,  providing the following expression
\begin{equation}
\label{eq:ICC}
dI_{\nu~\alpha} = \left( \displaystyle\frac{\nu}{\nu_{\mbox{\tiny{c}}}} \right)^\alpha~dI_{\mbox{\tiny{c}}~\alpha} \quad \left[\frac{\mbox{W}}{\mbox{m$^2$\,sr\,Hz}}\right] \ .
\end{equation}

A colour correction to a modified-blackbody of the form $dI_\nu \propto \nu\hspace{1pt}^\beta B_\nu(\nu,T)$, with $B_\nu$ as defined above, is given by
\begin{equation}
\label{eq:ICCMBB}
dI_{\nu~\beta} = \left[ \displaystyle\frac{\nu^\beta~B_\nu(\nu,T)}{\nu_{\mbox{\tiny{c}}}^{~\beta}~B_\nu(\nu_{\mbox{\tiny{c}}},T)} \right]~dI_{\mbox{\tiny{c}}~\beta} \quad \left[\frac{\mbox{W}}{\mbox{m$^2$\,sr\,Hz}}\right] \ .
\end{equation}

For molecular rotational transitions, such as the CO $J$=1$\rightarrow$0, \dots, $J$=9$\rightarrow$8 transitions, the desired specific intensity term is an effective brightness temperature, $\Delta T_{\mbox{\scalebox{0.75}{CO}}}$, in units of K\,km\,s$^{-1}$. For a Doppler line profile, $\nu$ is equal to $\nu_{\mbox{\scalebox{0.75}{CO}}}(1 + \varv /c)^{-1}$, which is closely approximated by $\nu_{\mbox{\scalebox{0.75}{CO}}}(1 - \varv /c)$ for $\varv << c$.  The intensity can be said to be distributed across a narrow velocity distribution, $d\varv$
\footnote{The variable $\varv$ is used to denote velocity in units of km\,s$^{-1}$, while the variable $\nu$ is used to denote frequency in units of Hz (or equivalent).}, such that the integral over all velocities yields the temperature-velocity effective brightness. To relate the effective brightness across a narrow frequency range, $d\nu$, to frequency units rather than those of velocity, the relationship $d\nu/d\varv = -\nu_{\mbox{\scalebox{0.75}{CO}}}/c$ is used, i.e.,
\begin{equation}
\label{eq:COBT1}
\left.\Delta T_{\mbox{\scalebox{0.75}{CO}}}\right|_{\nu} = \displaystyle\frac{\Delta T_{\mbox{\scalebox{0.75}{CO}}} \left(\displaystyle\frac{\nu_{\mbox{\scalebox{0.75}{CO}}}}{c}\right)}{d\nu} \quad \mbox{[K]} \ .
\end{equation}
It is important to note that the effective brightness temperature, i.e., $\Delta T_{\mbox{\scalebox{0.75}{CO}}}$, is given in units of K~km~s$^{-1}$ while this brightness distributed across a defined velocity interval ($d\varv$), or a defined frequency range ($d\nu$), i.e., $\left.\Delta T_{\mbox{\scalebox{0.75}{CO}}}\right|_{\nu}$, is given in units of K\@.  The CO transition intensity is thus given by the following relation
\begin{equation}
\label{eq:COBT2}
\begin{array}{rcl}
dI_{\mbox{\scalebox{0.75}{CO}}} & = & \left(\left.\Delta T_{\mbox{\scalebox{0.75}{CO}}}\right|_{\nu}\right)b_{\mbox{\scalebox{0.6}{RJ}}}' \\
 & & \\
 & = & \left( \Delta T_{\mbox{\scalebox{0.75}{CO}}} \right)\left(\displaystyle\frac{\nu_{\mbox{\scalebox{0.75}{CO}}}}{c ~d\nu}\right) b_{\mbox{\scalebox{0.6}{RJ}}}' \quad \left[\frac{\mbox{W}}{\mbox{m$^2$\,sr\,Hz}}\right]
\end{array} \ .
\end{equation}
The Rayleigh-Jeans (RJ) approximation temperature derivative used above is given as follows
\begin{equation}
\label{eq:IRJ}
\begin{array}{rcl}
b_{\mbox{\scalebox{0.6}{RJ}}}' & = & \left.\displaystyle\frac{\partial B_{\nu~\mbox{\scalebox{0.6}{RJ}}}(T,\nu)}{\partial T} \right|_{\mbox{\scalebox{0.6}{RJ}}} \\
 & & \\
 & = & \displaystyle\frac{2 \nu^2 k }{c^2}  \quad \left[\frac{\mbox{W}}{\mbox{m$^2$\,sr\,Hz\,K$_{\mbox{\scalebox{0.6}{RJ}}}$}}\right] \\
\end{array} \ .
\end{equation}
As the CO transitions occur at discrete frequencies, with line widths much narrower than the spectral resolution of the detector spectral transmission profiles\footnote{Low-order rotational CO line widths are in the tens to few hundreds of km\,s$^{-1}$ (i.e., a few MHz to hundreds of MHz), see e.g., \cite{COwidth}. The highest spectral resolution of the bandpass data is 0.5\,GHz.}, the CO intensity integral can be approximated by a delta function at each corresponding CO frequency, i.e.,
\begin{equation}
\label{eq:COInt}
\int \!\!d\nu \tau(\nu)dI_{\mbox{\scalebox{0.75}{CO}}} \cong \tau(\nu_{\mbox{\scalebox{0.75}{CO}}})\left( \Delta T_{\mbox{\scalebox{0.75}{CO}}} \right)\left(\displaystyle\frac{\nu_{\mbox{\scalebox{0.75}{CO}}}}{c}\right) \left.b_{\mbox{\scalebox{0.6}{RJ}}}'\right|_{\nu_{\mbox{\scalebox{0.75}{CO}}}} \quad \left[\frac{\mbox{W}}{\mbox{m$^2$\,sr}}\right] \ .
\end{equation}

\subsubsection{Coefficient Formulae}
\label{sec:UcFccMath2}

The \Planck\ spectral response conversion coefficients are therefore given by the following ratios
\begin{equation}
\label{UC:KCMB_MJysr} U\mbox{({\tiny K$_{\mbox{\scalebox{0.75}{CMB}}}$\,to\,MJy\,sr$^{-1}$})} = \displaystyle\frac{\displaystyle\int\!\! d\nu ~\tau(\nu) b_\nu'}{\displaystyle\int\!\! d\nu ~\tau(\nu) (\nu_{\mbox{\tiny{c}}} / \nu)} \times 10^{20} \quad \left[ \frac{\mbox{MJy}/\mbox{sr}}{\mbox{K$_{\mbox{\scalebox{0.6}{CMB}}}$}} \right] \ , 
\end{equation}
\begin{equation}
\begin{array}{l}
\label{UC:KCMB_ySZ} 
U\mbox{({\tiny K$_{\mbox{\scalebox{0.75}{CMB}}}$\,to\,y$_{\mbox{\scalebox{0.75}{SZ}}}$})} = \\
\displaystyle\frac{\displaystyle\int\!\! d\nu ~\tau(\nu) b_\nu'}{\displaystyle\int \!\! d\nu \left.\left\{ ~\tau(\nu) \left(b_\nu'\right)(T)\left[\left(\displaystyle\frac{h\nu}{k T}\right)\displaystyle\frac{\exp{[h\nu/(k T)]} + 1}{\exp{[h\nu/(k T)]} - 1} - 4\right] \right|_{T_{\mbox{\scalebox{0.6}{CMB}}}}\right\}} \\
 \\
\quad \quad \quad \left[ \frac{\mbox{1}}{\mbox{K$_{\mbox{\scalebox{0.6}{CMB}}}$}} \right]
\end{array}
  \ ,
\end{equation}
\begin{equation}
\label{UC:CC} U\mbox{({\tiny IRAS\,to\,$\alpha$})} = \displaystyle\frac{\displaystyle\int\!\! d\nu ~\tau(\nu) (\nu_{\mbox{\tiny{c}}} / \nu)}{\displaystyle\int\!\! d\nu ~\tau(\nu) (\nu / \nu_{\mbox{\tiny{c}}})^\alpha} \quad \left[ \frac{\mbox{Hz}}{\mbox{Hz}} \right] \ ,
\end{equation}
\begin{equation}
\label{UC:CCMBB} U\mbox{({\tiny IRAS\,to\,$\beta$,\,$T$})}
 = \displaystyle\frac{\displaystyle\int\!\! d\nu ~\tau(\nu) (\nu_{\mbox{\tiny{c}}} / \nu)}{\displaystyle\int\!\! d\nu ~\tau(\nu) (\nu / \nu_{\mbox{\tiny{c}}})^\beta\left[\displaystyle\frac{\nu^\beta~B_\nu(\nu,T)}{\nu_{\mbox{\tiny{c}}}^{~\beta}~B_\nu(\nu_{\mbox{\tiny{c}}},T)}\right]} \quad \left[ \frac{\mbox{Hz}}{\mbox{Hz}} \right] \ ,
\end{equation}
and
\begin{equation}
\label{UC:CO} U\mbox{({\tiny CO})} = \displaystyle\frac{\tau(\nu_{\mbox{\scalebox{0.75}{CO}}})\left(\displaystyle\frac{\nu_{\mbox{\scalebox{0.75}{CO}}}}{c}\right) \left.b_{\mbox{\scalebox{0.6}{RJ}}}'\right|_{\nu_{\mbox{\scalebox{0.75}{CO}}}}}{\displaystyle\int\!\! d\nu ~\tau(\nu) b_\nu'} \quad \left[\displaystyle\frac{\mbox{K$_{\mbox{\scalebox{0.6}{CMB}}}$}}{\mbox{K$_{\mbox{\scalebox{0.6}{RJ}}}$\,km\,s$^{-1}$}}\right] \ .
\end{equation}


\subsubsection{HFI Unit Conversion and Colour Correction Coefficients}
\label{sec:HFICoeffs}

\begin{figure}
\centering
\includegraphics[width=88mm]{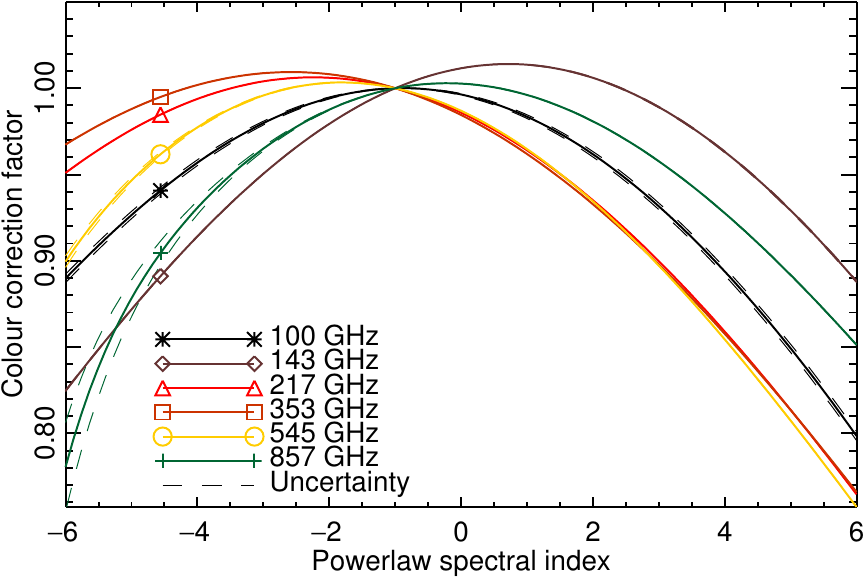}
\caption{\label{fig:CC} Colour correction coefficients for varying spectral index for the HFI band-average spectra.  Similar data 
for individual detectors and DetSet subsets are available in \cite{planck2013-p28}.}
\end{figure}

This section presents unit conversion and colour correction coefficients resultant from the above relations and HFI detector spectra.  Similar values for the LFI may be found in \citet[e.g., Table 8]{planck2013-p02}.
In addition to processing data from individual detectors, HFI data are processed to provide band-average frequency maps, and sub-band average 
frequency maps.  The sub-band frequency maps are comprised of three sets, DetSet1, DetSet2, and SWBs (i.e., spider-web bolometers only).  These groupings for the HFI detectors are summarized in Table \ref{tab:DetSet}.  Results presented here are restricted to the average spectra; similar data for individual detectors is available in \cite{planck2013-p28}.  Table \ref{tab:UCTb} provides multiplicative unit conversion coefficients to go from MJy\,sr$^{-1}$ to K$_{\mbox{\tiny{b}}}$ brightness temperature\footnote{This unit conversion is independent of spectral index, and the brightness temperature does not imply a Rayleigh-Jeans spectral profile.}. Table \ref{tab:UC} provides sample unit conversion and colour correction coefficients for the HFI band-average and sub-band-average spectra; including coefficients for CMB, SZ, and dust (with assumed $\alpha$~=~4 SED profile) sources. Figure \ref{fig:CC} illustrates the variation in colour correction coefficients for each of the HFI bands as a function of power-law spectral index. Colour correction coefficients for planets within our Solar system (i.e., HFI calibration sources) have also been determined.  Table \ref{tab:planetCC} provides these coefficients for the band-average and sub-band-average spectra; these are needed for the HFI beam calibration \citep{planck2013-p03c}.  The CO conversion coefficients are provided in Sect.~\ref{sec:CO} and in \cite{planck2013-p28}.  

The multiplicative unit conversion and colour correction coefficients are to be used as follows.  Take, for example, a dust region within a \Planck\ 100\,GHz band-average map with an estimated intensity of 10\,K$_{\mbox{\scalebox{0.6}{CMB}}}$\@.  To convert this intensity to an equivalent specific intensity in MJy\,sr$^{-1}$ (IRAS), the original intensity should be multiplied by the unit conversion coefficient of 244.1 (see Table~\ref{tab:UC}) to obtain a brightness intensity of 2441 MJy\,sr$^{-1}$.  If the dust can be approximated to follow a SED spectral profile with $\alpha$~=~4, then the colour correction would be applied by multiplying the 2441 MJy\,sr$^{-1}$ by the colour correction coefficient of 0.8938 to convert \emph{from} $\alpha$~=~$-$1 \emph{to} $\alpha$~=~4; yielding 2182 MJy\,sr$^{-1}$.  The fourth column of Table~\ref{tab:UC} provides the combined unit conversion and colour correction as the product of the second and third columns.  If the same 10\,K$_{\mbox{\scalebox{0.6}{CMB}}}$ intensity were instead found within a 100\,GHz DetSet1 map, then the resulting colour-corrected dust intensity would be 2176\,MJy\,sr$^{-1}$.  To colour correct \emph{from} $\alpha$~=~4 \emph{to} $\alpha$~=~$-$1, one would divide the intensity by the colour correction coefficient found in the third column of Table~\ref{tab:UC}.  


\begin{table}[tmb]                 
\begin{center}
\begingroup
\newdimen\tblskip \tblskip=5pt
\caption{HFI flux density to brightness temperature unit conversion coefficients.  As defined in the text, these coefficients are the same for every detector within a given frequency band.}                          
\label{tab:UCTb}                            
\nointerlineskip
\vskip -3mm
\footnotesize
\setbox\tablebox=\vbox{
   \newdimen\digitwidth 
   \setbox0=\hbox{\rm 0} 
   \digitwidth=\wd0 
   \catcode`*=\active 
   \def*{\kern\digitwidth}
   \newdimen\signwidth 
   \setbox0=\hbox{+} 
   \signwidth=\wd0 
   \catcode`!=\active 
   \def!{\kern\signwidth}
   \def\leaderfi1{\leaders\hbox to 5pt{\hss.\hss}\hfil}
%
\halign{\hbox to 1.0in{#\leaderfil}\tabskip=1em&
  \tabskip=2em\hfil#\hfil\tabskip=0em plus 1em\cr 
\noalign{\doubleline}
\omit Band$\quad$\,[GHz]& $U_C$ [K$_{\mbox{\tiny{b}}}$MJy$^{-1}$\,sr]\cr                                    
\noalign{\vskip 3pt\hrule\vskip 5pt}
100& 0.0032548074**\cr
143& 0.0015916707**\cr
217& 0.00069120334*\cr
353& 0.00026120163*\cr
545& 0.00010958025*\cr
857& 0.000044316316\cr
\noalign{\vskip 5pt\hrule\vskip 3pt}}}
\endPlancktable                 
\endgroup
\end{center}
\end{table}                        

\begin{figure}
\begin{center}
\includegraphics[width=88mm]{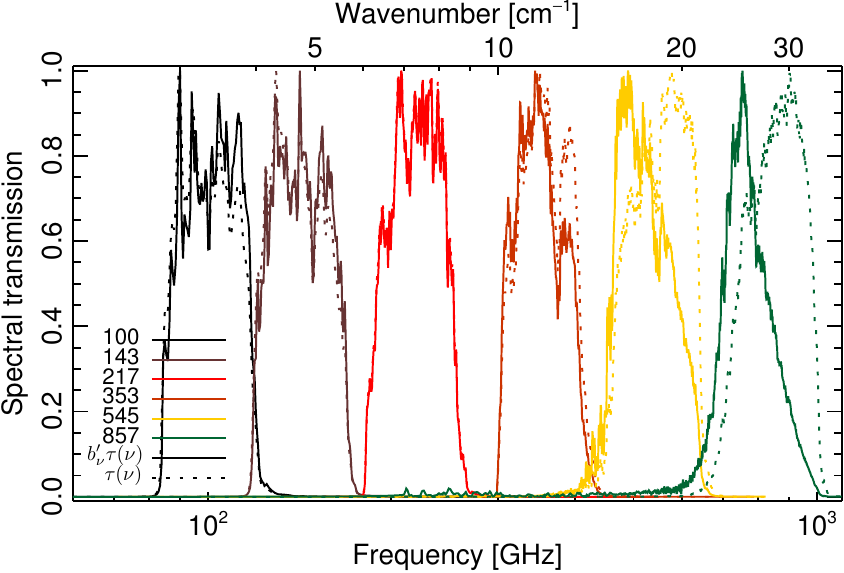}
\caption{\label{fig:CMBprod} Product of the the HFI band-average spectra, $\tau(\nu)$, with the CMB dipole spectral profile, i.e., $b_\nu'\tau(\nu)$ (Eq.\ \ref{eq:InuDeriv}) shown using solid curves.  The nominal spectra are shown as dotted curves for reference.}
\end{center}
\end{figure}

\begin{figure}
\begin{center}
\includegraphics[width=88mm]{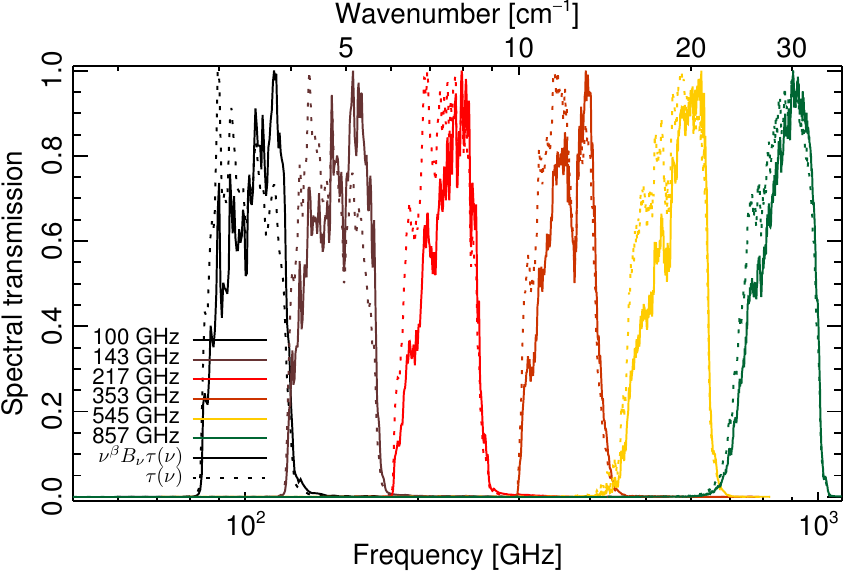}
\caption{\label{fig:dustprod} Product of the band-average spectra with an example dust spectrum ($T_d = $18\,K, $\beta_d = $1.5, Eq.\ \ref{eq:ICCMBB}) shown using solid curves.  The nominal spectra are shown as dotted curves for reference.}
\end{center}
\end{figure}

To demonstrate the relative contribution of the various regions of the spectral bands on the CMB dipole signal, Fig.\ \ref{fig:CMBprod} illustrates the $b_\nu'\tau(\nu)$ product for the band-average spectra (similar results are found for the individual detector spectra).  As discussed in Sect.~\ref{sec:UcCCPhil}, it is important to note that for the 857 GHz channel, and partly for the 545 GHz channel, the unit conversion integral (Eq.\ \ref{eq:InuDeriv} and \ref{UC:KCMB_MJysr}) is dominated by the low-frequency portion of the band.  Any residual systematics in the transmission spectra may cause undesired errors, e.g., in the conversion of 857\,GHz (or 545\,GHz) data to units of K$_{\mbox{\scalebox{0.6}{CMB}}}$\@.  It is preferred, for example, to convert the lower frequency channel data to MJy\,sr$^{-1}$ when needed, rather than converting the 857 GHz data into K$_{\mbox{\scalebox{0.6}{CMB}}}$ units.  For comparison, Fig.\ \ref{fig:dustprod} illustrates the product of the band-average spectra with a sample dust spectrum, with a modified blackbody of dust temperature, $T_d = $18\,K and $\beta_d = $1.5 (see Eq.\ \ref{eq:ICCMBB}).  

\begin{table*}[tmb]                 
\begin{center}
\begingroup
\newdimen\tblskip \tblskip=5pt
\caption{Example HFI unit conversion and colour correction coefficients for the band-average and sub-band-average spectra.}                          
\label{tab:UC}                            
\nointerlineskip
\vskip -3mm
\footnotesize
\setbox\tablebox=\vbox{
   \newdimen\digitwidth 
   \setbox0=\hbox{\rm 0} 
   \digitwidth=\wd0 
   \catcode`*=\active 
   \def*{\kern\digitwidth}
   \newdimen\signwidth 
   \setbox0=\hbox{+} 
   \signwidth=\wd0 
   \catcode`!=\active 
   \def!{\kern\signwidth}
   \def\leaderfi1{\leaders\hbox to 5pt{\hss.\hss}\hfil}
%
\halign{\hbox to 1.0in{#\leaderfil}\tabskip=1em&
  \tabskip=2em\hfil#\hfil\tabskip=2em&
  \tabskip=2em\hfil#\hfil\tabskip=2em&
  \tabskip=2em\hfil#\hfil\tabskip=2em&
  \tabskip=2em\hfil#\hfil\tabskip=0em\cr 
\noalign{\doubleline}
\omit Detector& $U_{\mbox{\tiny{C~IRAS}}}$& $C_{\mbox{\tiny{C~dust}}}$& $(U_{\mbox{\tiny{C~IRAS}}})(C_{\mbox{\tiny{C~dust}}})$& $U_{\mbox{\tiny{C~SZ}}}$\cr
\noalign{\vskip 3pt}
\omit Set& [MJy\,sr$^{-1}$\,K$_{\mbox{\scalebox{0.6}{CMB}}}^{-1}$]& & [MJy\,sr$^{-1}$\,K$_{\mbox{\scalebox{0.6}{CMB}}}^{-1}$]& [K$_{\mbox{\scalebox{0.6}{CMB}}}^{-1}$]\cr
\noalign{\vskip 3pt\hrule\vskip 5pt}
100-avg&      244.1** $\pm$ 0.3**& 0.8938* $\pm$ 0.0019*& 218.2** $\pm$ 0.3**& -0.24815* $\pm$ 0.00007*\cr
100-DetSet1&  244.9** $\pm$ 0.4**& 0.889** $\pm$ 0.003**& 217.6** $\pm$ 0.4**& -0.24833* $\pm$ 0.00010*\cr
100-DetSet2&  243.8** $\pm$ 0.4**& 0.896** $\pm$ 0.003**& 218.4** $\pm$ 0.4**& -0.24807* $\pm$ 0.00009*\cr
\noalign{\vskip 4pt}
143-avg&      371.74* $\pm$ 0.07*& 0.9632* $\pm$ 0.0004*& 358.04* $\pm$ 0.07*& -0.35923* $\pm$ 0.00006*\cr
143-DetSet1&  365.03* $\pm$ 0.15*& 1.0058* $\pm$ 0.0009*& 367.15* $\pm$ 0.15*& -0.35398* $\pm$ 0.00011*\cr
143-DetSet2&  369.30* $\pm$ 0.13*& 0.9773* $\pm$ 0.0008*& 360.93* $\pm$ 0.13*& -0.35743* $\pm$ 0.00010*\cr
143-SWBs&     378.58* $\pm$ 0.14*& 0.9238* $\pm$ 0.0008*& 349.74* $\pm$ 0.14*& -0.36446* $\pm$ 0.00011*\cr
\noalign{\vskip 4pt}
217-avg&      483.690 $\pm$ 0.012& 0.85895 $\pm$ 0.00011& 415.465 $\pm$ 0.012& *5.152*** $\pm$ 0.006***\cr
217-DetSet1&  480.36* $\pm$ 0.02*& 0.88411 $\pm$ 0.00016& 424.69* $\pm$ 0.02*& *7.212*** $\pm$ 0.019***\cr
217-DetSet2&  480.314 $\pm$ 0.019& 0.88235 $\pm$ 0.00017& 423.804 $\pm$ 0.019& *7.046*** $\pm$ 0.018***\cr
217-SWBs&     486.331 $\pm$ 0.018& 0.84069 $\pm$ 0.00015& 408.855 $\pm$ 0.018& *4.236*** $\pm$ 0.006***\cr
\noalign{\vskip 4pt}
353-avg&      287.450 $\pm$ 0.009& 0.85769 $\pm$ 0.00011& 246.543 $\pm$ 0.009& *0.161098 $\pm$ 0.000011\cr
353-DetSet1&  289.620 $\pm$ 0.012& 0.88255 $\pm$ 0.00015& 255.606 $\pm$ 0.012& *0.163757 $\pm$ 0.000014\cr
353-DetSet2&  287.967 $\pm$ 0.013& 0.86548 $\pm$ 0.00014& 249.229 $\pm$ 0.013& *0.160904 $\pm$ 0.000014\cr
353-SWBs&     286.786 $\pm$ 0.011& 0.84997 $\pm$ 0.00014& 243.759 $\pm$ 0.011& *0.160456 $\pm$ 0.000013\cr
\noalign{\vskip 4pt}
545-avg&      *58.04* $\pm$ 0.03*& 0.85444 $\pm$ 0.00016& *49.59* $\pm$ 0.03*& *0.06918* $\pm$ 0.00003*\cr
545-DetSet1&  *58.02* $\pm$ 0.03*& 0.8513* $\pm$ 0.0002*& *49.39* $\pm$ 0.03*& *0.06924* $\pm$ 0.00004*\cr
545-DetSet2&  *58.06* $\pm$ 0.05*& 0.8612* $\pm$ 0.0003*& *50.00* $\pm$ 0.05*& *0.06905* $\pm$ 0.00005*\cr
\noalign{\vskip 4pt}
857-avg&      **2.27* $\pm$ 0.03*& 0.9276* $\pm$ 0.0002*& **2.09* $\pm$ 0.03*& *0.0380** $\pm$ 0.0004**\cr
857-DetSet1&  **2.26* $\pm$ 0.03*& 0.9231* $\pm$ 0.0003*& **2.08* $\pm$ 0.03*& *0.0380** $\pm$ 0.0005**\cr
857-DetSet2&  **2.27* $\pm$ 0.04*& 0.9333* $\pm$ 0.0003*& **2.11* $\pm$ 0.04*& *0.0380** $\pm$ 0.0005**\cr
\noalign{\vskip 5pt\hrule\vskip 3pt}}}
\endPlancktablewide                 
\endgroup
\end{center}
\end{table*}                        

\begin{figure}
\begin{center}
\includegraphics[width=88mm]{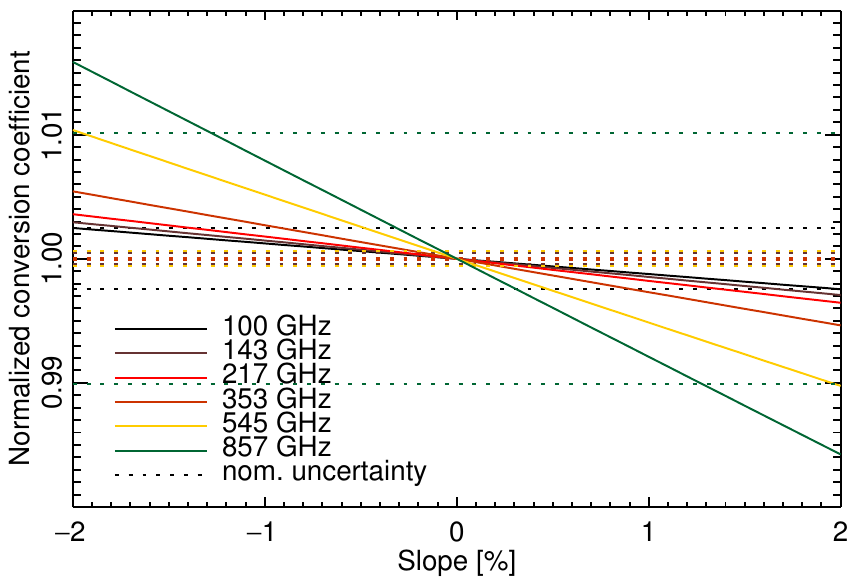}
\caption{\label{fig:UcCCslope} Variation of combined unit conversion and colour correction with a hypothetical systematic spectral error as described in the text.  The slope range $\in[-\mbox{2\,\%},\mbox{2\,\%}]$ corresponds to the value of the linear error at the $\nu_{\mbox{\tiny{c}}} \pm 15$\% band edges. The horizontal dotted lines represent the nominal uncertainty (nom.) corresponding to the 0\% slope case.}
\end{center}
\end{figure}

Another investigation was conducted in order to understand the effect of a hypothetical systematic bias in the spectral response data.  This study involved scaling the transmission spectra by a scaling term with a linear dependence on frequency, such that it is unity valued at the nominal band centre (i.e., 100, 143, 217, 353, 545, and 857\,GHz for the respective bands), is centred on the nominal band centre frequency, and has a linear deviation towards a specified value, $m_{\mbox{\tiny{ref}}}$, at the $\nu_{\mbox{\tiny{c}}}\pm 15$\,\% band edges.  Fig.~\ref{fig:UcCCslope} illustrates the normalized variation of a combined unit conversion and colour correction for each of the HFI bands, over a range of linear slopes spanning $m_{\mbox{\tiny{ref}}}\in[-\mbox{2\,\%},\mbox{2\,\%}]$.  The selected example illustrates a conversion from K$_{\mbox{\scalebox{0.6}{CMB}}}$ to MJy\,sr$^{-1}$ and a colour correction from $\alpha$~=~$-$1 to the dust profile described above.  The spectral uniformity of the reference bolometer used in characterizing the HFI detector spectral response (see \citealt{planck2013-p28}) is estimated to be at the level of 1\,\%; this is the  motivation behind the type of systematic distortion introduced in this study.  The figure demonstrates that a systematic reference spectral flatness error, as described above, results in biased unit conversion and colour correction coefficients.  The introduced coefficient bias has a magnitude that scales linearly with the slope of the spectral flatness systematic error introduced.  The 857\,GHz channel fluctuates the most due to the dominance of the unit conversion on the low-frequency region of the band, and the dominance of the selected colour correction on the high-frequency portion of the band, as illustrated in Figs.~\ref{fig:CMBprod} and \ref{fig:dustprod}.

\subsubsection{Unit Conversion and Colour Correction Software Tools}
\label{sec:UcCCcode}

A set of software tools, written in the Interactive Data Language (IDL) has been developed for distribution with the HFI detector spectra and \Planck\ data.
This tool package, herein the {\tt UcCC} package, uses the transmission spectra provided in a {\tt .fits} file format, and computes the unit conversion and colour 
correction factors using the relations derived above.  The {\tt UcCC} tools may be used to determine colour corrections for a variety of spectral profiles, 
including powerlaw and modified blackbody as described above as well as user defined source spectra.  these tools also provide coefficient uncertainty 
as an optional output.  The {\tt UcCC} code package may be obtained from the PLA, with further details on their use 
provided in \cite{planck2013-p28}.

\begin{table*}[bmt]                 
\reftabstrt 
\begingroup
\newdimen\tblskip \tblskip=5pt
\caption{Planet colour correction coefficients for the HFI band-average and sub-band-average spectra.}                          
\label{tab:planetCC}                            
\nointerlineskip
\vskip -3mm
\footnotesize
\setbox\tablebox=\vbox{
   \newdimen\digitwidth 
   \setbox0=\hbox{\rm 0} 
   \digitwidth=\wd0 
   \catcode`*=\active 
   \def*{\kern\digitwidth}
   \newdimen\signwidth 
   \setbox0=\hbox{+} 
   \signwidth=\wd0 
   \catcode`!=\active 
   \def!{\kern\signwidth}
   \def\leaderfi1{\leaders\hbox to 5pt{\hss.\hss}\hfil}
%
\halign{\hbox to 0.85in{#\leaderfil}\tabskip=1em&
  \tabskip=2em\hfil#\hfil\tabskip=2em plus 4em minus 1em&
  \tabskip=2em\hfil#\hfil\tabskip=2em plus 4em minus 1em&
  \tabskip=2em\hfil#\hfil\tabskip=2em plus 4em minus 1em&
  \tabskip=2em\hfil#\hfil\tabskip=2em plus 4em minus 1em&
  \tabskip=2em\hfil#\hfil\tabskip=0em\cr 
\noalign{\doubleline}
\omit Band& Mars& Jupiter& Saturn& Uranus& Neptune\cr                                    
\noalign{\vskip 3pt\hrule\vskip 5pt}
%
100-avg&        0.9613* $\pm$ 0.0017*& 0.962** $\pm$ 0.005**& 0.963** $\pm$ 0.004**& 0.9692* $\pm$ 0.0011*& 0.9741* $\pm$ 0.0010*\cr
100-DetSet1&    0.9581* $\pm$ 0.0019*& 0.959** $\pm$ 0.005**& 0.960** $\pm$ 0.004**& 0.9663* $\pm$ 0.0013*& 0.9713* $\pm$ 0.0013*\cr
100-DetSet2&    0.9625* $\pm$ 0.0019*& 0.963** $\pm$ 0.005**& 0.964** $\pm$ 0.004**& 0.9704* $\pm$ 0.0013*& 0.9752* $\pm$ 0.0013*\cr
\noalign{\vskip 4pt}
143-avg&        1.0069* $\pm$ 0.0009*& 1.008** $\pm$ 0.004**& 1.008** $\pm$ 0.003**& 1.0122* $\pm$ 0.0002*& 1.0127* $\pm$ 0.0002*\cr
143-DetSet1&    1.0333* $\pm$ 0.0009*& 1.034** $\pm$ 0.004**& 1.034** $\pm$ 0.003**& 1.0355* $\pm$ 0.0004*& 1.0364* $\pm$ 0.0004*\cr
143-DetSet2&    1.0153* $\pm$ 0.0009*& 1.016** $\pm$ 0.004**& 1.016** $\pm$ 0.003**& 1.0197* $\pm$ 0.0003*& 1.0203* $\pm$ 0.0003*\cr
143-SWBs&       0.9799* $\pm$ 0.0009*& 0.981** $\pm$ 0.004**& 0.982** $\pm$ 0.003**& 0.9882* $\pm$ 0.0003*& 0.9884* $\pm$ 0.0003*\cr
\noalign{\vskip 4pt}
217-avg&        0.9355* $\pm$ 0.0002*& 0.9376* $\pm$ 0.0016*& 0.9529* $\pm$ 0.0010*& 0.94897 $\pm$ 0.00004& 0.96195 $\pm$ 0.00004\cr
217-DetSet1&    0.9523* $\pm$ 0.0002*& 0.9542* $\pm$ 0.0016*& 0.9687* $\pm$ 0.0010*& 0.96368 $\pm$ 0.00007& 0.97572 $\pm$ 0.00007\cr
217-DetSet2&    0.9515* $\pm$ 0.0002*& 0.9534* $\pm$ 0.0016*& 0.9679* $\pm$ 0.0010*& 0.96305 $\pm$ 0.00007& 0.97572 $\pm$ 0.00007\cr
217-SWBs&       0.9216* $\pm$ 0.0002*& 0.9239* $\pm$ 0.0016*& 0.9398* $\pm$ 0.0010*& 0.93669 $\pm$ 0.00006& 0.95021 $\pm$ 0.00006\cr
\noalign{\vskip 4pt}
353-avg&        0.93365 $\pm$ 0.00005& 0.9385* $\pm$ 0.0004*& 0.94145 $\pm$ 0.00014& 0.94870 $\pm$ 0.00004& 0.93676 $\pm$ 0.00004\cr
353-DetSet1&    0.94972 $\pm$ 0.00006& 0.9540* $\pm$ 0.0004*& 0.95681 $\pm$ 0.00015& 0.96253 $\pm$ 0.00005& 0.95080 $\pm$ 0.00006\cr
353-DetSet2&    0.93749 $\pm$ 0.00006& 0.9420* $\pm$ 0.0004*& 0.94466 $\pm$ 0.00015& 0.95155 $\pm$ 0.00005& 0.94023 $\pm$ 0.00006\cr
353-SWBs&       0.92806 $\pm$ 0.00006& 0.9331* $\pm$ 0.0004*& 0.93620 $\pm$ 0.00015& 0.94397 $\pm$ 0.00005& 0.93183 $\pm$ 0.00006\cr
\noalign{\vskip 4pt}
545-avg&        0.93603 $\pm$ 0.00010& 0.92946 $\pm$ 0.00009& 0.81721 $\pm$ 0.00008& 0.95148 $\pm$ 0.00009& 0.96409 $\pm$ 0.00009\cr
545-DetSet1&    0.93449 $\pm$ 0.00011& 0.92691 $\pm$ 0.00011& 0.81500 $\pm$ 0.00010& 0.95027 $\pm$ 0.00011& 0.96262 $\pm$ 0.00011\cr
545-DetSet2&    0.93930 $\pm$ 0.00017& 0.93488 $\pm$ 0.00017& 0.82192 $\pm$ 0.00015& 0.95405 $\pm$ 0.00016& 0.96720 $\pm$ 0.00017\cr
\noalign{\vskip 4pt}
857-avg&        0.98273 $\pm$ 0.00013& 0.99918 $\pm$ 0.00014& 0.99923 $\pm$ 0.00014& 0.99210 $\pm$ 0.00013& 0.99815 $\pm$ 0.00013\cr
857-DetSet1&    0.98017 $\pm$ 0.00018& 0.99703 $\pm$ 0.00019& 0.99690 $\pm$ 0.00019& 0.99005 $\pm$ 0.00018& 0.99605 $\pm$ 0.00018\cr
857-DetSet2&    0.98607 $\pm$ 0.00020& 1.0020* $\pm$ 0.0002*& 1.0023* $\pm$ 0.0002*& 0.9948* $\pm$ 0.0002*& 1.0009* $\pm$ 0.0002*\cr
\noalign{\vskip 5pt\hrule\vskip 3pt}}}
\endPlancktablewide                 
\endgroup
\reftabend 
\end{table*}                        

\subsection{Error Propagation}
\label{sec:erMC}

The uncertainty of the HFI detector spectral response and band-average spectral response was propagated to the coefficient factors described above.  For each of the correction factors shown (Eq.~\ref{UC:KCMB_MJysr}--\ref{UC:CO}), the measured spectral transmission profile, $\tau(\nu)$, is used along with the associated spectral uncertainty.  An additional uncorrelated uncertainty was repeatedly introduced to the spectrum under consideration, where a Gaussian noise distribution in frequency space was weighted by the uncertainty in $\tau(\nu)$, with the resultant spectrum plus noise then used in the calculation of the desired coefficient.  The uncertainty values for coefficients shown in this work correspond to the statistical evaluation of 10\,000 trials in each instance.  These uncertainties reflect only the propagation of the spectral uncertainty into the coefficient and may not represent the full uncertainty in every case.  An evaluation of some potential sources of systematic uncertainty is included in Sect.~\ref{sec:Disc} below.  


\subsection{Band-average Coefficient Maps} 
\label{sec:CoeffMaps}

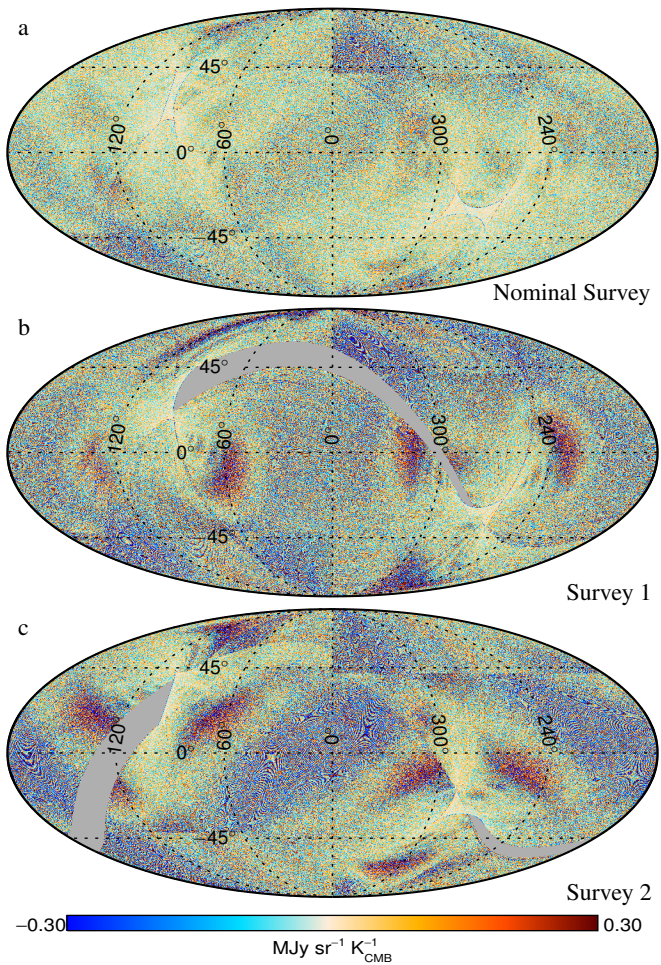
\begin{figure}
\begin{center}
\begin{tabular}{c}
\begin{overpic}[trim= 0 25 0 0, clip, width=88mm]{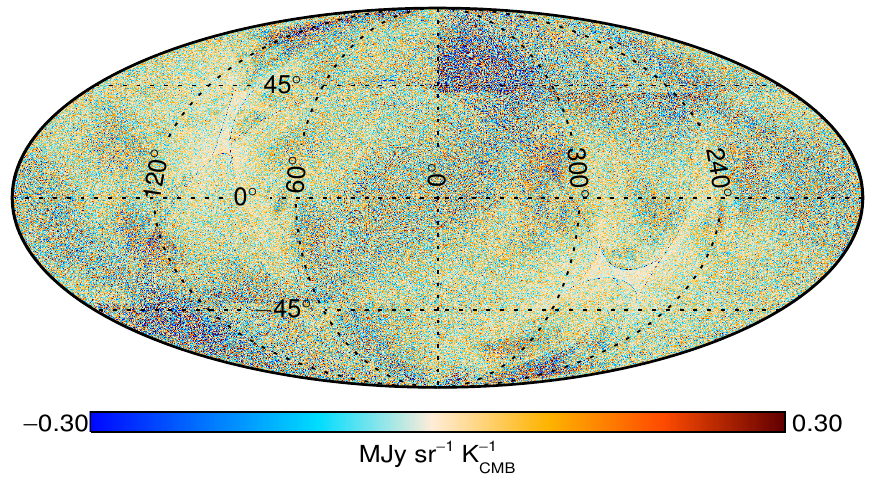} 
\put(3,40){a}
\put(74,0){Nominal Survey}
\end{overpic}\\
\begin{overpic}[trim= 0 25 0 0, clip, width=88mm]{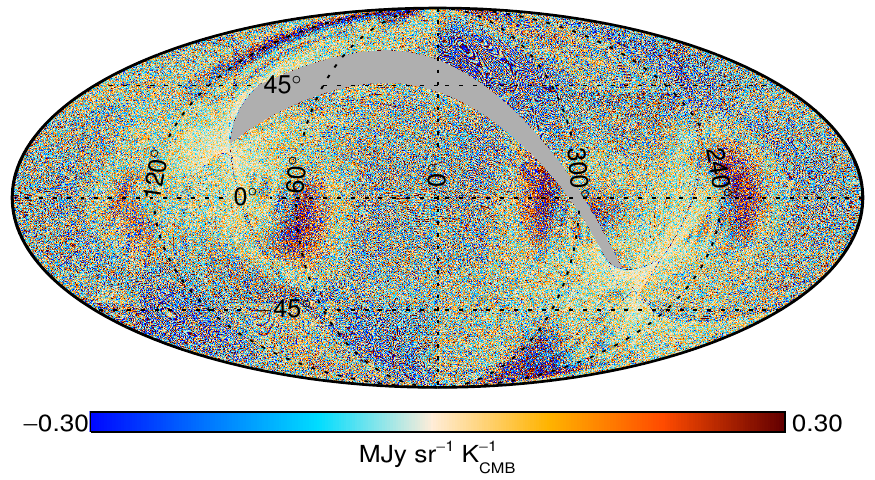} 
\put(3,40){b}
\put(85,0){Survey 1}
\end{overpic}\\
\begin{overpic}[trim= 0 0 0 0, clip, width=88mm]{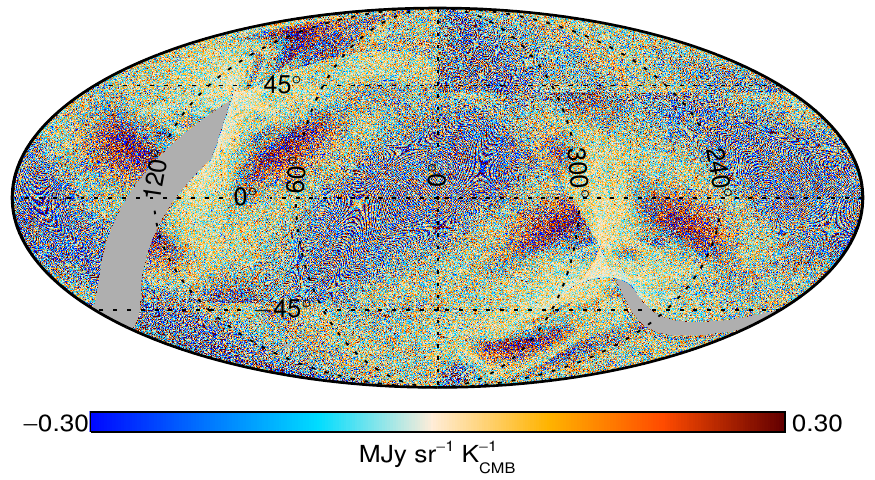} 
\put(3,50){c}
\put(85,10){Survey 2}
\end{overpic}\\
\end{tabular}
\caption{\label{fig:BandAvgMaps} 353\,GHz band-average dust unit conversion and colour correction coefficient maps.  The maps show the deviation of the coefficient about the median.  The medians in each case are 278.970 (a), 278.960 (b), and 278.970 (c) MJy\,sr$^{-1}$\,K$_{\mbox{\tiny{CMB}}}^{-1}$.}
\end{center}
\end{figure}

Using the $W_{m,i}(\theta,\phi)$ maps described above (Eq.~\ref{eq:HitNoiseNorm}), band-average spectra may be computed for every pixel of a sky map (rather than integrating the relative weights across the map).  Thus, for frequency channel maps, coefficient sky maps may be generated for individual surveys and combinations of surveys.  For individual detectors, the response is expected to be constant across the sky, but for channel average data, the sky coverage and relative noise causes variations in the proportional averaging.  An example band-average coefficient map is shown in Fig.~\ref{fig:BandAvgMaps}, where the 353 GHz combined K$_{\mbox{\scalebox{0.6}{CMB}}}$ to MJy\,sr$^{-1}$ (IRAS) unit conversion and IRAS to dust colour corrections (see Sect.~\ref{sec:DustCC}) are shown for the nominal and individual surveys.  

\renewcommand{\tabcolsep}{0em}

These coefficient maps can be used to investigate differences between surveys, and the coefficient variation across a region of interest can be compared to the magnitude of other sources of error, providing a probe of the effect of scan strategy, integration time, and relative intra-band detector noise levels on map consistency and unit conversion coefficients.  Histograms of the coefficient sky maps provide a verification of the Monte Carlo derived uncertainty estimates in much the same way that the weight factor histograms verify the band-average scaling factors (see Fig.\ \ref{fig:Sfullwhist}).  For coefficient map histogram distributions that are relatively narrow, i.e., of the same width as the Monte Carlo simulation based uncertainty estimates, data processing and analysis may be simplified by using a single constant conversion coefficient in place of a coefficient map.  While these maps are not being provided under the current data release, the discussed histograms have been used to validate the use of the scalar unit conversion and colour correction coefficients (as opposed to using the discussed coefficient maps).  The maps within Fig.~\ref{fig:BandAvgMaps} demonstrate the variation of the band-average unit conversion and colour correction coefficient to be small: typically less than 0.1\,\%.  Histograms of these maps are shown below (see Sect.\ \ref{sec:DustCC} and Fig.~\ref{fig:DustUcCChist}). This discussion will continue in greater detail within future work as these magnitudes of coefficient fluctuations at the map level are expected to become more important with polarization analysis.  


\section{Discussion}
\label{sec:Disc}

This section evaluates the HFI detector and band-average spectral response data, and the associated unit conversion and colour correction algorithms and coefficients.  This is done through comparison of these data with HFI flight data.  HFI detector bandpass mismatch, i.e., the relative difference between an individual detector spectrum and its band-average counterpart, can be compared with variations between individual detector and channel average detector results.  Examples include CO, SZ, dust, etc.  Out of band signal rejection is verified through comparisons with the HFI zodiacal light observations, where any out-of-band sky signal received would be evident in the data.   As the progress of the HFI polarization data analysis advances, comparisons between spectral response predicted polarization leakage and observed intensity to polarization leakage in HFI maps can also be used to verify the accuracy of the HFI spectral response data.  



\subsection{Out-of-band Signal Rejection}
\label{sec:OOB}

Each of the HFI bands has a filter stack of 5 low-pass filters with varying cut-off frequencies in order to achieve suitable out-of-band
signal rejection. In addition, the 545 and 857 bands also have a high-pass filter dictating the cut-on frequency, which also serves to provide some rejection at much higher frequencies. When the transmission of the individual filters comprising each HFI filter stack was measured, a series of FTS scans were recorded with external low pass filters up to 650 \cm\ (approximately 19.5\,THz, 15.4 $\mu$m).  This was not done as a single measurement but a series of experiments to allow the transmission in-band to be measured to higher resolution and SNR\@.  For each of the low pass-filters, the transmission at the high-frequency end dropped to about 10$^{-4}$ to 10$^{-6}$ beyond the cut-off frequency.  Given that there are 5 filters in a stack, the out-of-band rejection is very high.  The spectral transmission for all frequencies greater than 3\,THz is less than 10$^{-15}$ for all HFI bands, reaching as low as 10$^{-25}$ or 10$^{-30}$ in some cases.  This level of out-of-band signal rejection is confirmed by the zodiacal light data described below.

\subsubsection{Zodiacal Light Verification}
\label{sec:Zodi}

\begin{figure}[htbp]
  \centering
  \includegraphics{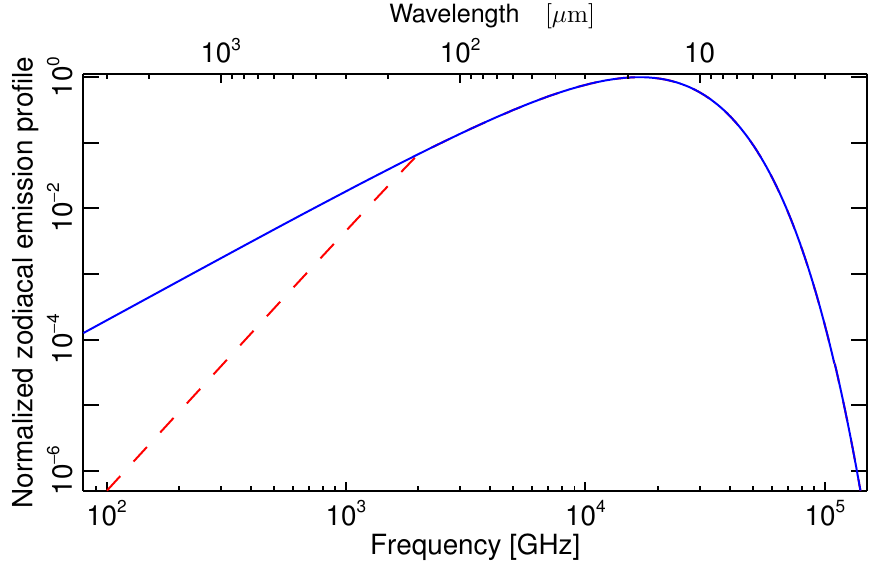}
  \caption{The spectrum of the diffuse zodiacal cloud in the ecliptic plane. The solid (blue) curve 
  represents the spectrum assuming uniform cloud emissivity of unity. The dashed (red) line shows the expected $\lambda^{-2}$ reduced emissivity proportionality for frequencies below 2\,THz (i.e., wavelengths longer than 150\,$\mu$m).}
  \label{fig:ZLeakLimit}
 \end{figure}

  
As mentioned above, optical filters were used to prevent out-of-band light from impinging on the HFI detectors and registering as signal. While these filters are thoroughly characterized and tested before flight, it is of interest to try to confirm, as well as possible, their behavior in situ. In this section, the HFI observations of the zodiacal light, specifically 100\,GHz data, are used to place an upper limit on out-of-band spectral contribution, i.e., spectral leaks.  The spectral profile of zodiacal emission is well understood (e.g., \citealt{kelsall1998}), with zodiacal emission much brighter in the mid-infrared than in the millimeter wavelength range and very little zodiacal-correlated signal observed at 100\,GHz \citep{fixsendwek2002}.  While the diffuse zodiacal cloud is observed in the higher-frequency HFI channels \citep{planck2013-pip88}, it is not detected within the lower frequency HFI channels. A model zodiacal light spectral emission profile is shown in Fig.\,\ref{fig:ZLeakLimit}, with a dashed line representing the reduced emissivity for frequencies less than 2\,THz, as expected. 

To set limits on a hypothetical leak at short wavelengths, the solid curve in Fig.\,\ref{fig:ZLeakLimit} is frequency integrated to find the expected zodiacal cloud emission flux density.  To estimate the expected in-band flux density, an in-band emissivity of approximately 0.041 (see \citealt{planck2013-pip88}) is coupled with the zodiacal emission profile at an effective blackbody temperature of 286\,K ($Z_{100~\mathrm{GHz}}$ below).  This is combined with the nominal 100 GHz channel bandwidth ($\Delta\nu_{100~\mathrm{GHz}}$ below, approximately 33\,GHz), where the product yields the expected in-band zodiacal signal.  This product is then divided by the total zodiacal flux density estimated by the frequency integration of the zodiacal spectral profile. Thus, an upper limit on any 100\,GHz high frequency spectral leakage, $L_{\mbox{\scalebox{0.6}{U}}}$, is given by:
\begin{equation}
\label{eq:ZodiLeak}
\begin{array}{rcl}
  L_{\mbox{\scalebox{0.6}{U}}}
  & < &
  {
   \displaystyle\frac{Z_{100~\mathrm{GHz}}\cdot \Delta\nu_{100~\mathrm{GHz}}}
   {\displaystyle\int_{\nu_1}^{\nu_2} d\nu\cdot I\left(\nu\right)}
  }
  \\
  &   &  \\ 
  & < &
  1.2\cdot 10^{-8}
\end{array}
\ . 
\end{equation}
 The frequency range used in the above integration is 3--150\,THz (equivalent to 100--2\,$\mu$m).  This integration region covers the dominant signal and is well above the 100 GHz band cut-off. Differences in the integration between this range and larger ones result in changes to the spectral leak upper limit at the sub-percent level, and thus do not affect the calculation at the desired accuracy. 

The lack of any \Planck\ detection of the diffuse zodiacal cloud signal at 100\,GHz \citep{planck2013-pip88} indicates high out-of-band zodiacal signal rejection, which in turn places confidence in the overall out-of-band signal rejection of the HFI detector optical filters.  Through this analysis, out-of-band rejection has been shown to be greater than $10^8$ for the HFI 100\,GHz detectors. Less stringent limits can be set on smaller subsets of the band including wavelengths greater than 100\,$\mu$m. 


\subsection{Bandpass Mismatch}
\label{sec:mismatch}

Variations of the spectral response of individual detectors from the band-average response can be compared with similar variations in HFI data (e.g., individual detector maps compared to band-average maps).  An example of this is the observation of CO emission within some of the HFI bands.  As CO emission is intrinsically narrow-band, differences in CO sensitivity for different HFI detectors are easily compared to the HFI spectral response variations.  Other components of the sky signal may also be used to perform similar comparisons (e.g., SZ, polarization leakage, dust, etc.).  

Polarization leakage is an effect where spectral mismatch between detectors in the same frequency band induces a reduction or loss of signal within an unpolarized intensity map and an over-estimation of the corresponding polarization map.  This effect, which is a result of the assumptions required within the mapmaking algorithms, can also cause polarization to intensity leakage; but this is of less concern, however, due to relatively weak amplitudes of polarization signals,  coupled with the low levels of leakage expected.  While a discussion of the polarization aspects of the HFI spectral response will be withheld until the release of \Planck\ polarization data, the concept is introduced here to provide context to the current work as the same effect that causes polarization leakage allows a sky-based estimate of the respective colour correction and unit conversion coefficients (see Sect.~\ref{sec:DustCC}).  Spectral mismatch is discussed in the context of the BICEP instrument in \cite{BICEP2011}.

\subsubsection{CO Bandpass Verification}
\label{sec:CO}

Analysis of HFI data, using component separation methods that include a CO emission component (\citealt{planck2013-p03a} and \citealt{planck2013-p06}), has provided all-sky maps of CO emission for the first three CO rotational transitions, i.e., CO $J$=1$\rightarrow$0, CO $J$=2$\rightarrow$1, and CO $J$=3$\rightarrow$2.  Select molecular cloud regions with well-known CO emission properties provide external validation of the HFI CO maps.  One of the validation observations used was the Dame Milky Way survey \citep{dame2001}, which observed the CO $J$=1$\rightarrow$0 transition.  As the HFI CO maps are the result of component separation performed on maps in K$_{\mbox{\scalebox{0.6}{CMB}}}$ units, and the external CO observations are typically available in units of velocity integrated brightness temperature, i.e., K\,km\,s$^{-1}$, this allows for a sky-based estimate of the CO unit conversion coefficients from units of K\,km\,s$^{-1}$ to units of K$_{\mbox{\scalebox{0.6}{CMB}}}$.  These coefficient estimates are then compared against the bandpass-based coefficients, i.e., those based on the pre-flight measured spectral transmission data and the respective unit conversion relation above (Eq.~\ref{UC:CO}).  Another benefit of comparing the bandpass and sky-based CO results is the potential for improved understanding of systematic uncertainties in the bandpass data.

While a detailed discussion regarding the derivation of the sky-based CO unit conversion coefficients is found elsewhere \citep{planck2013-p03a}, this work discusses only those details relevant to the comparison of the sky and bandpass CO coefficients.  The bandpass CO coefficients provide a direct conversion from differential CMB emission to that of CO, with the caveat that a precise knowledge of the CO unit conversion coefficients from the bandpass data is difficult to obtain in that the spectral resolution of the bandpass data is much broader than any observed CO emission features.  Thus, the bandpass CO coefficients are an estimate based upon under-resolved spectroscopic measurements.  The sky-based CO coefficients, on the other hand, provide a relative conversion coefficient based on the variation in spectral transmission between detectors within a common frequency band.  

The linear combination method of CO signal extraction used to obtain the HFI CO products involves using a weighted sum derived to maximize the contrast between the desired component and its residuals (see \citealt{hurier2010milca} for details).  As part of the use of this method in the CO extraction, the Dame data set is used as a calibration template for the weighted map sums within a frequency band (i.e., the \Planck\ 100, 217, or 353 GHz bands).  

As demonstrated in Eqs.~1 and 5 of \cite{planck2013-p03a}, the weighted coefficient for an individual detector signal is given by $w_i F_i$, where $w_i$ is the {\tt MILCA} relative weight, and $F_i$ is the CO unit conversion coefficient. For the sky-based CO coefficients, the coefficient estimate is based upon the correlation of the CO component separation output with the Dame survey; thus, the relative CO and CMB weighting of the bandpass spectra (i.e., the numerator and denominator of Eq.~\ref{UC:CO}) are determined indirectly, without the bandpass data. 

\begin{figure}
\begin{tabular}{c}
\begin{overpic}[width=88mm, trim=0 12 0 0,clip]{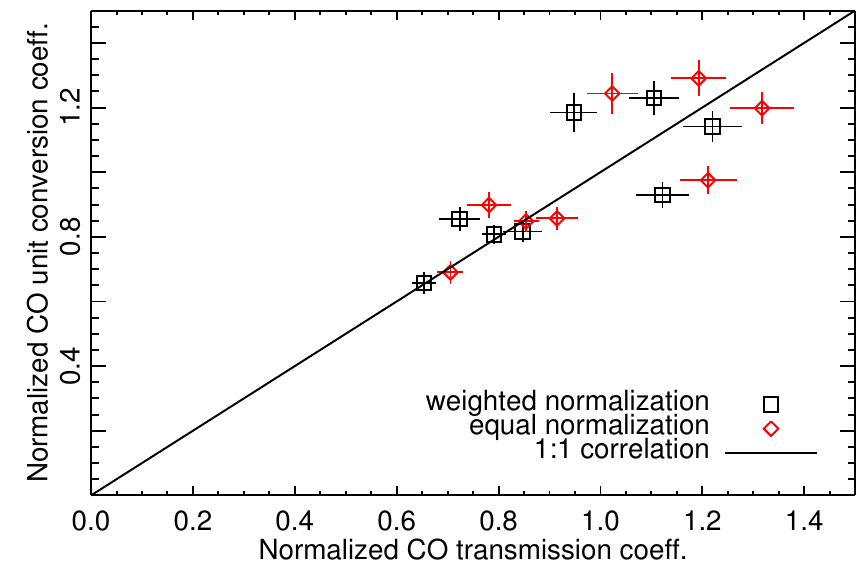} 
\put(13,56){a}
\end{overpic} \\
\begin{overpic}[width=88mm, trim=0 12 0 0, clip]{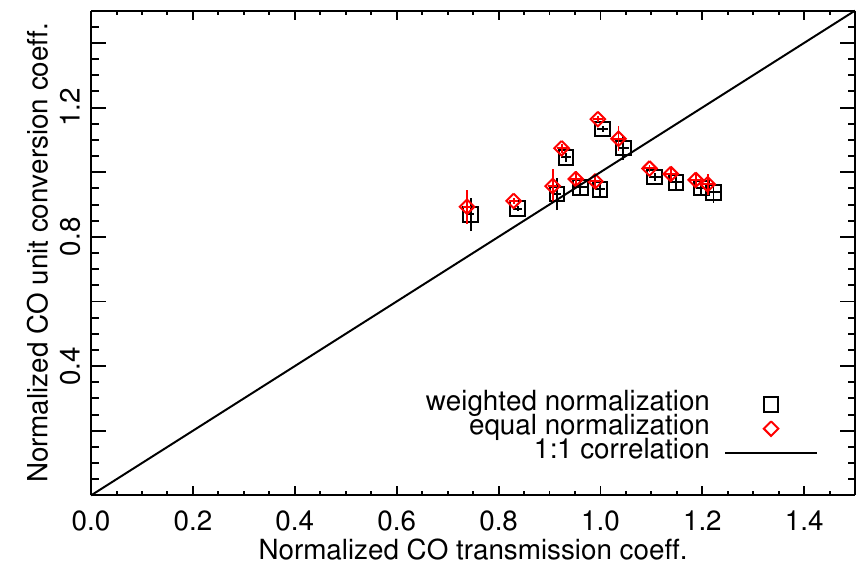} 
\put(13,56){b}
\end{overpic} \\
\begin{overpic}[width=88mm]{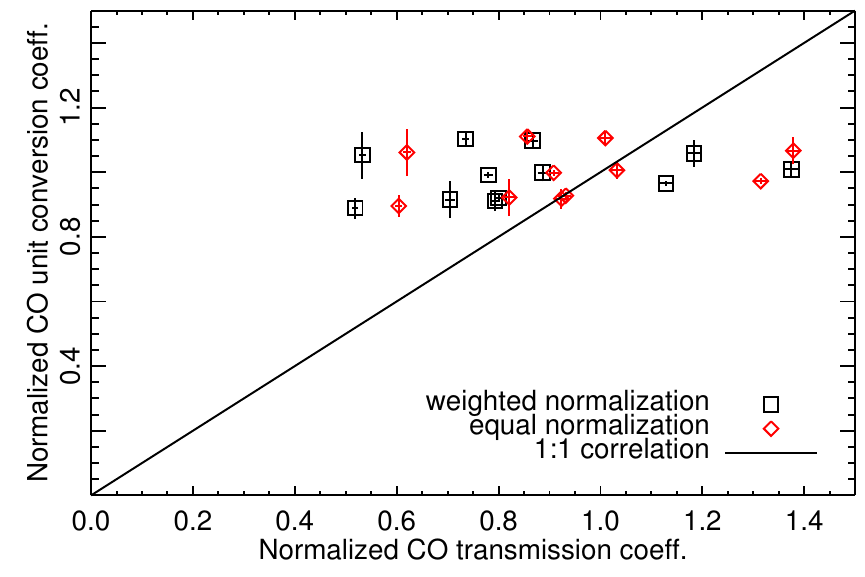} 
\put(13,61){c}
\end{overpic} 
\end{tabular}
\caption{\label{fig:CO_uc_trans} A comparison of the relative spectral transmission at the CO transition frequency with the relative CO unit conversion coefficient (see Eq.~\ref{UC:CO}) for the 100 (a), 217 (b), and 353 (c) GHz bands.  Each plot shows a uniform normalization (diamonds) where each detector coefficient contributes to the mean equally, a weighted normalization (squares) where the normalization is detector noise weighted, and a reference line to demonstrate deviations of the data points from a perfect one-to-one correlation.}
\end{figure}

To illustrate the difference between the relative transmission at the CO rotational transition frequencies and the relative unit conversion for the CO transitions (i.e., coefficients based on either only the numerator or all of Eq.~\ref{UC:CO}), Fig.\ \ref{fig:CO_uc_trans} compares these two sets of parameters. Although the two relative values demonstrate a correlation, they do not demonstrate perfect agreement.  This comparison also varies noticably with the normalization scheme used.  It is therefore important that CO unit conversion coefficients account for both the relative differences in spectral transmission at the CO rotational transition frequencies as well as the relative spectral transmission variations within the CMB signal over the entire band of a given detector.


Through modeling of CO emission in a molecular cloud as a function of CO rotation temperature, it can be shown that the relative line intensities vary with CO rotation temperature, where the lower order transitions dominate for very low temperatures (e.g., less than 10\,K).  As the relative line intensity for rotational CO transitions, e.g., the CO $J$=2$\rightarrow$1 and CO $J$=1$\rightarrow$0 transitions, varies with temperature, and the lowest CO transition is used in the Dame HFI CO map calibration, the uncertainty of the HFI CO maps is expected to vary with CO temperature. This is consistent with the uncertainty varying with the observed CO intensity.  Thus, in comparing the sky and bandpass CO coefficients, it is important to note that the relative intensities of CO emission change with CO rotational temperature and that for the higher order rotational transitions, the sky CO coefficients are a generalized scalar value representing a dynamic quantity.  

The sky-based coefficients do not provide a conversion that includes an absolute calibration, but represent the relative levels between detectors of the same frequency band; the absolute calibration comes from the external data sources used for validation.  This is a result of the linear combinations used in combining detector signals in such a way as to enhance certain spectral components while reducing other spectral contaminants \citep{hurier2010milca}.  

Another difference between the sky (e.g., Dame) and bandpass CO coefficients is the isotopic content of the calibration source emission.  The sky-based coefficients provide an estimate of CO emission from all isotopic species, while the bandpass CO coefficients are calculated for isotopic contributions individually (i.e., the emission frequencies vary with isotope).  The bandpass coefficients for individual isotopes could be combined, provided information regarding the isotopic ratios were available for a given region.  Thus, there is a normalization and rescaling step needed to compare the two values, and a complete comparison is unconstrained at present without additional calibration data.  

Additionally, the sky-based coefficients are based on unpolarized intensity, so the PSB detector pairs are combined in the form $(a + b)/2$ to reduce the influence of any polarized signal.  Any uncertainty in the data introduced as a result of this step must be propagated through to the sky CO coefficients.  

For these reasons the sky and bandpass CO coefficients are expected to show a correlation, but are not expected to exhibit perfect agreement.  This is particularly true of an absolute comparison; relative differences should be proportionate, but the absolute values of the bandpass and sky coefficients are not expected to be directly compatible \citep{planck2013-p03a}.

\begin{figure}
\begin{center}
\includegraphics{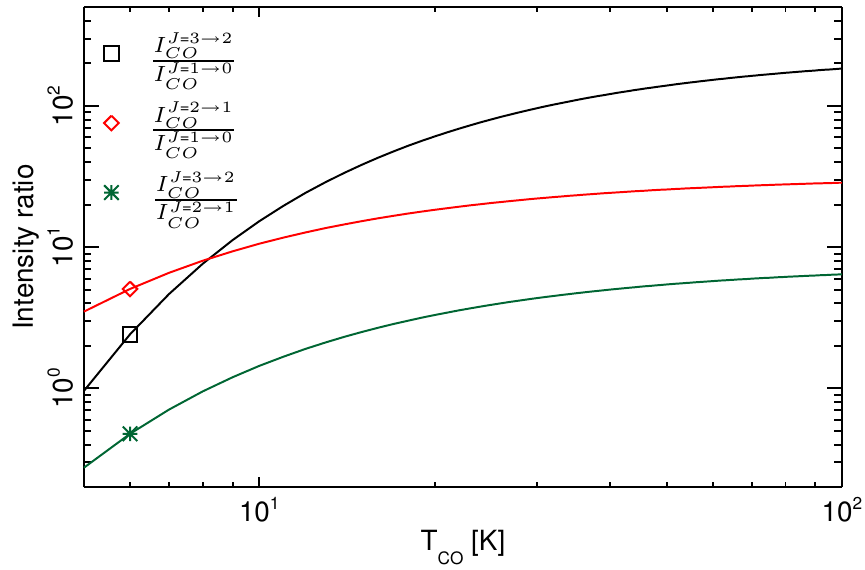}
\caption{\label{fig:COratios} Comparison of the relative CO specific intensity for the first three rotational transitions, for a varied CO 
rotation temperature.  The normalization employed removes any dependence on the column density.  Only one data point is illustrated for each curve to keep the plot legible and provide the desired curve identification.}
\end{center}
\end{figure}

As demonstrated in Eq.\ \ref{UC:CO}, the relative spectral transmission over a very narrow frequency range, in principle a delta function, is needed in order to obtain a CO unit conversion coefficient.  This ideally requires knowledge of the HFI spectral response to much finer spectral resolution than is available.  Each of the other unit conversion and colour correction calculations are based on relatively broad spectral features that are not dominated by uncertainties within a single spectral bin.  Thus, it is difficult to ensure that the nominal transmission and uncertainty for one spectral bin, used in the context of CO emission, are consistent with the values provided within the detector bandpass data.  

\begin{figure}
\begin{center}
\begin{tabular}{c}
\begin{overpic}[width=88mm]{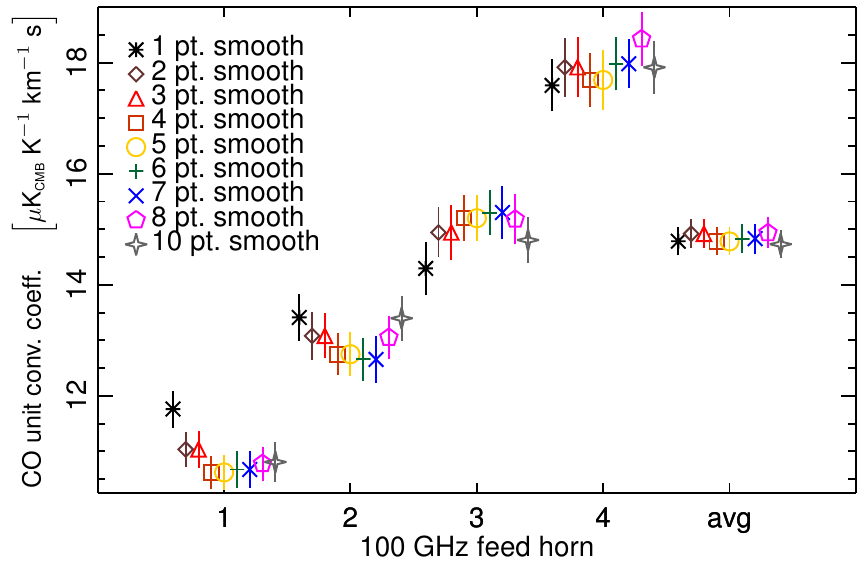} 
\put(12,63){a}
\end{overpic} \\
\begin{overpic}[width=88mm]{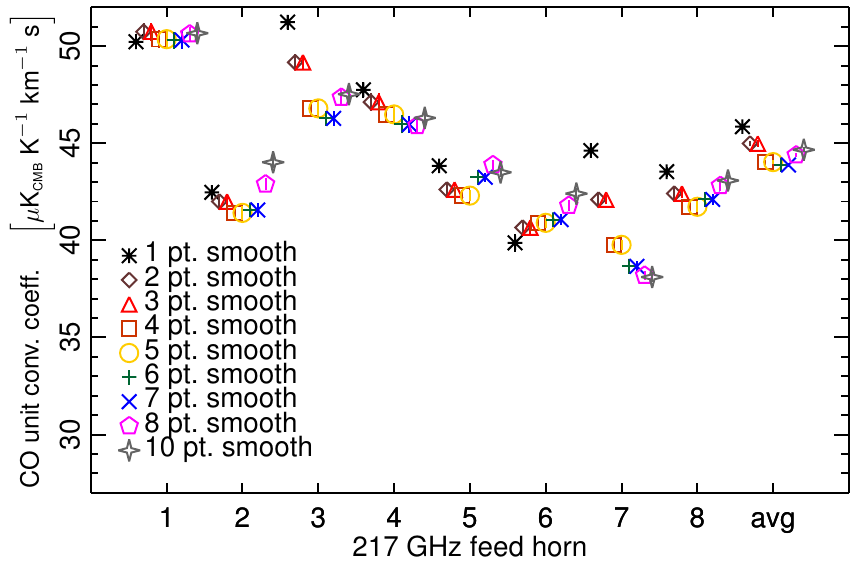} 
\put(12,62){b}
\end{overpic} \\
\begin{overpic}[width=88mm]{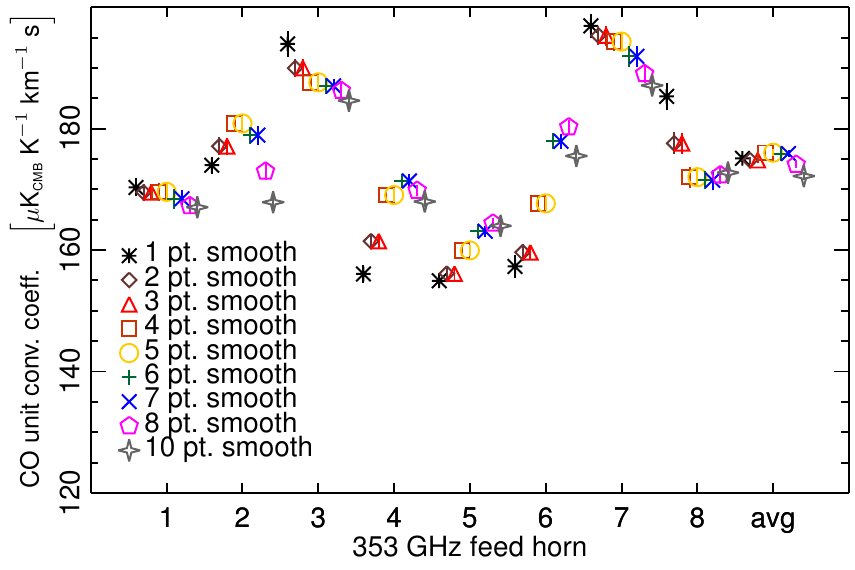} 
\put(12,62){c}
\end{overpic} 
\end{tabular}
\caption{\label{fig:COsm} Comparison of the CO unit conversion coefficients with a spectral re-binning factor applied to the unit conversion algorithm (Eq.~\ref{UC:CO}) spectral input, for the 100 (a), 217 (b), and 353 (c) GHz detectors.}
\end{center}
\end{figure}

Efforts were taken to understand the accuracy of the bandpass CO coefficients in light of the differences observed between the sky and bandpass CO coefficients. In order to investigate the effects of the CO transmission uncertainty and interpolation errors on the CO conversion coefficients, the bandpass CO coefficients were repeatedly calculated after varying levels of smoothing were applied to the spectral data.  The input spectra were smoothed in steps of one resolution element, starting with one (i.e., no smoothing) and ending with a spectral bin width of ten.  The smoothing results in noise averaging within an increasingly broad spectral bin width, so the intrinsic spectral transmission uncertainty is reduced, but the interpolation error may be increased.  Fig.~\ref{fig:COsm} illustrates the changes in CO coefficients upon introduction of this spectral smoothing.  The results shown here are for SWB detectors and combined PSB pairs, i.e., $(a+b)/2$. This is to facilitate comparison with the sky CO coefficients, which are computed in a similar fashion. It is evident that the spread in coefficient values with this smoothing factor is greater than the coefficient uncertainty in many cases.  Thus, the bandpass CO coefficients have been revised to include this result.  Within the \cite{planck2013-p03a} paper, this is reflected as an increase in the error bars of the bandpass CO coefficients.

\begin{figure}
\begin{center}
\begin{tabular}{c}
\begin{overpic}[width=88mm]{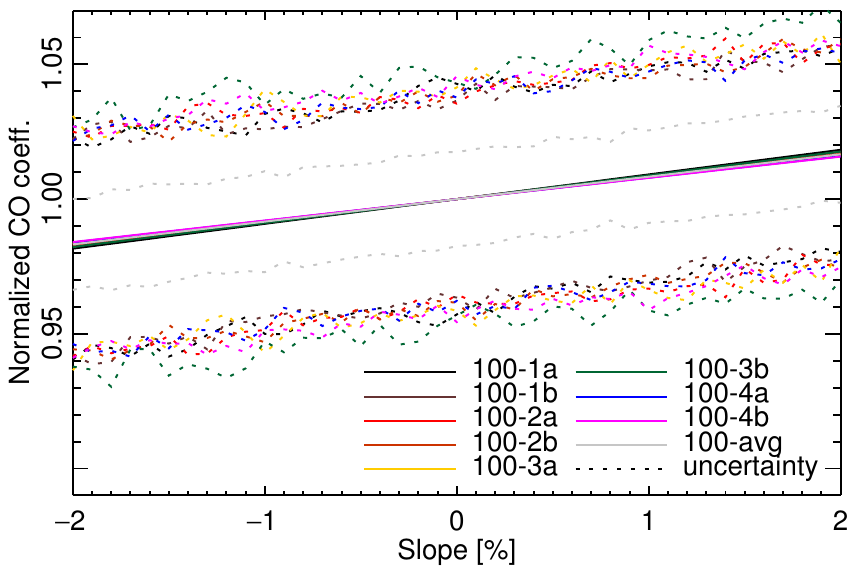} 
\put(10,62){a}
\end{overpic} \\
\begin{overpic}[width=88mm]{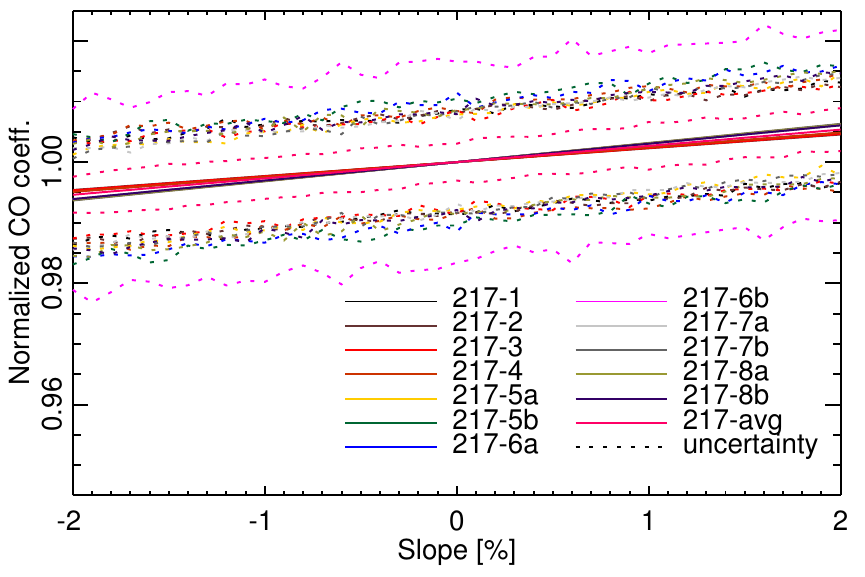} 
\put(10,62){b}
\end{overpic} \\
\begin{overpic}[width=88mm]{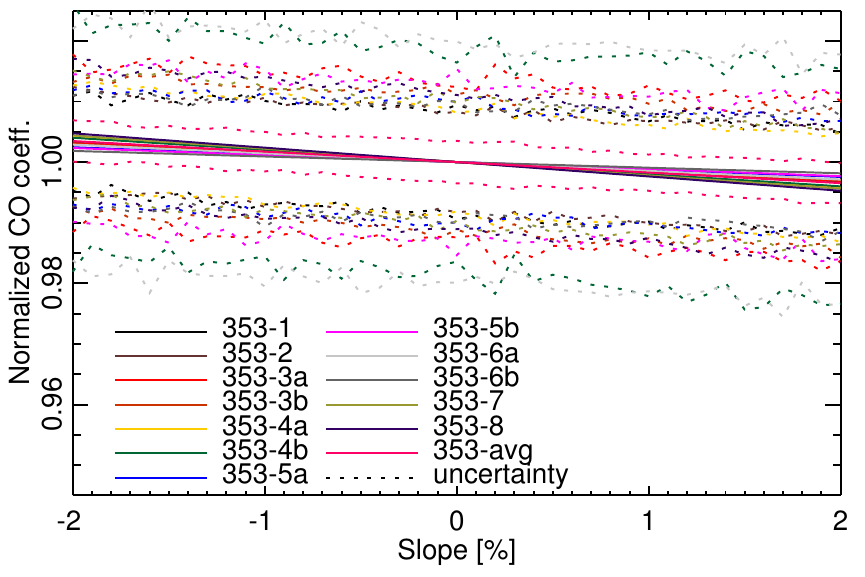}
\put(10,62){c}
\end{overpic} 
\end{tabular}
\caption{\label{fig:COslope} Comparison of CO unit conversion coefficients with an additional linear scaling of the input spectra, for the 100 (a), 217 (b), and 353 (c) GHz detectors.}
\end{center}
\end{figure}

Another check performed on the bandpass CO coefficients was the introduction of a linear slope to the detector spectra, as described in 
Sect.~\ref{sec:HFICoeffs} and demonstrated in Fig.\ \ref{fig:UcCCslope} for a combined unit conversion and dust colour correction.  As was found to be the 
case above, the addition of a linear slope to the spectra resulted in a linear change in the CO coefficients.  In this instance the changes observed in the CO 
coefficients were well within the quoted uncertainty, so the uncertainty is not underestimated in this respect.  Fig.~\ref{fig:COslope} illustrates the shift in CO coefficients caused by the introduction of a linear scaling of the spectra.  

\begin{figure}
\begin{center}
\begin{tabular}{c}
\begin{overpic}[width=88mm]{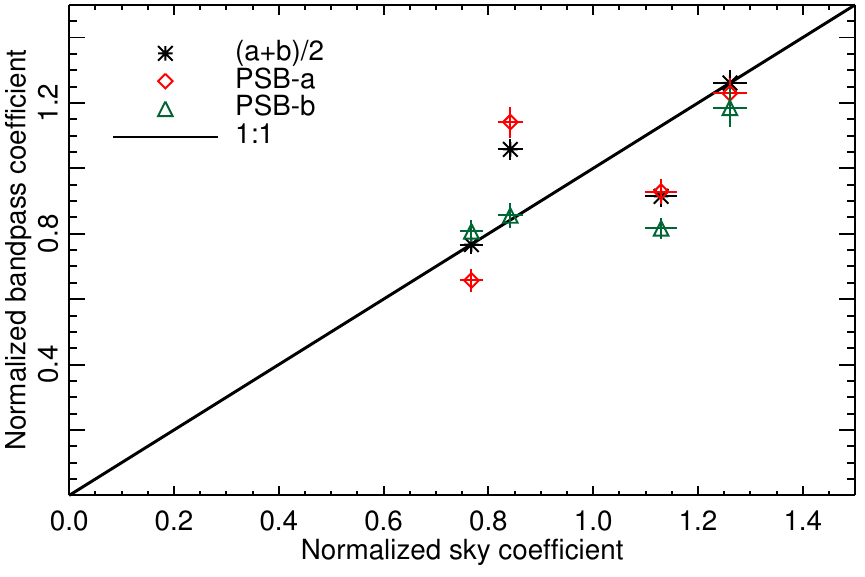} 
\put(11,62){a}
\end{overpic} \\
\begin{overpic}[width=88mm]{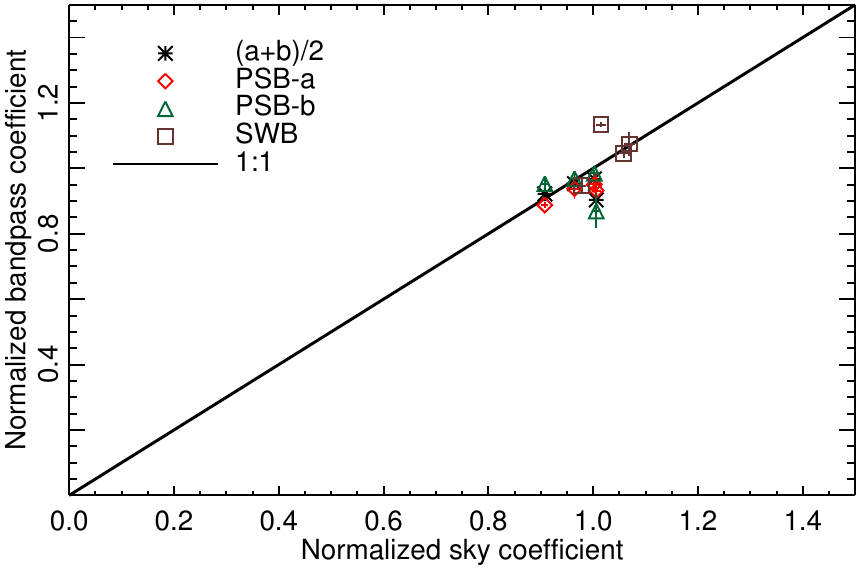} 
\put(11,62){b}
\end{overpic} \\
\begin{overpic}[width=88mm]{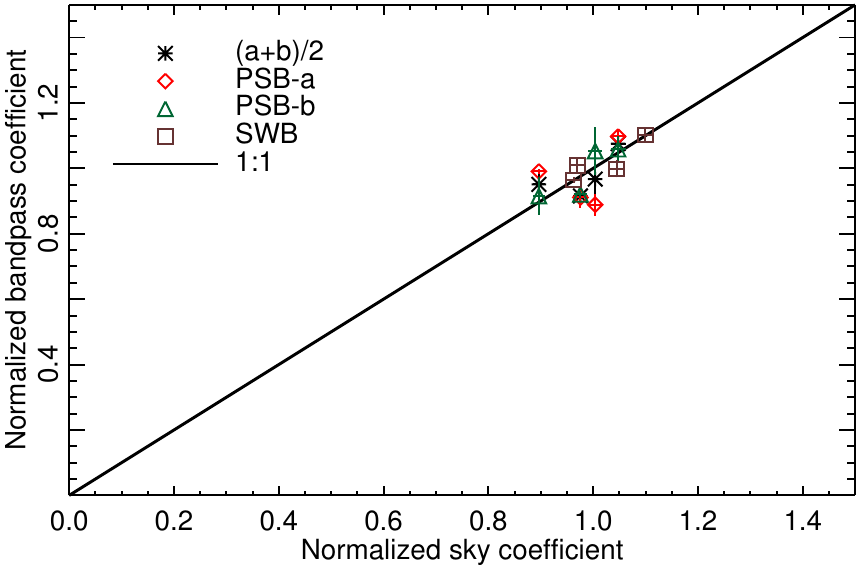} 
\put(11,62){c}
\end{overpic} 
\end{tabular}
\caption{\label{fig:CO_rel} Relationship between the CO unit conversion coefficients based on the CO-sky maps and the spectral response data for the CO $J$=1$\rightarrow$0 transition within the 100 GHz HFI band (a), 
the CO $J$=2$\rightarrow$1 transition within the 217 GHz HFI band (b),
and the CO $J$=3$\rightarrow$2 transition within the 353 GHz HFI band (c).}
\end{center}
\end{figure}

\begin{figure}
\begin{center}
\begin{tabular}{c}
\begin{overpic}[width=88mm,trim=0 0 0 2,clip]{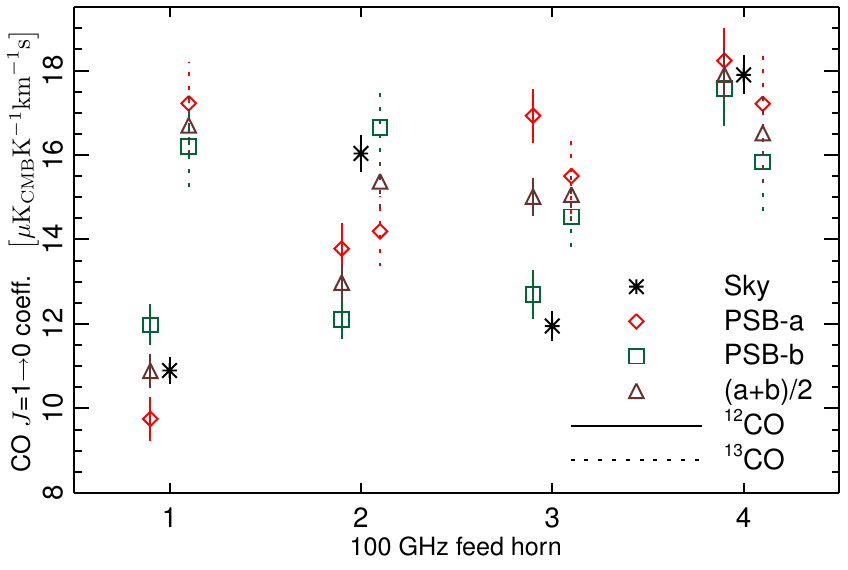} 
\put(12,63){a}
\end{overpic} \\
\begin{overpic}[width=88mm,trim=0 0 0 2,clip]{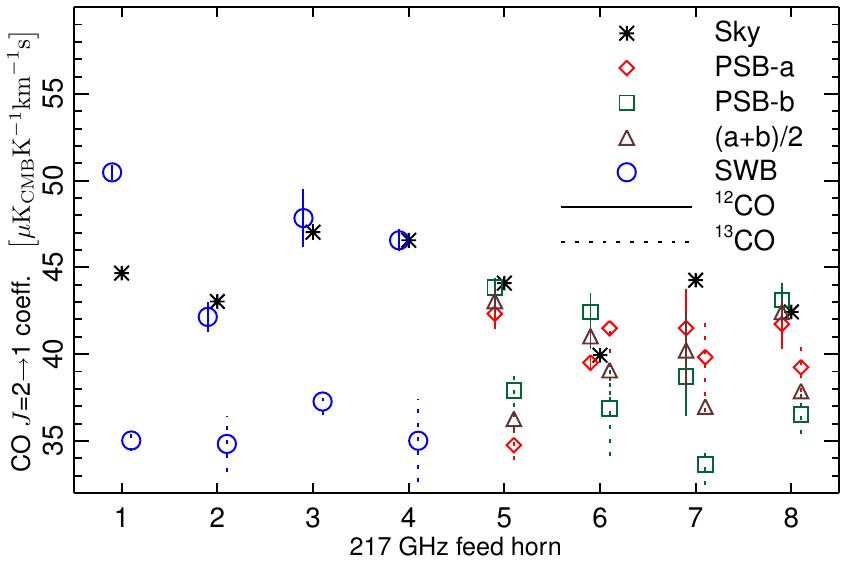} 
\put(12,63){b}
\end{overpic} \\
\begin{overpic}[width=88mm,trim=0 0 0 2,clip]{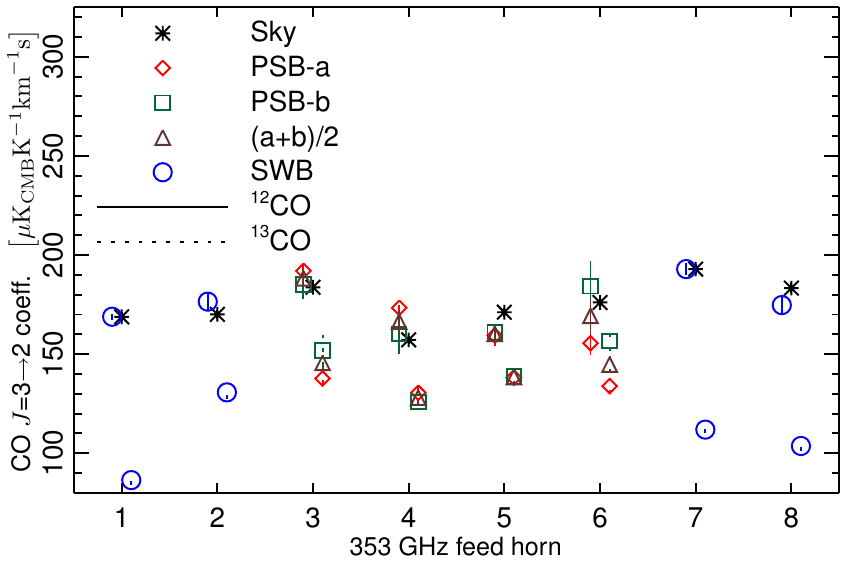} 
\put(12,63){c}
\end{overpic} \\
\end{tabular}
\caption{\label{fig:CO_BPscale} Comparison of CO unit conversion coefficients, scaled about the bandpass band-average coefficient amplitudes for the sky and bandpass coefficients.  This scaling allows comparison with both the $^{12}$CO and $^{13}$CO bandpass coefficients; the bandpass isotopic coefficients are horizontally offset in opposing directions for clarity.}
\end{center}
\end{figure}

The variation between the sky-based and bandpass-based CO unit conversion coefficients is demonstrated in Figs.~\ref{fig:CO_rel}--\ref{fig:CO_BPscale} for the HFI 100, 217, and 353\,GHz channels. The first figure indicates the correlation between the normalized sky and bandpass coefficients, while the second shows the sky and bandpass coefficients grouped together about their respective feed\,horns.  Plotting the sky coefficients on this scale required scaling the relative sky coefficients about the bandpass coefficient mean.  This was done to allow comparison with the $^{13}$CO coefficients as well as the $^{12}$CO coefficients (Fig.~\ref{fig:CO_BPscale} also includes the $^{13}$CO coefficients although these were not used in the sky coefficient scaling).  The $^{13}$CO coefficients are expected to have a stronger influence on the 100\,GHz channel than for the higher frequency channels \citep{planck2013-p03a}.  The plots indicate the detector type as either the individual PSB detectors (a and b separately), the combined PSB detectors ($(a + b)/2$), or the SWB detectors. Although this comparison is under-constrained, and the two sets of coefficients do not represent exactly the same quantity, as has been discussed above, there is a general agreement between the sky-based and bandpass-based CO coefficients.  The CO analysis also required a dust template as part of the component separation; thus, a sky-based estimate of HFI dust colour correction coefficients was a by-product of this analysis.  There is excellent agreement found between the bandpass-based and CO-sky-based dust colour correction coefficients; more details on this are provided in Sect.~\ref{sec:DustCC}, with the results presented in Fig.~\ref{fig:DustCC}.  

There remain differences in the sky- and bandpass-based CO unit conversion coefficients.  The analysis of CO data in the 2013 \Planck\ data release is based upon the sky-based CO coefficients.  The differences between the two approaches are acceptable for present analysis, but are increasingly important for future work, including the analysis of \Planck\ polarization data. Thus, while there is a correlation between the two approaches, a better understanding of these differences is required to gain a deeper understanding of the \Planck\ data. This comparison presents the current standing of a work in progress.




\subsubsection{Dust Colour Correction Bandpass Verification}
\label{sec:DustCC}

As a result of work investigating polarization leakage, a study was conducted to estimate the on-sky integration of SEDs for dusty regions directly using flight data, i.e., without using the spectral transmission data.  This was done to investigate the compatibility of the two methods.  

To derive the dust spectral mismatch for a given HFI frequency band, dust colour correction coefficients for each individual detector 
must be known for the bolometers within the given band.  Differences between the dust content of individual bolometer maps 
and the corresponding frequency channel map can be used to estimate these coefficients.  In performing this analysis, care must be taken to understand 
all of the spectral components included in the maps (e.g., CMB, free-free, CO, etc.), and calibration errors, beam errors and beam differences, and any 
residual polarization signal within the bolometer intensity maps.  

To avoid polarization bias, the study was restricted to areas surrounding either of the ecliptic poles, where there is a large variety in the crossing angles for 
multi-scan observations of a given region with a single detector.  Stokes $I$, $Q$, and $U$ maps were produced for individual detectors in addition to the 
standard $I$, $Q$, and $U$ frequency channel maps nominally produced by the data pipelines.  

\begin{figure}
\begin{center}
\begin{tabular}{ccc}
\begin{overpic}[trim= 0 10 0 0, clip,width=29mm]{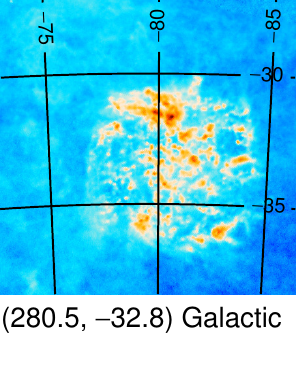}
\put(3,90){a}
\end{overpic}&
\begin{overpic}[trim= 0 10 0 0, clip,width=29mm]{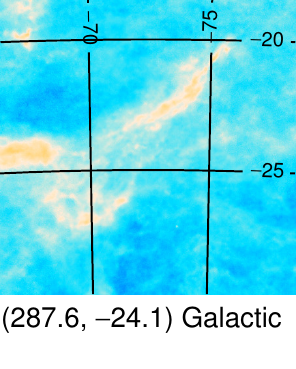} 
\put(3,90){b}
\end{overpic}&
\begin{overpic}[trim= 0 10 0 0, clip,width=29mm]{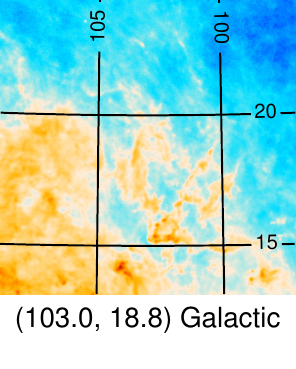} 
\put(3,90){c}
\end{overpic}\\
\multicolumn{3}{c}{\includegraphics[trim= 21 1 16 272, clip,width=88mm]{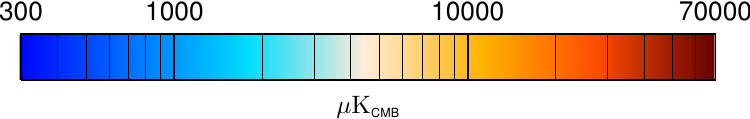}} 
\end{tabular}
\caption{\label{fig:DustTCMB} HFI 353\,GHz band-average intensity maps for the LMC (a), CHA (b), and NEP (c) ecliptic polar sky regions introduced above.}
\end{center}
\end{figure}

\begin{figure}
\begin{center}
\begin{tabular}{cccc}
 & LMC & CHA & NEP \\
\rotatebox{90}{\hspace{6.5mm}Full Survey}&
\begin{overpic}[trim= 0 20 0 0, clip,width=28mm]{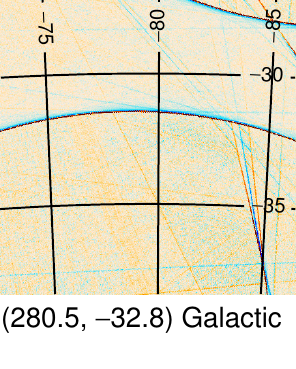}
\put(3,90){a}
\end{overpic}&
\begin{overpic}[trim= 0 20 0 0, clip,width=28mm]{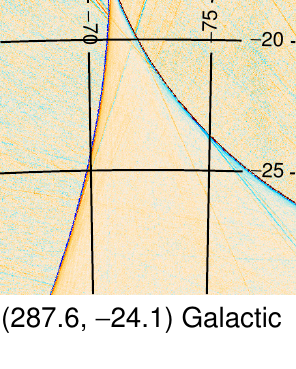} 
\put(3,90){b}
\end{overpic}&
\begin{overpic}[trim= 0 20 0 0, clip,width=28mm]{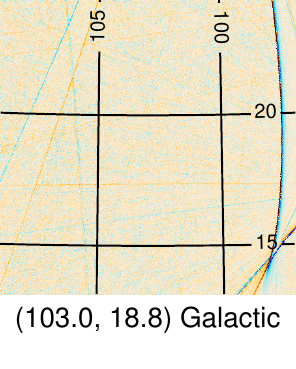} 
\put(3,90){c}
\end{overpic}\\
\rotatebox{90}{\hspace{3.5mm}Nominal Survey}&
\begin{overpic}[trim= 0 20 0 0, clip,width=28mm]{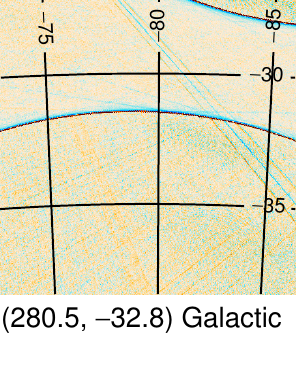}
\put(3,90){d}
\end{overpic}&
\begin{overpic}[trim= 0 20 0 0, clip,width=28mm]{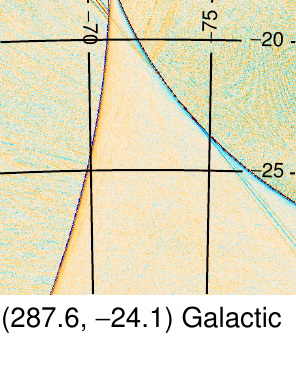} 
\put(3,90){e}
\end{overpic}&
\begin{overpic}[trim= 0 20 0 0, clip,width=28mm]{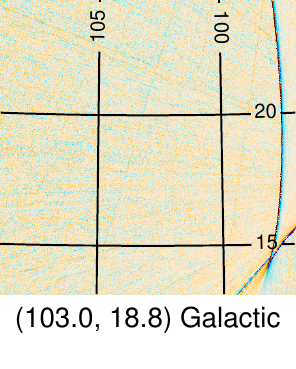} 
\put(3,90){f}
\end{overpic}\\
\rotatebox{90}{\hspace{8.5mm}Survey 1}&
\begin{overpic}[trim= 0 20 0 0, clip,width=28mm]{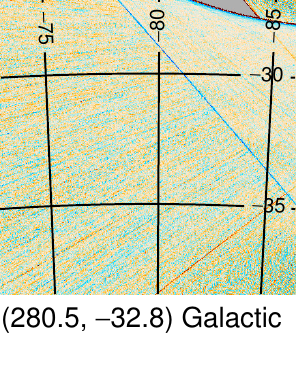}
\put(3,90){g}
\end{overpic}&
\begin{overpic}[trim= 0 20 0 0, clip,width=28mm]{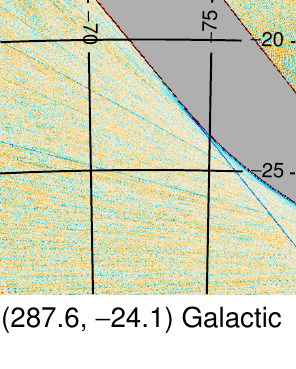} 
\put(3,90){h}
\end{overpic}&
\begin{overpic}[trim= 0 20 0 0, clip,width=28mm]{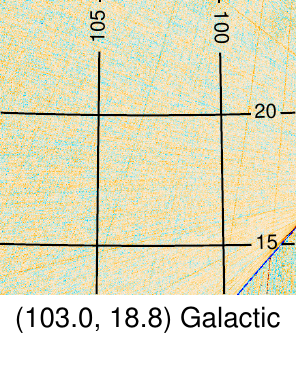} 
\put(3,90){i}
\end{overpic}\\
\rotatebox{90}{\hspace{11.5mm}Survey 2}&
\begin{overpic}[trim= 0 10 0 0, clip,width=28mm]{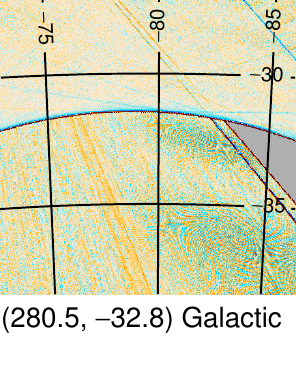}
\put(3,90){j}
\end{overpic}&
\begin{overpic}[trim= 0 10 0 0, clip,width=28mm]{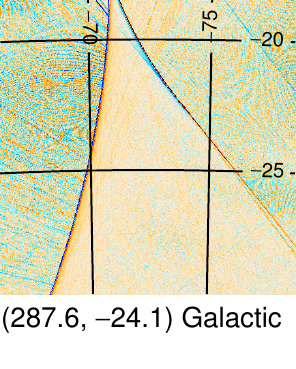} 
\put(3,90){k}
\end{overpic}&
\begin{overpic}[trim= 0 10 0 0, clip,width=28mm]{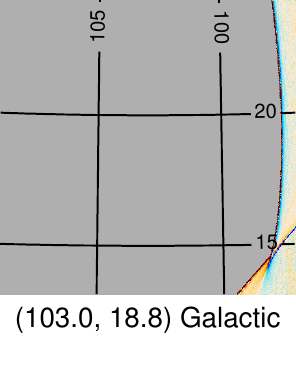} 
\put(3,90){l}
\end{overpic}\\
\multicolumn{4}{c}{\includegraphics[trim= 11 15 17 256, clip,width=88mm]{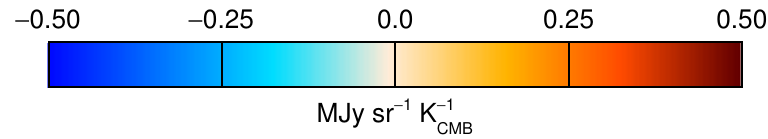}} 
\end{tabular}
\caption{\label{fig:DustUcCCmaps} 353 GHz maps of band-average unit conversion and dust colour correction (see Sect.~\ref{sec:CoeffMaps}).  
The maps show the deviation in the combined unit conversion and colour correction from the full-sky map median values. 
The full-sky median coefficient values, i.e., offsets, used for the Full (a,b,c), Nominal (d,e,f), First (g,h,i), and Second (j,k,l) surveys are: 278.980, 278.970, 278.960, and 
278.970\,MJy~sr$^{-1}$~K$_{\mbox{\tiny{CMB}}}$, respectively.  The figure columns represent the ecliptic polar maps for the LMC (a, d, g, and j), CHA (b, e, h, and k), and NEP (c, f, i, and l) regions of the sky.}
\end{center}
\end{figure}

\begin{figure}
\begin{center}
\begin{tabular}{c}
\begin{overpic}[trim = 0 12 0 0, clip, width=88mm]{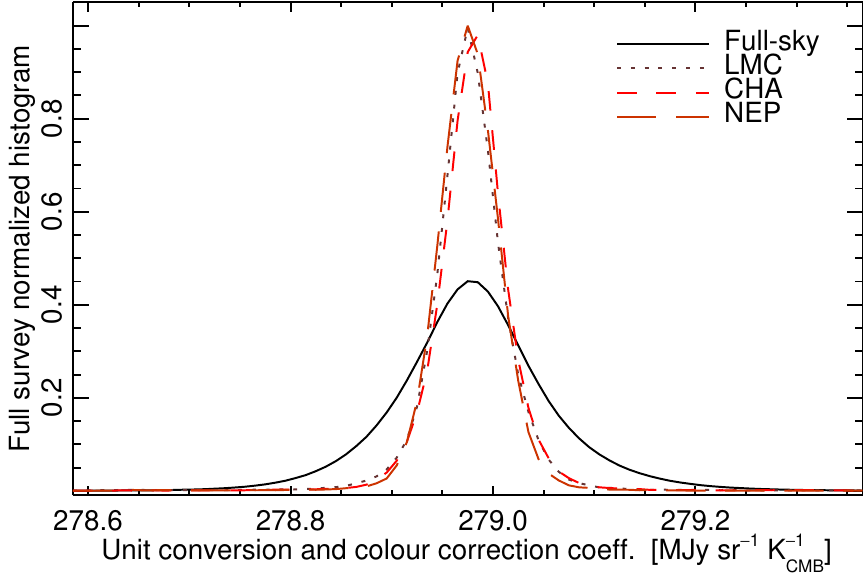}
\put(11,56){a}
\end{overpic}\\
\begin{overpic}[trim = 0 12 0 0, clip, width=88mm]{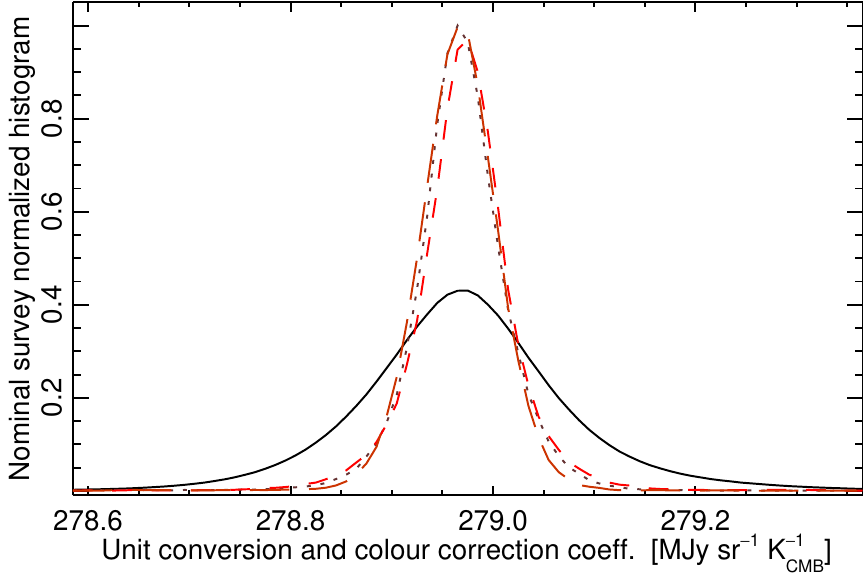}
\put(11,56){b}
\end{overpic}\\
\begin{overpic}[trim = 0 12 0 0, clip, width=88mm]{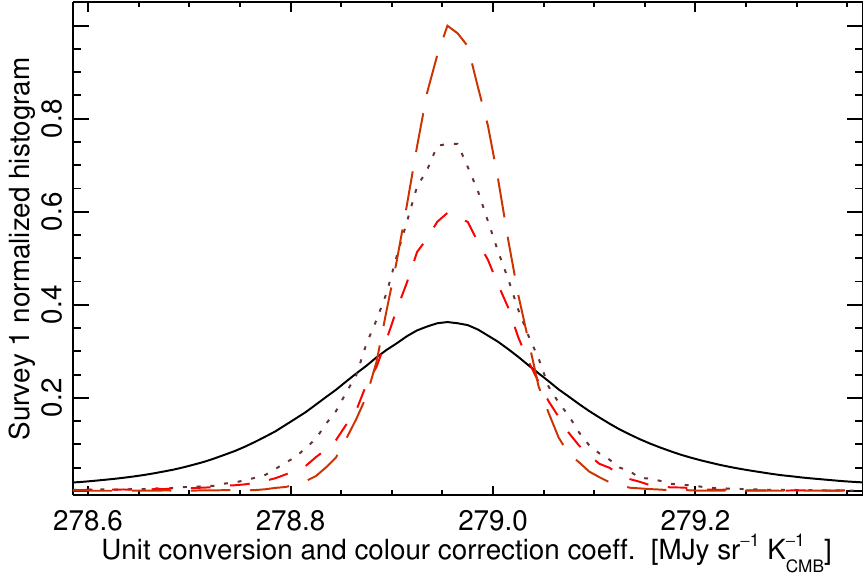}
\put(11,56){c}
\end{overpic}\\
\begin{overpic}[trim = 0 0 0 0, clip, width=88mm]{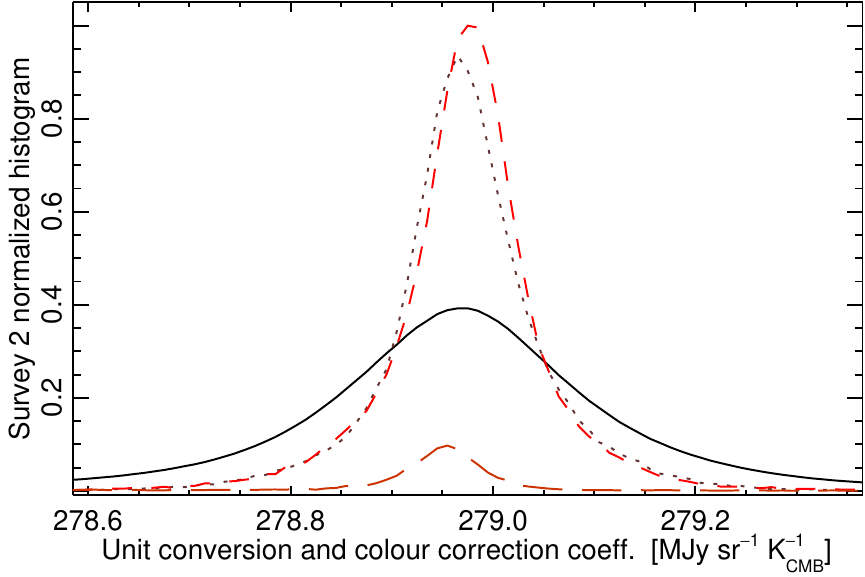}
\put(11,61){d}
\end{overpic}\\
\end{tabular}
\caption{\label{fig:DustUcCChist} Histograms of the 353 GHz combined unit conversion and dust colour correction maps corresponding to the Full survey (a), Nominal survey (b), survey 1 (c) and survey 2 (d) data.  The plots include histograms based on the full sky as well as the LMC, CHA, and NEP regions.}
\end{center}
\end{figure}

\begin{figure} 
\begin{center}
\begin{tabular}{c}
\begin{overpic}[width=88mm]{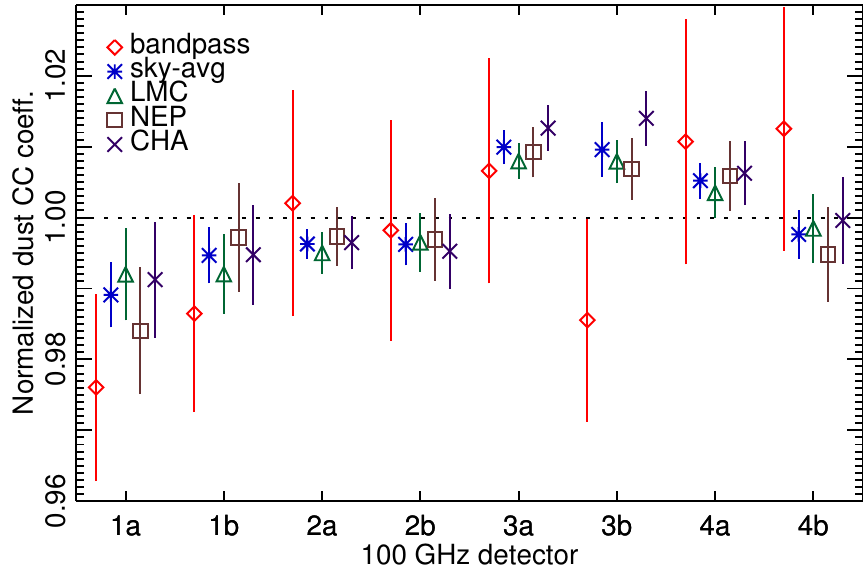} 
\end{overpic} \\
\begin{overpic}[width=88mm]{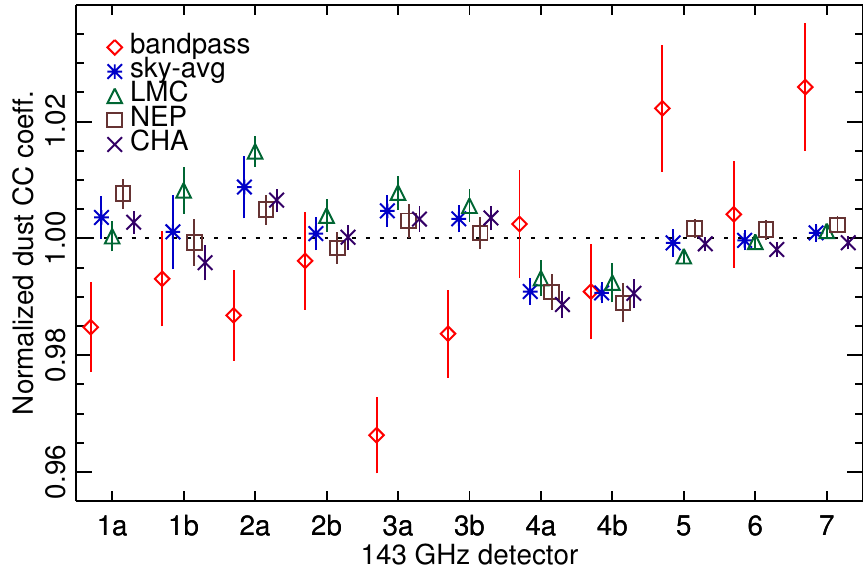} 
\end{overpic} \\
\begin{overpic}[width=88mm]{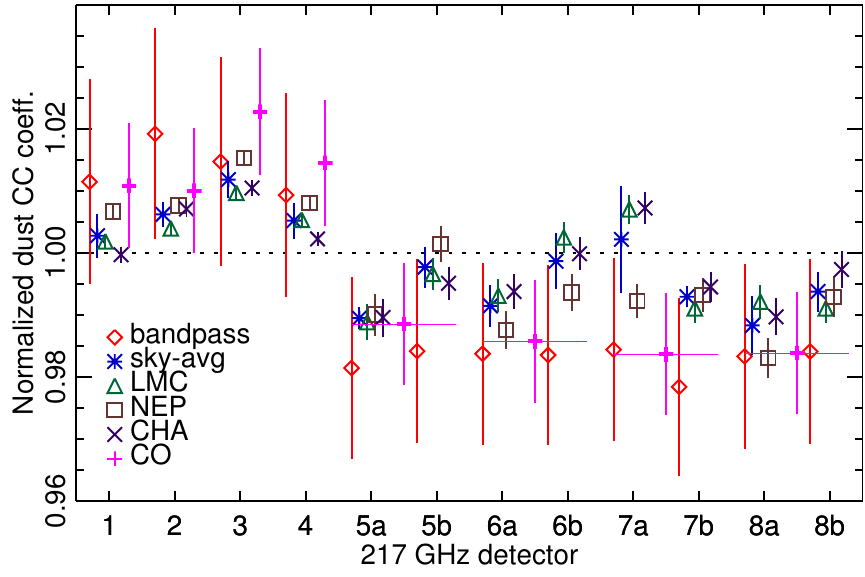} 
\end{overpic} \\
\begin{overpic}[width=88mm]{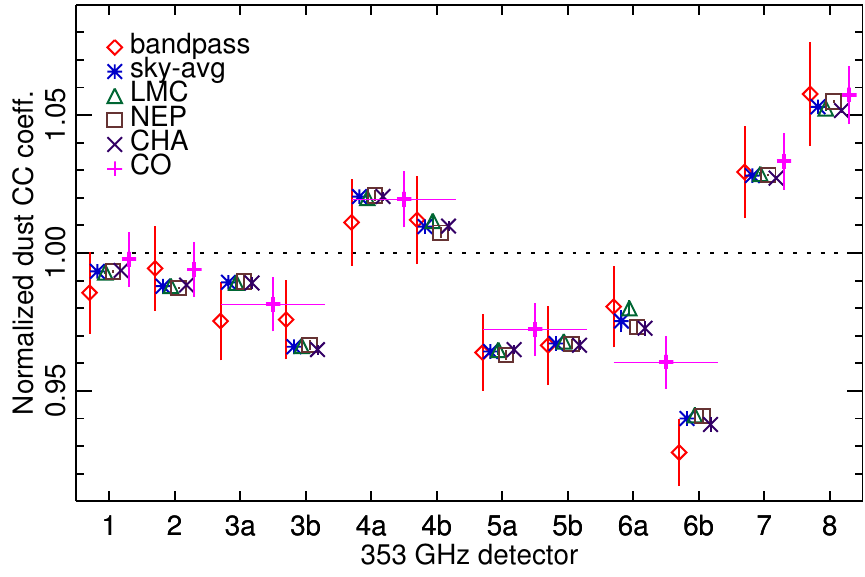} 
\end{overpic} 
\end{tabular}
\caption{\label{fig:DustCC} Comparison of the sky and bandpass based dust unit conversion and colour correction coefficients for the HFI 100\,GHz--353\,GHz spectral bands (top--bottom).}
\end{center}
\end{figure}

From the ecliptic polar maps, there were three regions selected for this study, two near the South pole and one near the North pole, all specifically selected for their relative component and content differences.  The Large Magellanic Cloud (LMC, ($l$,$b$)=(280\pdeg5, -32\pdeg8), \citealt{planck2011-6.4b}) was selected as a familiar target with known dust properties.  Another source selected was a region slightly offset from the LMC in the Chamealeon constellation (CHA, ($l$,$b$)=(287\pdeg6, $-$24\pdeg1)); the final region used in this study was near the North ecliptic pole (NEP, ($l$,$b$)=(103\pdeg0, 18\pdeg8)).  Further details on \Planck\ dust observations and analysis are provided in \cite{planck2013-p06b}. For each of the HFI frequency bands, and all of the individual detectors, square maps of 13\deg\ width, with 4\arcm\ pixel resolution, were extracted from the $I$, $Q$, and $U$ maps described above.  The dust SED properties of the LMC from a previous \Planck\ publication are $T_{\mbox{\scalebox{0.6}{D}}} = 21.0 \pm 1.9$\,K and $\beta = 1.48 \pm 0.25$ \citep{planck2011-6.4b}; as the sky-based dust colour correction coefficients for all sources remain consistent (see Fig.~\ref{fig:DustCC}), the LMC dust properties and uncertainties mentioned above are used to determine the bandpass coefficients included in this section of this work, along with their respective uncertainty.  Fig.\ \ref{fig:DustTCMB} provides the HFI observations over these regions at 353 GHz\@.  The corresponding band-average unit conversion and colour correction coefficient maps (see Sect.~\ref{sec:CoeffMaps}) for each band were used to verify the uniformity of the expected band-average colour correction across these regions of the sky.  While Fig.~\ref{fig:BandAvgMaps} provides examples of full-sky band-average coefficient maps, Fig.~\ref{fig:DustUcCCmaps} provides a similar example over the three specified sky regions used in this dust colour correction coefficient comparison.  The changing level of coverage between surveys is especially apparent for the NEP region.  Histograms of these band-average coefficients for the full sky and dusty regions are provided in Fig.~\ref{fig:DustUcCChist}.



\renewcommand{\tabcolsep}{0.1em}

\renewcommand{\tabcolsep}{0.1em}

A component separation is performed to isolate CMB, free-free, CO, and dust emission from the 100, 143, 217, and 353 GHz frequency channels. Regions 
of high free-free and high CO emission are masked in the dust maps, leaving residual maps primarily containing CMB and dust.  The CMB signal is 
removed by subtracting the frequency-band map from the individual detector maps.  This step is based on the fact that the CMB dipole is the calibration source for each of these detectors; therefore, it is assumed that the CMB component of each detector map is consistant with that of the band-average map.  Let $U_D$ represent the dust colour correction for the frequency-band map of a given channel, and let $U_{D~i}$ represent the individual bolometer dust colour correction coefficient.  The intensity map, $I$, for the channel will thus be given by $I = C + U_DD$, where $C$ is the CMB component, and $D$ is the dust component (note: $C$ and $D$ 
represent the actual observable intensity, not only that observed by a given detector). The detector intensity map will be given by $I_i = C + U_{D~i}D$.  The regression 
coefficient resultant from the correlation of the difference map $I_i - I$ with the channel map $I$ can be used to derive the {\it relative} dust 
colour correction coefficient as follows:
\begin{eqnarray}
\label{eq:SkyDustCCDiff1}
I_i - I & = & (U_{D~i} - U_D)D \ ,\\
\label{eq:SkyDustCCDiff2}
\mbox{Corr}\left(I_i - I, I\right) & = & \displaystyle\frac{U_{D~i} - U_D}{U_D} \ ,\\
\label{eq:SkyDustCC}
\displaystyle\frac{U_{D~i}}{U_D} & = & \mbox{Corr}\left(I_i - I, I\right) + 1 \ .
\end{eqnarray}
Thus the relative dust colour correction coefficient may be determined for each individual HFI detector by employing the above relation.  
It is important to note that the dust colour correction coefficients in the above expression convert from the dust spectral profile to 
the CMB spectral profile; in the notation of the HFI unit conversion and colour correction syntax, this is equivalent to the inverse (i.e., $U_D^{-1}$)
of a colour correction from the dust spectral profile to a power-law spectral index of $\alpha$~=~$-$1 {\it followed} by a unit conversion from $\alpha$~=~$-$1 to differential CMB temperature units.  

Fig.\ \ref{fig:DustCC} illustrates the excellent agreement between the sky-based dust colour correction coefficients 
and those based on the HFI spectral response (i.e., Eq.\ \ref{UC:KCMB_MJysr} and \ref{UC:CCMBB}) for the 100, 217, and 353 GHz spectral bands.  This level of agreement is not found amongst the 143 GHz detectors, however.  The source of this discrepancy is under investigation.  It should be noted that the dust emission is much stronger at higher frequencies, and is thus less dominant at 143 and 100 GHz\@. A dust colour correction coefficient originating from the CO extraction \citep{planck2013-p03a} is also shown in Fig.~\ref{fig:DustCC} for the 217 and 353\,GHz bands. The oversized horizontal bars on some of the CO-extraction-based data points indicate that the coefficient was derived for the PSB a/b pair of detectors, not the detectors individually.  For the 217 and 353\,GHz coefficients, the bandpass and CO values are in excellent agreement, even where the dust coefficients from this study appear to diverge from the bandpass values slightly.  The exact causes of these variations remains under study for a future data release.  


The uncertainty in the figures for the sky-based dust coefficients is based on an absolute calibration uncertainty of 0.5\,\% for each band.  The bandpass 
based coefficient uncertainties are based upon the quoted uncertainties of $T_{\mbox{\scalebox{0.6}{D}}}$ and $\beta$ of 1.9\,K and 0.25, respectively, as well as the spectral uncertainty associated with each spectral response profile.  The CO coefficient uncertainties are based upon a 1\,\% relative calibration uncertainty.

%

\subsubsection{Sunyaev-Zeldovich Bandpass Verification}
\label{sec:SZ}

The Sunyaev-Zeldovich (SZ) effect \citep{SZ1980} has a characteristic signature in the millimetre and 
submillimetre domain. With a changing sign at 217\,GHz, it is
quite different from the CMB anisotropy spectrum and other power-law spectra typical of these frequencies.
Consequently, any bandpass leakage should be clearly identifiable in SZ
spectra. To isolate the SZ signature from other foregrounds and
the CMB anisotropies themselves, the 20 brightest clusters in the Planck
catalogue (\citealt{planck2013-p05}, \citealt{planck2013-p05a}, and \citealt{planck2013-p05b}) have been selected to provide 
lines-of-sight over which to investigate spectral mismatch leakage.
The lines-of-sight sources selected are A2319, A3266, RXC J1638.2-6420, A2219,
A2142, Coma, A366, A2255, A2029, A3186, A2218, A3158, A85, A3827,
A697, A1795, A644, A2204, A3628, and A3888. For each of the above sources, 
the integrated flux in individual HFI channel maps is measured 
(calibrated in thermodynamic temperature K$_{\mbox{\scalebox{0.6}{CMB}}}$ units) at the
position of each cluster. Using the nominal bandpass conversion
coefficients for each of the expected components, the integrated $y_{\mbox{\scalebox{0.6}{SZ}}}$ for each cluster 
is deduced from the residuals using a $\chi^2$ statistical analysis. Thus, an estimate of the SZ unit conversion 
from the $y_{\mbox{\scalebox{0.6}{SZ}}}$ Compton parameter to the $dT_{\mbox{\scalebox{0.6}{CMB}}}$ differential CMB temperature may be obtained for each 
cluster, and each frequency; without invoking the bandpass-based SZ unit conversion coefficients themselves (Eq.\ \ref{UC:KCMB_ySZ}, Table~\ref{tab:UC}).  
This analysis provides a consistency check of the bandpass-based SZ coefficients.  

Two methods have been used to perform this analysis. The first allows only a dust template
removal, fixed on the 857\,GHz channel, with coefficients deduced
outside each cluster. This first method does not make assumptions
about the colour of the background. The background removal was accomplished via subtraction of an average value taken on an
annulus surrounding the cluster, i.e., aperture photometry. This method has a low
signal-to-noise ratio because it is dominated by the CMB
anisotropy residuals. The second method uses the {\tt MILCA}
algorithm \citep{hurier2010milca} to subtract a component with a CMB
spectrum at all frequencies.  The average of these coefficients per frequency is shown in
Table \ref{tab:SZ} for the two methods and compared with the expected
coefficients based on the nominal band-average transmission spectra. Although the
SNR is $<$10 for the first method and $<$30 for the second method, the agreement is
remarkable. Thus, the SZ cluster data provides a pseudo-quantitative verification of the accuracy 
of the HFI detector spectra as gross spectral leakages would 
demonstrate themselves through this analysis. 


\begin{table}[tmb]                 
\begin{center}
\begingroup
\newdimen\tblskip \tblskip=5pt
\caption{Various $dT_{\mbox{\tiny{CMB}}}$ to $y_{\mbox{\tiny{SZ}}}$ unit conversion coefficients used to validate the ground-based HFI spectral response measurements with sky-based SZ results.}                          
\label{tab:SZ}                            
\nointerlineskip
\vskip -3mm
\footnotesize
\setbox\tablebox=\vbox{
   \newdimen\digitwidth 
   \setbox0=\hbox{\rm 0} 
   \digitwidth=\wd0 
   \catcode`*=\active 
   \def*{\kern\digitwidth}
   \newdimen\signwidth 
   \setbox0=\hbox{+} 
   \signwidth=\wd0 
   \catcode`!=\active 
   \def!{\kern\signwidth}
   \newdimen\perwidth 
   \setbox0=\hbox{.} 
   \perwidth=\wd0 
   \catcode`?=\active 
   \def?{\kern\perwidth}
   \def\leaderfi1{\leaders\hbox to 5pt{\hss.\hss}\hfil}
%
\halign{\hbox to 0.65in{#\leaderfil}\tabskip=2em plus 4em minus 1em&
  \tabskip=2em\hfil#\hfil\tabskip=2em plus 4em minus 1em&
  \tabskip=2em\hfil#\hfil\tabskip=2em plus 4em minus 1em&
  \tabskip=2em\hfil#\hfil\tabskip=0em\cr                           
\noalign{\doubleline}
\omit Band& $U_{\mbox{\tiny{c~BP}}}$& $U_{\mbox{\tiny{c~Ap.~Phot}}}$ &$U_{\mbox{\tiny{c~{\tt MILCA}}}}$\cr 
\omit [GHz]&  [K$_{\mbox{\scalebox{0.6}{CMB}}}^{-1}$]& [K$_{\mbox{\scalebox{0.6}{CMB}}}^{-1}$]& [K$_{\mbox{\scalebox{0.6}{CMB}}}^{-1}$]\cr
\noalign{\vskip 3pt\hrule\vskip 5pt}
100& $-$4.030 $\pm$ 0.018& *$-$4.7 $\pm$ 0.6& *$-$4.6* $\pm$ 0.2*\cr
143& $-$2.78* $\pm$ 0.04*& *$-$3.9 $\pm$ 0.6& *$-$3.00 $\pm$ 0.10\cr
217&   !0.19* $\pm$ 0.05*& *$-$1.3 $\pm$ 0.6& *$-$0.10 $\pm$ 0.10\cr
353&   !6.21* $\pm$ 0.11*&   *!4.3 $\pm$ 0.8&   *!5.6* $\pm$ 0.2*\cr
545&   14.46* $\pm$ 0.07*&   !17.4 $\pm$ 6?*&   !17?** $\pm$ 2?**\cr                                    
\noalign{\vskip 5pt\hrule\vskip 3pt}}}
\endPlancktablewide                 
\endgroup
\end{center}
\end{table}                        

\section{Conclusions}
The spectral response of the \Planck\ HFI detectors has been presented.  The derivation of the HFI band-average spectra has been presented; including photometric, noise, and sky coverage constituent scaling coefficients.  The scaling coefficients for the band-average spectra were compared with those resulting from individual HFI surveys, and it was demonstrated that the individual survey values converged to the combined survey average values used in the derivation of the spectral response data products.  Unit conversion and colour correction coefficient relations have been derived; this includes band-average coefficients and coefficient maps based on sky coverage and scan strategy.  The corresponding coefficients, uncertainties, and related unit conversion and colour correction software tools, are available within the \Planck\ legacy data archive \citep{planck2013-p28}.  The accuracy of the HFI spectral response has been verified using both ground-based component level data and, importantly, HFI flight data.  The defining requirement indicated in the original HFI calibration plan was knowledge of the spectral transmission of the individual detectors with uncertainties below 3\,\% for the low frequency channels (100, 143, and 217\,GHz) and below 1\,\% for the high frequency channels (353, 545, and 857\,GHz); all with a spectral resolution of better than 3\,GHz\@.  The spectral resolution requirement has been exceeded by more than a factor of five. It is possible to degrade the spectral resolution requirement to gain an improvement in the SNR, which allows the desired spectral transmission accuracy to be achieved in most cases (the absolute uncertainty for the 100\,GHz detectors is high due to the high noise levels of the reference bolometer at these frequencies, but the relative uncertainty even for the 100\,GHz detectors remains low).  Estimates of the out-of-band transmission profiles have been incorporated into the bandpass data products. Out-of-band signal attenuation is demonstrated to be better than $10^8$ through HFI observations of the diffuse zodiacal cloud \citep{planck2013-pip88}.  Good agreement has been demonstrated between sky and bandpass dust colour correction coefficients, as well as SZ unit conversion coefficients.  The sky and bandpass based CO coefficients have shown a correlation, yet a full comparison is unconstrained at this time due to differences between the two approaches.  The CO comparison is a work in progress, where improvements are expected as \Planck\ polarization data analysis progresses.  The HFI spectral response data, and associated data products, have been verified within the stated uncertainty using a variety of tests based on in-flight HFI observations.  



\begin{acknowledgements}
The development of \Planck\ has been supported by: ESA; CNES and CNRS/INSU-IN2P3-INP (France); ASI, CNR, and INAF (Italy); NASA and DoE (USA); STFC and UKSA (UK); CSIC, MICINN, JA and RES (Spain); Tekes, AoF and CSC (Finland); DLR and MPG (Germany); CSA (Canada); DTU Space (Denmark); SER/SSO (Switzerland); RCN (Norway); SFI (Ireland); FCT/MCTES (Portugal); and PRACE (EU). A description of the Planck Collaboration and a list of its members, including the technical or scientific activities in which they have been involved, can be found at \url{http://www.sciops.esa.int/index.php?project=planck&page=Planck_Collaboration}. The authors thank Emmanuel Lellouch, Raphael Moreno, and Matt Griffin for the provision of the planet model spectra used to generate HFI planet colour correction coefficients.  L.~Spencer acknowledges support from NSERC (Canada) and STFC (UK)\@. 
\end{acknowledgements}

\bibliographystyle{aa}

\bibliography{Planck_bib,P03d_extra}

\begin{thebibliography}{55}
\expandafter\ifx\csname natexlab\endcsname\relax\def\natexlab#1{#1}\fi

\bibitem[{Ade {et~al.}(2006)Ade, Pisano, Tucker, \& Weaver}]{Ade2006Filters}
Ade, P. A.~R., Pisano, G., Tucker, C., \& Weaver, S. 2006, Proc. SPIE, 6275,
  62750U

\bibitem[{{Ade} {et~al.}(2010){Ade}, {Savini}, {Sudiwala}, {Tucker},
  {Catalano}, {Church}, {Colgan}, {Desert}, {Gleeson}, {Jones}, {Lamarre},
  {Lange}, {Longval}, {Maffei}, {Murphy}, {Noviello}, {Pajot}, {Puget},
  {Ristorcelli}, {Woodcraft}, \& {Yurchenko}}]{ade2010}
{Ade}, P.~A.~R., {Savini}, G., {Sudiwala}, R., {et~al.} 2010, \aap, 520, A11+

\bibitem[{{Beichman} {et~al.}(1988){Beichman}, {Neugebauer}, {Habing}, {Clegg},
  \& {Chester}}]{IRASExSup}
{Beichman}, C.~A., {Neugebauer}, G., {Habing}, H.~J., {Clegg}, P.~E., \&
  {Chester}, T.~J., eds. 1988, {Infrared astronomical satellite (IRAS) catalogs
  and atlases. Volume 1: Explanatory supplement}, Vol.~1

\bibitem[{Bell(1972)}]{Bell}
Bell, R.~J. 1972, {Introductory Fourier Transform Spectroscopy} (New York:
  Academic Press)

\bibitem[{Bierman {et~al.}(2011)Bierman, Matsumura, Dowell, Keating, Ade,
  Barkats, Barron, Battle, Bock, Chiang, Culverhouse, Duband, Hivon, Holzapfel,
  Hristov, Kaufman, Kovac, Kuo, Lange, Leitch, Mason, Miller, Nguyen, Pryke,
  Richter, Rocha, Sheehy, Takahashi, \& Yoon}]{BICEP2011}
Bierman, E.~M., Matsumura, T., Dowell, C.~D., {et~al.} 2011, The Astrophysical
  Journal, 741, 81

\bibitem[{Born \& Wolf(1999)}]{BornWolf}
Born, M. \& Wolf, E. 1999, Principles of Optics: Electromagnetic Theory of
  Propagation, Interference and Diffraction of Light, seventh edn. (Cambridge
  University Press)

\bibitem[{Brault(1987)}]{Brault87}
Brault, J.~W. 1987, Mikrochimica Acta, 3, 215

\bibitem[{Catalano(2008)}]{ACATphd}
Catalano, A. 2008, PhD thesis, IAP, Paris, title Translation: Development of
  numerical models of the High Frequency Instrument (HFI) of Planck necessary
  for its operation

\bibitem[{Catalano {et~al.}(2006)Catalano, Coulias, Recouvreur, \&
  Lamarre}]{EFF}
Catalano, A., Coulias, A., Recouvreur, G., \& Lamarre, J. 2006, {HFI
  Calibration Report -- EFF Sequences}, Tech. Rep. 01-00, {Observatoire de
  Paris}, filename: EFF\_Calibration.pdf

\bibitem[{{Dame} {et~al.}(2001){Dame}, {Hartmann}, \& {Thaddeus}}]{dame2001}
{Dame}, T.~M., {Hartmann}, D., \& {Thaddeus}, P. 2001, \apj, 547, 792

\bibitem[{Davis {et~al.}(2001)Davis, Abrams, \& Brault}]{davisFTS}
Davis, S.~P., Abrams, M.~C., \& Brault, J.~W. 2001, Fourier Transform
  Spectroscopy, 1st edn. (Academic Press)

\bibitem[{{Fixsen}(2009)}]{fixsen2009}
{Fixsen}, D.~J. 2009, \apj, 707, 916

\bibitem[{{Fixsen} {et~al.}(1994){Fixsen}, {Cheng}, {Cottingham}, {Eplee},
  {Isaacman}, {Mather}, {Meyer}, {Noerdlinger}, {Shafer}, {Weiss}, {Wright},
  {Bennett}, {Boggess}, {Kelsall}, {Moseley}, {Silverberg}, {Smoot}, \&
  {Wilkinson}}]{Fixsen1994}
{Fixsen}, D.~J., {Cheng}, E.~S., {Cottingham}, D.~A., {et~al.} 1994, \apj, 420,
  445

\bibitem[{{Fixsen} \& {Dwek}(2002)}]{fixsendwek2002}
{Fixsen}, D.~J. \& {Dwek}, E. 2002, \apj, 578, 1009

\bibitem[{Forman {et~al.}(1966)Forman, Steel, \& Vanasse}]{For66}
Forman, M.~L., Steel, W.~H., \& Vanasse, G.~A. 1966, Journal of the Optical
  Society of America, 56, 59

\bibitem[{Fukui {et~al.}(2008)Fukui, Kawamura, Minamidani, Mizuno, Kanai,
  Mizuno, Onishi, Yonekura, Mizuno, Ogawa, \& Rubio}]{COwidth}
Fukui, Y., Kawamura, A., Minamidani, T., {et~al.} 2008, The Astrophysical
  Journal Supplement Series, 178, 56

\bibitem[{{G{\'o}rski} {et~al.}(2005){G{\'o}rski}, {Hivon}, {Banday},
  {Wandelt}, {Hansen}, {Reinecke}, \& {Bartelmann}}]{Healpix}
{G{\'o}rski}, K.~M., {Hivon}, E., {Banday}, A.~J., {et~al.} 2005, \apj, 622,
  759

\bibitem[{{Grainger}(2001)}]{GraingerSZ}
{Grainger}, W. 2001, PhD thesis, University of Cambridge

\bibitem[{Griffin {et~al.}(2013)Griffin, North, Amaral-Rogers, Bendo, Bock,
  Conley, Dowell, Ferlet, Glenn, Lim, Pearson, Pohlen, Schulz, Sibthorpe,
  Spencer, Swinyard, \& Valtchanov}]{GriffinCC}
Griffin, M., North, C., Amaral-Rogers, A., {et~al.} 2013, MNRAS, In Press,
  submitted Feb.~2013.

\bibitem[{{Holmes} {et~al.}(2008){Holmes}, {Bock}, {Crill}, {Koch}, {Jones},
  {Lange}, \& {Paine}}]{holmes2008}
{Holmes}, W.~A., {Bock}, J.~J., {Crill}, B.~P., {et~al.} 2008, \ao, 47, 5996

\bibitem[{Hurier {et~al.}(2010)Hurier, Hildebrandt, \&
  Macias-Perez}]{hurier2010milca}
Hurier, G., Hildebrandt, S., \& Macias-Perez, J. 2010, arXiv preprint
  arXiv:1007.1149

\bibitem[{Jet Propulsion~Laboratory(2004)}]{JPLspec}
Jet Propulsion~Laboratory, C. I. o.~T. 2004, Molecular Spectroscopy Catalog,
  http://spec.jpl.nasa.gov/

\bibitem[{{Kelsall} {et~al.}(1998){Kelsall}, {Weiland}, {Franz}, {Reach},
  {Arendt}, {Dwek}, {Freudenreich}, {Hauser}, {Moseley}, {Odegard},
  {Silverberg}, \& {Wright}}]{kelsall1998}
{Kelsall}, T., {Weiland}, J.~L., {Franz}, B.~A., {et~al.} 1998, \apj, 508, 44

\bibitem[{{Kompaneets}(1957)}]{Kompaneets1957}
{Kompaneets}, A.~S. 1957, Soviet Phys.-JETP, 4, 730

\bibitem[{{Lamarre} {et~al.}(2010){Lamarre}, {Puget}, {Ade}, {Bouchet},
  {Guyot}, {Lange}, {Pajot}, {Arondel}, {Benabed}, {Beney}, {Beno{\^i}t},
  {Bernard}, {Bhatia}, {Blanc}, {Bock}, {Br{\'e}elle}, {Bradshaw}, {Camus},
  {Catalano}, {Charra}, {Charra}, {Church}, {Couchot}, {Coulais}, {Crill},
  {Crook}, {Dassas}, {de Bernardis}, {Delabrouille}, {de Marcillac}, {Delouis},
  {D{\'e}sert}, {Dumesnil}, {Dupac}, {Efstathiou}, {Eng}, {Evesque},
  {Fourmond}, {Ganga}, {Giard}, {Gispert}, {Guglielmi}, {Haissinski},
  {Henrot-Versill{\'e}}, {Hivon}, {Holmes}, {Jones}, {Koch}, {Lagard{\`e}re},
  {Lami}, {Land{\'e}}, {Leriche}, {Leroy}, {Longval},
  {Mac{\'{\i}}as-P{\'e}rez}, {Maciaszek}, {Maffei}, {Mansoux}, {Marty}, {Masi},
  {Mercier}, {Miville-Desch{\^e}nes}, {Moneti}, {Montier}, {Murphy},
  {Narbonne}, {Nexon}, {Paine}, {Pahn}, {Perdereau}, {Piacentini}, {Piat},
  {Plaszczynski}, {Pointecouteau}, {Pons}, {Ponthieu}, {Prunet}, {Rambaud},
  {Recouvreur}, {Renault}, {Ristorcelli}, {Rosset}, {Santos}, {Savini},
  {Serra}, {Stassi}, {Sudiwala}, {Sygnet}, {Tauber}, {Torre}, {Tristram},
  {Vibert}, {Woodcraft}, {Yurchenko}, \& {Yvon}}]{lamarre2010}
{Lamarre}, J., {Puget}, J., {Ade}, P.~A.~R., {et~al.} 2010, \aap, 520, A9+

\bibitem[{{Maffei} {et~al.}(2010){Maffei}, {Noviello}, {Murphy}, {Ade},
  {Lamarre}, {Bouchet}, {Brossard}, {Catalano}, {Colgan}, {Gispert}, {Gleeson},
  {Haynes}, {Jones}, {Lange}, {Longval}, {McAuley}, {Pajot}, {Peacocke},
  {Pisano}, {Puget}, {Ristorcelli}, {Savini}, {Sudiwala}, {Wylde}, \&
  {Yurchenko}}]{maffei2010}
{Maffei}, B., {Noviello}, F., {Murphy}, J.~A., {et~al.} 2010, \aap, 520, A12+

\bibitem[{{Mather} {et~al.}(1994){Mather}, {Cheng}, {Cottingham}, {Eplee},
  {Fixsen}, {Hewagama}, {Isaacman}, {Jensen}, {Meyer}, {Noerdlinger}, {Read},
  {Rosen}, {Shafer}, {Wright}, {Bennett}, {Boggess}, {Hauser}, {Kelsall},
  {Moseley}, {Silverberg}, {Smoot}, {Weiss}, \& {Wilkinson}}]{Mather1994}
{Mather}, J.~C., {Cheng}, E.~S., {Cottingham}, D.~A., {et~al.} 1994, \apj, 420,
  439

\bibitem[{Murphy {et~al.}(2010)Murphy, Peacocke, Maffei, McAuley, Noviello,
  Yurchenko, Ade, Savini, Lamarre, Brossard, Colgan, Gleeson, Lange, Longval,
  Pisano, Puget, Ristorcelli, Sudiwala, \& Wylde}]{MurphyHorn}
Murphy, J.~A., Peacocke, T., Maffei, B., {et~al.} 2010, {Journal of
  Instrumentation}, 5, T04001

\bibitem[{Naylor {et~al.}(2009)Naylor, Gom, Jones, \& Spencer}]{SJVancouver}
Naylor, D.~A., Gom, B.~G., Jones, S.~C., \& Spencer, L.~D. 2009, in Spring
  Optics and Photonics Congress Fourier Transform Spectroscopy Topical Meeting
  (Optical Society of America), JTuB15

\bibitem[{Naylor \& Tahic(2007)}]{NaylorApod07}
Naylor, D.~A. \& Tahic, M.~K. 2007, J. Opt. Soc. Am. A, 24, 3644

\bibitem[{Nyquist(1928)}]{Nyquist1928}
Nyquist, H. 1928, Transactions of the American Institute of Electrical
  Engineers, 47, 617

\bibitem[{{Pajot} {et~al.}(2010){Pajot}, {Ade}, {Beney}, {Br{\'e}elle},
  {Broszkiewicz}, {Camus}, {Carab{\'e}tian}, {Catalano}, {Chardin}, {Charra},
  {Charra}, {Cizeron}, {Couchot}, {Coulais}, {Crill}, {Dassas}, {Daubin}, {de
  Bernardis}, {de Marcillac}, {Delouis}, {D{\'e}sert}, {Duret}, {Eng},
  {Evesque}, {Fourmond}, {Fran{\c c}ois}, {Giard}, {Giraud-H{\'e}raud},
  {Guglielmi}, {Guyot}, {Haissinski}, {Henrot-Versill{\'e}}, {Hervier},
  {Holmes}, {Jones}, {Lamarre}, {Lami}, {Lange}, {Lefebvre}, {Leriche},
  {Leroy}, {Macias-Perez}, {Maciaszek}, {Maffei}, {Mahendran}, {Mansoux},
  {Marty}, {Masi}, {Mercier}, {Miville-Deschenes}, {Montier}, {Nicolas},
  {Noviello}, {Perdereau}, {Piacentini}, {Piat}, {Plaszczynski},
  {Pointecouteau}, {Pons}, {Ponthieu}, {Puget}, {Rambaud}, {Renault},
  {Renault}, {Rioux}, {Ristorcelli}, {Rosset}, {Savini}, {Sudiwala}, {Torre},
  {Tristram}, {Vall{\'e}e}, {Veneziani}, \& {Yvon}}]{pajot2010}
{Pajot}, F., {Ade}, P.~A.~R., {Beney}, J., {et~al.} 2010, \aap, 520, A10+

\bibitem[{{Planck Collaboration ES}(2013)}]{planck2013-p28}
{Planck Collaboration ES}. 2013, {The Explanatory Supplement to the Planck 2013
  results} ({ESA})

\bibitem[{{Planck Collaboration I}(2013)}]{planck2013-p01}
{Planck Collaboration I}. 2013, Submitted to \aap, [arXiv:astro-ph/1303.5062]

\bibitem[{{Planck Collaboration II}(2013)}]{planck2013-p02}
{Planck Collaboration II}. 2013, Submitted to \aap, [arXiv:astro-ph/1303.5063]

\bibitem[{{Planck Collaboration VI}(2013)}]{planck2013-p03}
{Planck Collaboration VI}. 2013, Submitted to \aap, [arXiv:astro-ph/1303.5067]

\bibitem[{{Planck Collaboration VII}(2013)}]{planck2013-p03c}
{Planck Collaboration VII}. 2013, Submitted to \aap, [arXiv:astro-ph/1303.5068]

\bibitem[{{Planck Collaboration VIII}(2013)}]{planck2013-p03f}
{Planck Collaboration VIII}. 2013, Submitted to \aap,
  [arXiv:astro-ph/1303.5069]

\bibitem[{{Planck Collaboration XI}(2013)}]{planck2013-p01a}
{Planck Collaboration XI}. 2013, In preparation

\bibitem[{{Planck Collaboration XII}(2013)}]{planck2013-p06}
{Planck Collaboration XII}. 2013, Submitted to \aap, [arXiv:astro-ph/1303.5072]

\bibitem[{{Planck Collaboration XIII}(2013)}]{planck2013-p03a}
{Planck Collaboration XIII}. 2013, Submitted to \aap,
  [arXiv:astro-ph/1303.5073]

\bibitem[{{Planck Collaboration XIV}(2013)}]{planck2013-pip88}
{Planck Collaboration XIV}. 2013, Submitted to \aap, [arXiv:astro-ph/1303.5074]

\bibitem[{{Planck Collaboration XIX}(2011)}]{planck2011-7.0}
{Planck Collaboration XIX}. 2011, \aap, 536, A19

\bibitem[{{Planck Collaboration XVII}(2011)}]{planck2011-6.4b}
{Planck Collaboration XVII}. 2011, \aap, 536, A17

\bibitem[{{Planck Collaboration XXI}(2013)}]{planck2013-p05b}
{Planck Collaboration XXI}. 2013, Submitted to \aap, [arXiv:astro-ph/1303.5081]

\bibitem[{{Planck Collaboration XXIX}(2013)}]{planck2013-p05a}
{Planck Collaboration XXIX}. 2013, Submitted to \aap,
  [arXiv:astro-ph/1303.5089]

\bibitem[{{Planck Collaboration XXVIII}(2013)}]{planck2013-p05}
{Planck Collaboration XXVIII}. 2013, Submitted to \aap,
  [arXiv:astro-ph/1303.5088]

\bibitem[{{Planck Collaboration XXX}(2013)}]{planck2013-pip56}
{Planck Collaboration XXX}. 2013, Submitted to \aap, [arXiv:astro-ph/1309.0382]

\bibitem[{{Planck Collaboration XXXI}(2013)}]{planck2013-p06b}
{Planck Collaboration XXXI}. 2013, In preparation

\bibitem[{{Planck HFI Core Team}(2011)}]{planck2011-1.7}
{Planck HFI Core Team}. 2011, \aap, 536, A6

\bibitem[{{Rosset} {et~al.}(2010){Rosset}, {Tristram}, {Ponthieu}, {Ade},
  {Aumont}, {Catalano}, {Conversi}, {Couchot}, {Crill}, {D{\'e}sert}, {Ganga},
  {Giard}, {Giraud-H{\'e}raud}, {Ha{\"i}ssinski}, {Henrot-Versill{\'e}},
  {Holmes}, {Jones}, {Lamarre}, {Lange}, {Leroy}, {Mac{\'{\i}}as-P{\'e}rez},
  {Maffei}, {de Marcillac}, {Miville-Desch{\^e}nes}, {Montier}, {Noviello},
  {Pajot}, {Perdereau}, {Piacentini}, {Piat}, {Plaszczynski}, {Pointecouteau},
  {Puget}, {Ristorcelli}, {Savini}, {Sudiwala}, {Veneziani}, \&
  {Yvon}}]{rosset2010}
{Rosset}, C., {Tristram}, M., {Ponthieu}, N., {et~al.} 2010, \aap, 520, A13+

\bibitem[{{Rybicki} \& {Lightman}(1986)}]{1986RybickiLightman}
{Rybicki}, G.~B. \& {Lightman}, A.~P. 1986, {Radiative Processes in
  Astrophysics} (New York: John Wiley \& Sons)

\bibitem[{Shannon(1948)}]{Shannon1948}
Shannon, C.~E. 1948, The Bell System Technical Journal, 27, 379

\bibitem[{{Sunyaev} \& {Zeldovich}(1980)}]{SZ1980}
{Sunyaev}, R.~A. \& {Zeldovich}, I.~B. 1980, \araa, 18, 537

\bibitem[{{Zonca} {et~al.}(2009){Zonca}, {Franceschet}, {Battaglia}, {Villa},
  {Mennella}, {D'Arcangelo}, {Silvestri}, {Bersanelli}, {Artal}, {Butler},
  {Cuttaia}, {Davis}, {Galeotta}, {Hughes}, {Jukkala}, {Kilpi{\"a}},
  {Laaninen}, {Mandolesi}, {Maris}, {Mendes}, {Sandri}, {Terenzi}, {Tuovinen},
  {Varis}, \& {Wilkinson}}]{zonca2009}
{Zonca}, A., {Franceschet}, C., {Battaglia}, P., {et~al.} 2009, Journal of
  Instrumentation, 4, 2010

\end{thebibliography}


\raggedright

\end{document}